\documentclass[11pt]{article}
\usepackage{amsmath,amssymb,mathtools}
\usepackage{mathrsfs}
\usepackage{bbold}
\usepackage{color}
\usepackage{graphicx}
\usepackage{slashed}
\usepackage{cite}
\usepackage[small]{caption}
\usepackage{subcaption}
\usepackage{tikz}
\usepackage{setspace}

\definecolor{phaseBlue}{RGB}{31,102,184}
\definecolor{phaseOrange}{RGB}{204,85,20}

\newcommand*\de{\partial}
\newcommand*\al{\alpha}
\newcommand*\di{\delta}
\newcommand*\la{\lambda}
\newcommand*\si{\sigma}
\newcommand*\sgn{{\rm sgn}\,}
\newcommand*\ZZ{\mathbb{Z}}
\newcommand*\calM{\mathcal{M}}
\newcommand*\Tr{\mathop{\rm tr}}
\newcommand*\tr{\mathop{\rm tr}}

\makeatletter
\def\section{\@startsection {section}{1}{\z@}{-3.5ex plus -1ex minus
 -.2ex}{2.3ex plus .2ex}{\large\bf}}
\def\subsection{\@startsection{subsection}{2}{\z@}{-3.25ex plus -1ex
minus -.2ex}{1.5ex plus .2ex}{\normalsize\bf}}
\@addtoreset{equation}{section}
\makeatother

\usepackage[colorlinks,linkcolor=black,citecolor=blue,urlcolor=blue,linktocpage,pagebackref]{hyperref}
\allowdisplaybreaks
\graphicspath{{figures/}}
\begin{document}

\thispagestyle{empty}
\begin{flushright}
    \texttt{CERN-TH-2026-247}
\end{flushright}

\begin{center}

	\vspace*{-.6cm}

	\begin{center}

		\vspace*{1.1cm}

	{\centering \Large\textbf{Maxwell--Chern--Simons-Matter in AdS$_3$}}

	\end{center}

	\vspace{0.8cm}
	{\bf Dean Carmi$^{a,b,c}$, Riccardo Ciccone$^{a,c}$}

	\vspace{1.cm}

 ${}^a\!\!$
	{\em Department of Physics and\\ Haifa Research Center for Theoretical
Physics and Astrophysics,\\ University of Haifa, 31905 Haifa, Israel}

	\vspace{.3cm}
${}^b\!\!$
	{\em Department of Mathematics and Physics,\\ University of Haifa at Oranim,
36006 Kiryat Tivon, Israel}

	\vspace{.3cm}
 ${}^c\!\!$
	{\em Theoretical Physics Department,\\ CERN, 1211 Geneva 23, Switzerland}

\end{center}

\vspace{1cm}

\centerline{\bf Abstract}
\vspace{2 mm}
\begin{quote}
We study Maxwell--Chern--Simons theory on AdS$_3$ coupled to $N$ complex scalars or Dirac fermions, including sextic and Gross--Neveu interactions, respectively. First, we derive the bulk Maxwell--Chern--Simons propagator, clarifying how Dirichlet boundary conditions fix the contributions of the massive and flat sectors. Upon coupling to the matter sector, we work at large $N$ with $\kappa=k/N$ and $\alpha=Ne^2$ fixed, where $k$ is the Chern--Simons level. We determine the corresponding phase diagrams, and we resum matter insertions to obtain the photon propagator at leading order in large $N$, exactly in $\alpha$ and $\kappa$. Its poles determine the spin-one boundary spectrum, split into two helicity sectors by the Chern--Simons interaction. With Dirichlet boundary conditions, the unHiggsed scalar phase contains a protected chiral current, which is lifted by Higgsing. In the fermionic phases, the current persists for nonzero effective Chern--Simons level, with helicity determined by the induced level shift. At bulk conformal points, we extract large-$N$ BCFT data for the critical and tricritical scalar theories, as well as for the regular fermion and Gross--Neveu theories.
\end{quote}

\newpage

\tableofcontents\newpage

\section{Introduction}
Three-dimensional Abelian gauge theories provide continuum descriptions of superconducting transitions and deconfined quantum critical points \cite{Ginzburg:1950sr,Senthil:2003eed}, and enter particle-vortex \cite{Dasgupta:1981zz,Peskin:1977kp} and boson-fermion dualities \cite{Polyakov:1988md,Giombi:2011kc,Aharony:2012nh,Karch:2016sxi,Seiberg:2016gmd}. A central question is whether their transitions are continuous, with an interacting CFT governing the long-distance behavior, or whether fluctuations instead drive a first-order transition. A Chern--Simons interaction enriches this problem by allowing for topological phases and by giving the photon a topological mass \cite{Deser:1982vy}. When coupled to a gapped matter sector with unHiggsed gauge symmetry, the low-energy theory reduces to a pure Chern--Simons theory \cite{Witten:1988hf}, whereas gapless charged matter can support interacting Chern--Simons-matter critical points \cite{Chen:1993cd}.

A possible handle on low-energy properties of field theories can be obtained by placing them on a fixed AdS background. The AdS background is maximally symmetric, its radius $L$ sets the scale at which curvature affects the infrared dynamics, and the infinite volume still allows phase transitions and spontaneous symmetry breaking at finite $L$ \cite{Callan:1989em}, with flat space approached as $L\to\infty$. At the same time, AdS gives access to boundary conformal observables that depend on the bulk couplings and on $L$. For boundary conditions preserving the AdS isometries, bulk operators admit an expansion in boundary operators, whose correlators obey crossing symmetry; unlike in flat space, this holds also away from bulk criticality \cite{Paulos:2016fap,Carmi:2018qzm}. In the large-radius limit, boundary correlators determine flat-space scattering amplitudes \cite{Polchinski:1999ry,Penedones:2010ue}. At a bulk conformal point, the Weyl map to flat half-space identifies these observables with BCFT data. Distinct conformal boundary conditions then define different boundary universality classes, characterized by their spectra and bulk-to-boundary OPE coefficients \cite{Cardy:1984bb,McAvity:1995zd}, and bulk conformal symmetry further organizes bulk two-point functions in bulk conformal blocks, to be compared with the boundary-channel expansion \cite{McAvity:1995zd,Liendo:2012hy,Carmi:2018qzm}.

At bulk conformal points, calculations in AdS have yielded boundary data for scalar and fermionic vector models and gauge theories, including higher-loop results in the $\epsilon$-expansion \cite{Giombi:2020rmc,Giombi:2021cnr,Giombi:2025pxx,DeCesare:2026rup}. Away from criticality, large-$N$ resummations and Hamiltonian truncation give access to interacting spectra \cite{Carmi:2018qzm,Ankur:2023lum,Hogervorst:2021spa}, while analytic and numerical bootstrap methods constrain bulk RG flows and their boundary data \cite{Antunes:2021abs,Meineri:2023mps,Lauria:2023uca}. The variation of boundary data under bulk relevant deformations has also been formulated as a system of ordinary differential equations \cite{Loparco:2026fki,DeCesare:2026bor}.

Gauge theories in AdS have been studied in connection with mass generation, confinement, and symmetry breaking \cite{Aharony:2012jf,Copetti:2023sya,Ciccone:2024guw,Ciccone:2025dqx,DiPietro:2025ozw}, including confining flux tubes and their effective string description \cite{Gabai:2025hwf,Gabai:2026myo}, and the dressing and screening of charged operators \cite{Ankur:2026ylr}. In these theories, the boundary condition of the gauge field is an additional important choice. With Dirichlet boundary conditions, gauge transformations approaching a constant at the boundary act as a global symmetry, with a conserved boundary current $J$ dual to the bulk gauge field. In three bulk dimensions, the boundary operator $J^2$ is classically marginal, and the gauge interaction induces a nonzero beta function for its coupling, obstructing Dirichlet conformal boundary conditions \cite{Ankur:2023lum}. This is an instance of merger and annihilation of boundary fixed points, here occurring as the boundary dimension is lowered to two \cite{Kaplan:2009kr,Gorbenko:2018ncu,Lauria:2023uca}.

A nonzero Chern--Simons level changes this picture. With Dirichlet boundary conditions, the Maxwell--Chern--Simons gauge field decomposes into a massive propagating sector and a topological flat sector, associated at the boundary with a spin-one operator of helicity $-{\rm sgn}(k)$ and dimension above one, and a chiral current of helicity $+{\rm sgn}(k)$, respectively \cite{Andrade:2011sx}. Only one chirality of the current remains conserved, and the would-be marginal scalar built from both chiralities is replaced by an operator of dimension above two. Neumann boundary conditions, which gauge this boundary symmetry, are instead inconsistent at nonzero level because of the chiral anomaly, unless additional degrees of freedom are added to cancel it \cite{Callan:1984sa,Aitken:2017nfd}.

In this paper we study Maxwell--Chern--Simons theory on Euclidean AdS$_3$ at level $k$, coupled to $N$ complex scalars or Dirac fermions of unit charge. We take $N,|k|\to\infty$ with $\kappa=k/N$ and $\alpha=Ne^2$ fixed. The scalars have quartic and sextic singlet interactions, and the fermions a Gross--Neveu quartic interaction, supplemented by a potential up to cubic order for the corresponding auxiliary field. The flat-space phase structure and stability of these theories were analyzed in \cite{DiPietro:2023zqn}. We repeat the analysis in AdS: at fixed bulk couplings, each unitary stationary point of the large-$N$ effective potential defines an AdS-invariant boundary condition; the lightest scalar singlet is irrelevant at a local minimum and relevant at a maximum, becoming marginal when the two merge \cite{Kaplan:2009kr,Gorbenko:2018ncu}.

We construct the pure Maxwell--Chern--Simons propagator with Dirichlet boundary conditions, and use the parity-even and parity-odd spin-one harmonics of \cite{Costa:2014kfa,Bhattacharya:2025udq} to resum matter bubbles. This gives the dressed photon propagator exactly in $\alpha$ and $\kappa$ at leading order in large $N$, including the Higgs contribution when the scalar condenses. For fermions, we fix the local parity-odd ambiguity with a parity-preserving Pauli--Villars prescription. Together with the matter saddle point equations, the dressed correlators in the various AdS phases determine their scalar-singlet and spin-one boundary spectra, with operator dimensions and OPE weights extracted from poles and residues.

The resulting spectra distinguish Higgsing from the generation of a Chern--Simons term by massive fermions. In the unHiggsed scalar phase, the dimension-one chiral current remains protected, while the other spin-one operators form two helicity towers whose dimensions vary with the couplings; in the helicity sector of the massive Maxwell--Chern--Simons mode, the tower displays a resonance. Higgsing instead lifts the protected current, and the two massive vector modes present at weak coupling produce resonances in both helicity sectors. In the fermionic phases, the parity-odd polarization gives the familiar induced level shift \cite{Redlich:1983dv}, which in our conventions is $k_{\rm eff}=k+\eta N/2$, with $\eta=\pm1$ labeling the fermion boundary condition. For $k_{\rm eff}\ne0$, the protected current persists and its helicity is fixed by $\operatorname{sgn}(k_{\rm eff})$.

We then take the bulk conformal limits, sending $\alpha\to\infty$ and tuning the matter couplings to obtain the critical and tricritical scalar theories, the regular fermion theory, and the Gross--Neveu critical family. We characterize ordinary, special, and normal boundary conditions for scalars, and standard and alternate flavor-preserving boundary conditions for fermions, extending previous large-$N$ BCFT analyses of ungauged vector models \cite{Carmi:2018qzm,Herzog:2020lel,Giombi:2021cnr,Metlitski:2020cqy}. The bulk-channel expansions of scalar-singlet and topological-current correlators determine bulk normalizations and OPE coefficients, which agree across boundary conditions. The saddle values fix scalar one-point coefficients, while the scalar correlators determine boundary spectra and bulk-to-boundary OPE data, including those of the displacement operator. The spin-one boundary spectra follow from the dressed gauge propagators.

These results are obtained at strict large $N$. At subleading order, the running of the marginal bulk interactions restricts the conformal families through the combined requirements of vanishing beta functions and vacuum stability \cite{DiPietro:2023zqn}. Our analysis is also restricted to operators neutral under the topological symmetry $U(1)_{\rm top}$. Extending it to monopole operators, or to non-Abelian gauge groups, would give access to further observables relevant to three-dimensional dualities \cite{Aitken:2017nfd,Seiberg:2016gmd,Hsin:2016blu}.

The paper is organized as follows. Section~\ref{sec:MCS} discusses the boundary conditions and propagators of Maxwell--Chern--Simons theory in AdS$_3$. Sections~\ref{sec:scalar} and~\ref{sec:fermion} analyze scalar and fermionic matter, respectively, including the AdS phases, dressed correlators, and boundary spectra. Section~\ref{sec:conformal-points} presents the BCFT data at bulk conformal points, and section~\ref{sec:outlook} discusses open directions. Appendix~\ref{app:harmonic-functions} collects the harmonic functions and spectral identities, appendix~\ref{app:matter-bubbles} presents the matter bubble functions, while appendix~\ref{app:bulk-channel} explains the bulk-channel expansions used to extract bulk conformal data.

\section{Maxwell--Chern--Simons in AdS$_3$}
\label{sec:MCS}
In this section we discuss Maxwell--Chern--Simons theory in the absence of matter. We work on Euclidean AdS$_3$ throughout, so the boundary dimension is $d=2$. The action is
\begin{equation}
S_{\rm MCS}=\frac{1}{4e^2}\int d^3x\,\sqrt g\,F_{\mu\nu}F^{\mu\nu}-\frac{ik}{4\pi}\int d^3x\,\sqrt g\,\epsilon^{\mu\nu\rho}A_\mu\nabla_\nu A_\rho\,,
\end{equation}
where $k\in\mathbb Z$ is the Chern--Simons level,\footnote{The sign of the Chern--Simons term is chosen to reproduce the coefficient $+\frac{k}{4\pi}$ upon Wick rotation to Lorentzian signature.} and $\epsilon^{\mu\nu\rho}$ is the covariant Levi-Civita tensor. We use Poincaré coordinates $x^\mu=(x^i,z)$, with $z>0$ and $i=1,2$, in which
\begin{equation}
ds^2=L^2\frac{dz^2+\delta_{ij}dx^i dx^j}{z^2}=L^2\frac{dz^2+dw\,d\bar w}{z^2}\,.
\end{equation}
Here $w=x^1+ix^2$ and $\bar w=x^1-ix^2$ are complex coordinates along the boundary at $z=0$. We choose the orientation $\epsilon_{x^1x^2z}=L^3/z^3$ and set the AdS radius $L=1$ in the following.

The Chern--Simons coupling gives the bulk vector a mass without breaking gauge invariance. Defining $m=ke^2/(2\pi)$, the variation of the action gives the bulk equations
\begin{equation}
\nabla_\mu F^{\mu\nu}+\frac{im}{2}\epsilon^{\nu\rho\sigma}F_{\rho\sigma}=0\,,
\end{equation}
together with a boundary term that we discuss below. To identify the bulk degrees of freedom, it is convenient to introduce the dual field strength and the curl operator,
\begin{equation}
(*F)^\mu=\frac12\epsilon^{\mu\nu\rho}F_{\nu\rho}=\mathcal C[A]^\mu\,,\qquad \mathcal C[X]^\mu=\epsilon^{\mu\nu\rho}\nabla_\nu X_\rho\,.
\end{equation}
The Bianchi identity implies $\nabla_\mu(*F)^\mu=0$, while the equation of motion becomes
\begin{equation}
(\mathcal C-im)(*F)=(\mathcal C-im)\mathcal C[A]=0\,.
\end{equation}
Applying $\mathcal C+im$ gives the Proca equation for $*F$, with mass squared $m^2$. The first-order equation further selects one helicity, so the theory contains a single propagating bulk degree of freedom.

We decompose the gauge field as
\begin{equation}
A_\mu=\hat A_\mu+A_\mu^\flat\,,\qquad \nabla^\mu\hat A_\mu=0\,,\qquad \mathcal C[A^\flat]=0\,.
\end{equation}
Gauge transformations act on the flat connection as $A_\mu^\flat\to A_\mu^\flat+\nabla_\mu\Lambda$, leaving $\hat A_\mu$ invariant. This decomposition has a residual ambiguity associated with gradients of scalar harmonic functions, which we assign to $A^\flat$. Along solutions of the equations of motion, and for $m\neq0$, one has
\begin{equation}
(\mathcal C-im)\hat A=0\,.
\end{equation}
We refer to $\hat A$ and $A^\flat$ as the massive and flat sectors, respectively. The massive sector carries the propagating degree of freedom, while the flat sector is locally pure gauge but its boundary value supports a chiral current. We must therefore specify boundary conditions for both sectors.

\subsection{Boundary conditions}
\label{subsec:boundary-conditions}
We follow the exposition of \cite{Andrade:2011sx}, adapting it to our conventions. The variation of the action contains the boundary contribution
\begin{equation}\label{eq:deltaSMCSboundary}
\delta S_{\rm MCS}^{\partial}=\int_{z=z_0}d^2x\,\sqrt{h}\,\delta A_i\left[-\frac1{e^2}n_\mu F^{\mu i} + \frac{ik}{4\pi} n_\mu\epsilon^{\mu ij} A_j\right] =\int_{z=z_0}d^2x\,\delta A_i\left[-\frac{z}{e^2}F_{zi}+\frac{ik}{4\pi}\varepsilon^{ij}A_j\right]\,,
\end{equation}
where $h_{ij}=\delta_{ij}/z^2$ is the induced two-dimensional metric, with $\sqrt{h}=z^{-2}$; we introduced the inward-pointing vector $n^\mu=z\delta^\mu_z$ normal to the boundary; $\epsilon^{\mu ij}=z^3\delta^\mu_z\varepsilon^{ij}$, with $\varepsilon^{12}=1$; and the cutoff $z_0$ is eventually taken to zero. The massive and flat fields have asymptotic expansions
\begin{equation}
\hat A_i=z^{-|m|}\hat a_i+z^{|m|}\hat b_i+\cdots\,,\qquad A_i^\flat=a_i^\flat+\cdots\,.
\end{equation}

Before choosing which boundary modes to fix, we must remove a divergence in the variation. Using the on-shell relation $zF_{zi}=im\varepsilon_{ij}\hat A_j$, \eqref{eq:deltaSMCSboundary} contains a term mixing massive and flat sectors,
\begin{equation}
\delta S_{\rm MCS}^{\partial}\supset\frac{ik}{4\pi}\,\int_{z=z_0}d^2x\,\delta(\varepsilon^{ij}\hat A_iA_j^\flat)\,.
\end{equation}
For nonzero $\hat a_i$, this contains a divergent term proportional to $z_0^{-|m|}\delta(\varepsilon^{ij}\hat a_i a_j^\flat)$. We subtract it on-shell with the following boundary counterterm,
\begin{equation}
S_{\rm ct}=\frac{1}{2e^2}\int_{z=z_0}d^2x\,\sqrt h\,A_i n_\mu F^{\mu i}\,.
\end{equation}
The remaining variation separates into massive and flat terms. We select Dirichlet boundary conditions by further adding
\begin{equation}\label{eq:SDfin}
S_{\rm fin}^{\rm D}=-\frac{|k|}{8\pi}\int_{z=z_0}d^2x\,\sqrt h\,h^{ij}\left(\hat A_i\hat A_j+A_i^\flat A_j^\flat\right)\,.
\end{equation}
Writing $S_{\rm D}=S_{\rm MCS}+S_{\rm ct}+S_{\rm fin}^{\rm D}$, we obtain, on-shell,
\begin{equation}
\delta S_{\rm D}=\frac{k-|k|}{2\pi}\int d^2x\,\left(\hat A_w\,\delta\hat A_{\bar w}+A_{\bar w}^\flat\,\delta A_w^\flat\right)-\frac{k+|k|}{2\pi}\int d^2x\,\left(\hat A_{\bar w}\,\delta\hat A_w+A_w^\flat\,\delta A_{\bar w}^\flat\right)\,.
\end{equation}
For $k>0$, we fix the massive source $\hat a_w$ and the flat-sector source $a_{\bar w}^\flat$; for $k<0$, we fix $\hat a_{\bar w}$ and $a_w^\flat$. We set these sources to zero in the following.\footnote{For $|m|\geq1$, generic sources require further local counterterms to cancel divergences from the derivative falloffs $z^{2n-|m|}$, including logarithms at integer $|m|$. For $k>0$ and $0<|m|<2$, $|m|\neq1$, the additional term is $S'_{\rm ct}=\frac{1}{e^2(1-|m|)}\int_{z=z_0}d^2x\,z^2\hat A_w\partial_{\bar w}^2\hat A_w$, which vanishes as $z_0\to0$ for $|m|<1$; for $k<0$, exchange $w$ and $\bar w$. At $|m|=1$, its coefficient becomes $2\log(\mu z_0)/e^2$, with $\mu$ the RG scale. Higher-derivative counterterms are needed for $|m|\geq2$. All these terms are local in the sources and vanish at zero sources.} The surviving modes define the boundary operators through
\begin{equation}
\hat A_i(x,z)\underset{z\to0}{\sim}e^2z^{\Delta_V-1}V_i^{-\varsigma}(x)\,,\qquad A_i^\flat(x,z)\underset{z\to0}{\sim}\frac{2\pi}{|k|}J_i^\varsigma(x)\,,
\end{equation}
where the superscript denotes the helicity, namely $s=h-\bar h=\pm1$ for a boundary spin-one operator with weights $(h,\bar h)$, and we introduced
\begin{equation}
    \varsigma \equiv \sgn (k)\,.
\end{equation}
Explicitly, the massive sector $\hat A_\mu$ corresponds to a boundary spin-one operator $V_i^{-\varsigma}$ of dimension $\Delta_V=1+|m|$. Its helicity $-\varsigma$ is fixed by the equations of motion, and its dimension lies strictly above the spin-one unitarity bound for nonzero $k$. The flat sector corresponds to a conserved current $J_i^\varsigma$ of dimension $\Delta_J=1$. For $k>0$, its only nonzero component is $J_w^+$, and flatness implies $\partial_{\bar w}J_w^+=0$. For $k<0$, the current is instead antiholomorphic, with $\partial_wJ_{\bar w}^-=0$.

Neumann boundary conditions would promote the source of $J_i^\varsigma$ to a dynamical boundary gauge field, thereby gauging a boundary chiral $U(1)$ symmetry. They therefore require additional boundary degrees of freedom that cancel the corresponding chiral anomaly \cite{Callan:1984sa,Aitken:2017nfd}, which we do not introduce. At $k=0$ this obstruction is absent, and pure Maxwell theory admits Neumann, but not Dirichlet, boundary conditions, as the latter disappear through merger and annihilation of boundary fixed points due to the marginal operator $J^2$ \cite{Ankur:2023lum}. Correspondingly, in our case the Dirichlet boundary condition has no smooth $k\to0$ limit: as $m\to0$ the boundary scalar built from $J_i^\varsigma$ and $V_i^{-\varsigma}$, of dimension $2+|m|$, approaches marginality. The alternative quantization of the massive sector, sometimes also called Neumann, instead exchanges its source and response and gives a spin-one operator of dimension $1-|m|$, below the unitarity bound \cite{Andrade:2011sx}.

\subsection{Bulk propagators}
\label{subsec:bulk-propagators}
We add the covariant gauge-fixing term
\begin{equation}
S_{\rm gf}=\frac{1}{2\xi e^2}\int d^3x\,\sqrt g\,(\nabla_\mu A^\mu)^2\,,
\end{equation}
and normalize the propagator as
\begin{equation}
\langle A_\mu(x)A_{\nu'}(x')\rangle=e^2G_A(x,x')_{\mu\nu'}\,.
\end{equation}
Its equation of motion is
\begin{equation}
(\mathcal K_\xi G_A)_{\mu\nu'}=\left[-\delta_\mu{}^\rho(\nabla^2+2)+\left(1-\frac1\xi\right)\nabla_\mu\nabla^\rho-im\epsilon_\mu{}^{\lambda\rho}\nabla_\lambda\right]G_{A,\rho\nu'}=g_{\mu\nu'}\delta\,,
\end{equation}
where $\mathcal K_\xi$ is the equation-of-motion operator,
\begin{equation}
\mathcal K_\xi=\mathcal C(\mathcal C-im)-\frac1\xi\nabla\,\mathrm{Div}\,.
\end{equation}
An AdS$_3$-invariant vector propagator contains two parity-even structures and one parity-odd structure,
\begin{equation}\label{eq:GAdeco}
G_{A,\mu\nu'}=-g_0(u)\nabla_\mu\nabla_{\nu'}u+g_1(u)\nabla_\mu u\,\nabla_{\nu'}u+g_2(u)\mathcal E_{\mu\nu'}\,,
\end{equation}
with
\begin{equation}
u=\frac{|w-w'|^2+(z-z')^2}{2zz'}\,,\qquad \mathcal E_{\mu\nu'}=\epsilon_\mu{}^{\lambda\rho}\nabla_\lambda\nabla_{\nu'}u\,\nabla_\rho u\,.
\end{equation}
We also use the geodesic distance $r=\operatorname{arcosh}(1+u)$. In these variables, the scalar propagator is
\begin{equation}\label{eq:GscalarAdS3}
G_\Delta(u)=\frac{e^{-(\Delta-1)r}}{4\pi\sinh r}\,.
\end{equation}

We construct $G_A$ using the scalar and transverse spin-one harmonics introduced in \cite{Costa:2014kfa,Bhattacharya:2025udq}, which we review in appendix~\ref{app:harmonic-functions}. In AdS$_3$, the helicity combinations $\Omega_\nu^\pm=\Omega_\nu^E\pm\Omega_\nu^O$ of the parity-even and parity-odd transverse harmonics diagonalize the curl, \eqref{eq:helicity-harmonics}, and satisfy the completeness relation \eqref{eq:completeness-AdS3}, $\int_{\mathbb R}d\nu\,\Omega_\nu^\pm=I-Q$, with $I_{\mu\nu'}=g_{\mu\nu'}\delta(x,x')$ and $Q_{\mu\nu'}=\nabla_\mu\nabla_{\nu'}G_2(x,x')$. Inverting the kinetic operator gives
\begin{equation}\label{eq:GAspectral}
G_A=\int_Cd\nu\,\gamma_+(\nu)\,\Omega_\nu^++\int_{\mathbb R}d\nu\,\gamma_L(\nu)\,\nabla\nabla'\Omega_\nu\,,
\end{equation}
where
\begin{equation}
\gamma_+(\nu)=\frac1{\nu(\nu-im)}\,,\qquad \gamma_L(\nu)=\frac{\xi}{(\nu^2+1)^2}\,.
\end{equation}
The contour $C$ follows the real axis from left to right, with a prescription at the origin that will be determined below by the boundary conditions. We first keep the prescription general.

To isolate its effect, we relate the helicity-plus integral to the even/odd representation,
\begin{equation}
\int_Cd\nu\,\gamma_+\Omega_\nu^+=\int_{\mathbb R}d\nu\,\left(\gamma_E\Omega_\nu^E+\gamma_O\Omega_\nu^O\right)+c\,\frac{\pi}{m}\Omega_0^E\,,
\end{equation}
with
\begin{equation}
\gamma_E(\nu)=\frac1{\nu^2+m^2}\,,\qquad \gamma_O(\nu)=\frac{im}{\nu(\nu^2+m^2)}\,,
\end{equation}
and $c=+1$, $0$, $-1$ when $C$ passes above the origin, takes its principal value, or passes below it, respectively. Indeed, $\gamma_+=\gamma_E+\gamma_O$, and the difference between the two integrands is $\gamma_E\Omega_\nu^O+\gamma_O\Omega_\nu^E$. This is odd in $\nu$, with residue $i\Omega_0^E/m$ at the origin. The even/odd integrand is regular there because $\Omega_\nu^O$ vanishes linearly with $\nu$. The bitensor added by the contour prescription is
\begin{equation}
\Omega_0^E=-\frac1{4\pi^2}\nabla\nabla'(r\coth r)\,,\qquad {\rm Div}\,\Omega_0^E=\mathcal C\,\Omega_0^E=\mathcal K_\xi \Omega_0^E=0\,.
\end{equation}

To evaluate the spectral integral, we express the harmonics through Proca propagators, \eqref{eq:OmegaE-Proca} and \eqref{eq:OmegaO-Proca}, where $\Pi_\Delta$ satisfies $\bigl[\mathcal C^2+(\Delta-1)^2\bigr]\Pi_\Delta=I$. Substituting the harmonic decomposition into the even/odd spectral representation gives
\begin{equation}\label{eq:MCS-split-integral}
\int_{\mathbb R}d\nu\,\left(\gamma_E\Omega_\nu^E+\gamma_O\Omega_\nu^O\right)=\frac{i}{2\pi}\int_{\mathbb R}d\nu\,\frac{\nu+\frac{im}{\nu}\mathcal C}{\nu^2+m^2}\left(\Pi_{1+i\nu}-\Pi_{1-i\nu}\right)\,.
\end{equation}
We close the contour in the lower half-plane for the term containing $\Pi_{1+i\nu}$ and in the upper half-plane for the term containing $\Pi_{1-i\nu}$. The residues at the massive poles $\nu=\pm i|m|$ give
\begin{equation}
\widehat G=\left(1+\frac{\mathcal C}{im}\right)\Pi_{1+|m|}\,.
\end{equation}

The two Proca terms in~\eqref{eq:MCS-split-integral} are separately singular at the origin. Their common prescription, together with
\begin{equation}
\Pi_{1+i\nu}=-\frac{Q}{\nu^2}-\frac{i\pi}{\nu}\Omega_0^E+O(1)\,,\qquad \mathcal C\Pi_\Delta=G_\Delta'(u)\mathcal E\,,
\end{equation}
gives
\begin{equation}
\int_{\mathbb R}d\nu\,\left(\gamma_E\Omega_\nu^E+\gamma_O\Omega_\nu^O\right)=\widehat G+G_{\rm CS}-\frac{Q}{m^2}\,,\qquad G_{\rm CS}=\frac{i}{m}G_1'(u)\mathcal E\,.
\end{equation}

The longitudinal integral in~\eqref{eq:GAspectral} is
\begin{equation}\label{eq:GApar}
G_A^\parallel=\xi\nabla\nabla'(G_2\star G_2)=\xi \int_{\mathbb{R}}d\nu\,\frac{\nabla \nabla'\Omega_\nu}{(\nu^2+1)^2}\,,
\end{equation}
where $(F\star G)(x,x')=\int d^3y\,\sqrt{g_y}\,F(x,y)G(y,x')$ denotes the convolution.

The Proca equation and the completeness relations imply
\begin{equation}
(\mathcal C-im)\widehat G=\frac{I}{im}\,,\qquad \mathrm{Div}\,\widehat G=-\frac{1}{m^2}\nabla'\delta\,,\qquad \mathcal C G_{\rm CS}=\frac{i}{m}(I-Q)\,.
\end{equation}
Together with $\mathcal K_\xi G_A^\parallel=Q$ and $\mathcal K_\xi Q=\frac1\xi \nabla \nabla'\delta$, these identities give
\begin{equation}
\mathcal K_\xi\left(\widehat G+G_{\rm CS}-\frac{Q}{m^2}+G_A^\parallel\right)=I\,.
\end{equation}
The Chern--Simons contribution is essential as it supplies the source term, and $\widehat G$ alone does not invert the gauge kinetic operator.

The remaining homogeneous contribution is fixed by the boundary chirality. For $k>0$, $G_{\rm CS}+\pi\Omega_0^E/m$ has vanishing $\bar w$ boundary component; for $k<0$, $G_{\rm CS}-\pi\Omega_0^E/m$ has vanishing $w$ component.\footnote{The principal-value prescription, as in \cite{Bhattacharya:2025udq}, keeps both boundary chiralities in the flat sector, one of which leads to a negative-norm state irrespective of the sign of $k$, incompatibly with having a unitary conformal boundary condition.} The contour therefore passes above the origin for $\varsigma=+1$ and below it for $\varsigma=-1$, that is, it implements the prescription
\begin{equation}\label{eq:gamma-plus-prescription}
\gamma_+(\nu)=\frac1{(\nu+i\varsigma0)(\nu-im)}\,,
\end{equation}
and
\begin{equation}
G_A^{\rm D}=\widehat G+G_{\rm CS}-\dfrac{Q}{m^2}+\varsigma\dfrac{\pi}{m}\Omega_0^E+G_A^\parallel\,.
\end{equation}

Let us give the explicit form of the propagator, according to the decomposition \eqref{eq:GAdeco}. We first write the coefficient functions of $\widehat G$,
\begin{equation}
\begin{aligned}
\hat g_0&=\frac{e^{-|m|r}}{4\pi m^2\sinh^3r}\left(|m|\cosh r\,\sinh r+m^2\sinh^2r+1\right)\,,\\
\hat g_1&=\frac{e^{-|m|r}}{4\pi m^2\sinh^5r}\left(|m|\sinh r(\sinh^2r+3)+\cosh r(m^2\sinh^2r+3)\right)\,,\\
\hat g_2&=\frac{ie^{-|m|r}}{4\pi m\sinh^3r}(|m|\sinh r+\cosh r)\,.
\end{aligned}
\end{equation}
The Chern--Simons contribution has only a parity-odd coefficient,
\begin{equation}
g_2^{\rm CS}=\frac{i}{m}G_1'(u)=-\frac{i\cosh r}{4\pi m\sinh^3r}\,.
\end{equation}
The full coefficient functions are then
\begin{equation}
\begin{aligned}
g_0^{\rm D}={}&\hat g_0-\frac{1}{4\pi m^2\sinh^3r}+\frac{\varsigma}{4\pi m}\left(\frac{\cosh r}{\sinh^2r}-\frac{r}{\sinh^3r}\right)+\frac{\xi(r-\sinh r\,e^{-r})}{8\pi\sinh^3r}\,,\\
g_1^{\rm D}={}&\hat g_1-\frac{3\cosh r}{4\pi m^2\sinh^5r}+\frac{\varsigma}{4\pi m}\left(\frac{\cosh^2r+2}{\sinh^4r}-\frac{3r\cosh r}{\sinh^5r}\right)\\
&+\frac{\xi(3r\cosh r-3\sinh r+\sinh^2r\,e^{-r})}{8\pi\sinh^5r}\,,\\
g_2^{\rm D}={}&\hat g_2+g_2^{\rm CS}\,.
\end{aligned}
\end{equation}
The on-shell projectors $\widehat{\mathcal P}=\mathcal C/(im)$ and $\mathcal P^\flat=1-\mathcal C/(im)$ separate the two sectors. Acting on $G_A^{\rm D}$, they give
\begin{equation}
\frac{\langle\hat A\hat A\rangle}{e^2}=\widehat G-\frac{I}{m^2}\,,\qquad \frac{\langle\hat A A^\flat\rangle}{e^2}=\frac{I-Q}{m^2}\,,\qquad \frac{\langle A^\flat A^\flat\rangle}{e^2}=G_{\rm CS}+\varsigma\frac{\pi}{m}\Omega_0^E+\frac{Q-I}{m^2}+G_A^\parallel\,.
\end{equation}
At separated points, $\widehat G$ is therefore the massive-sector correlator, while the mixed correlator is contact plus longitudinal and vanishes after taking curls. Field-strength correlators thus probe $\widehat G$, up to contact terms, while exchange between conserved bulk currents also retains $G_{\rm CS}$.

\subsection{Boundary limits and correlators}
\label{subsec:boundary-limits}
We now extract the bulk-to-boundary propagators and boundary two-point functions, using the operator normalizations specified in section~\ref{subsec:boundary-conditions}. The limits are taken separately in the massive and flat sectors.

It is convenient to introduce embedding coordinates. Euclidean AdS$_3$ is the hypersurface $X^2=-1$, $X^0>0$, in $\mathbb R^{3,1}$, and a boundary point $y^i$ is a null ray, $P^2=0$. We parametrize them as
\begin{equation}
X^A=(X^+,X^-,X^i)=\frac1z(1,z^2+x^ix_i,x^i)\,,\qquad P^A=(1,y^iy_i,y^i)\,,\qquad X^\pm=X^0\pm X^3\,.
\end{equation}
Bulk indices are contracted with an auxiliary polarization vector $W$, satisfying $W^2=X\cdot W=0$. With the orientation $\epsilon_{0123}=1$ and the physical-space polarizations $W^A=\partial_\mu X^A\,w^\mu$, $W'^A=\partial_{\nu'}X'^A\,w'^{\nu'}$, one has
\begin{equation}
\epsilon(X,W,X',W')\equiv\epsilon_{ABCD}X^AW^BX'^CW'^D=\mathcal E_{\mu\nu'}w^\mu w'^{\nu'}\,.
\end{equation}
Analogously, boundary indices are contracted with polarization vectors $Z$, with $Z^2=P\cdot Z=0$.

The massive bulk-to-boundary propagator is
\begin{equation}
\langle V_i^{-\varsigma}(y)\hat A_\mu(x,z)\rangle=\lim_{z'\to0}(z')^{1-\Delta_V}\widehat G_{\mu i'}(x,z;y,z')\,.
\end{equation}
Using $\widehat G=(1+\mathcal C/(im))\Pi_{\Delta_V}$, we obtain
\begin{equation}
\langle V^{-\varsigma}(P,Z)\hat A(X,W)\rangle=\mathcal C_{\Delta_V,1}\frac{(-2P\cdot X)(Z\cdot W)+2(P\cdot W)(X\cdot Z)+2i\varsigma\epsilon(P,Z,X,W)}{(-2P\cdot X)^{\Delta_V+1}}\,,
\end{equation}
where
\begin{equation}\label{eq:CDeltaJ}
    \mathcal C_{\Delta,J}=\frac{\Delta+J-1}{2\pi(\Delta-1)}\,.
\end{equation}
Taking the second point to the boundary yields, in complex coordinates,
\begin{equation}
\begin{aligned}
\langle V_w^+(w,\bar w)V_w^+(0)\rangle&=-\frac{\mathcal C_{\Delta_V,1}}{e^2}\frac1{w^{\Delta_V+1}\bar w^{\Delta_V-1}}\,,\\
\langle V_{\bar w}^-(w,\bar w)V_{\bar w}^-(0)\rangle&=-\frac{\mathcal C_{\Delta_V,1}}{e^2}\frac1{w^{\Delta_V-1}\bar w^{\Delta_V+1}}\,,
\end{aligned}
\end{equation}
with the first realized for $k<0$ and the second for $k>0$. The flat-sector boundary correlator similarly depends on $\varsigma$. With the chosen normalization of $J_i^\varsigma$, we find
\begin{equation}
\langle J^\varsigma(P,Z)A^\flat(X,W)\rangle=\frac1{2\pi}\frac{(-2P\cdot X)(Z\cdot W)+2(P\cdot W)(X\cdot Z)-2i\varsigma\epsilon(P,Z,X,W)}{(-2P\cdot X)^2}\,.
\end{equation}
Its expression in physical space is a pure gradient: for example, for $k>0$,
\begin{equation}
\langle J_w^+(y)A^\flat_\mu(x,z)\rangle=-\frac1{2\pi}\partial_{y^w}\partial_{x^\mu}\log\!\left(|x-y|^2+z^2\right)\,.
\end{equation}
Taking the remaining point to the boundary gives
\begin{equation}
\langle J_w^+(w)J_w^+(0)\rangle=-\frac{|k|}{4\pi^2w^2}\,,\qquad \langle J_{\bar w}^-(\bar w)J_{\bar w}^-(0)\rangle=-\frac{|k|}{4\pi^2\bar w^2}\,,
\end{equation}
for the respective signs of $k$.

\section{Scalar matter}
\label{sec:scalar}
In this section we study Maxwell--Chern--Simons theory coupled to scalars,
\begin{equation}
    S_{\rm sMCS}=S_{\rm MCS} + S_\Phi\,,
\end{equation}
with
\begin{equation}
    S_\Phi=\int d^3 x\sqrt{g}\left[|D_\mu\Phi_i|^2+\mu^2 |\Phi_i|^2+ V_{\rm int}\left(|\Phi_i|^2\right)\right]\,,
\end{equation}
where $\Phi_i$, $i=1,\dots,N$, are complex scalars, $D_\mu \Phi_i \equiv \de_\mu\Phi_i -i A_\mu\Phi_i$, and the interaction potential reads
\begin{equation}
    V_{\rm int}(t)= \frac \lambda {2N}t^2+\frac h{3N^2}t^3\,.
\end{equation}
Let us also introduce for later convenience the function
\begin{equation}
     U_{\rm int}(t)\equiv \frac 1N V_{\rm int}(Nt)= \frac\la 2 t^2 +\frac h3 t^3\,.
\end{equation}

We need to specify boundary conditions. For a free scalar of mass squared $\mu^2$ in AdS$_3$, a solution of the free Klein--Gordon equation behaves near the boundary as
\begin{equation}
    \Phi(x,z)\sim z^{\Delta_+}\hat \phi_+(x)+ z^{\Delta_-}\hat\phi_-(x)\,,
\end{equation}
with $\Delta_\pm =1\pm \sqrt{1+\mu^2}$. Standard quantization sets $\hat\phi_-=0$. For $-1<\mu^2<0$, alternate quantization, $\hat\phi_+=0$, is also consistent, with $\Delta_->0$ above the unitarity bound \cite{Breitenlohner:1982jf,Klebanov:1999tb}. At the Breitenlohner--Freedman (BF) bound $\mu^2=-1$, the two falloffs are replaced by $z$ and $z\log z$, and we set the logarithmic coefficient to zero. The scalar propagator is then \eqref{eq:GscalarAdS3} with $\Delta=\Delta_\phi\in\{\Delta_+,\Delta_-\}$, respectively. At large $N$ the effective mass is fixed by the saddle, and we retain both quantizations, see section~\ref{subsec:scalar-ads-phase-diagram}.

This theory has global symmetry $G=PSU(N)\times U(1)_{\rm top}$, where $PSU(N)\cong U(N)/U(1)$ rotates the scalars and $U(1)_{\rm top}$ is the magnetic symmetry of the gauge theory, with
\begin{equation}
    J_{\mu}^{\rm top}=\frac i{4\pi}\epsilon_{\mu\nu\rho}F^{\nu\rho}\,,\qquad \nabla^\mu J_\mu^{\rm top}=0\,.
\end{equation}
At the boundary, there is in addition a $U(1)$ symmetry generated by the chiral current $J_i^\varsigma$.

We are interested in large-$N$ physics, in particular in the limit
\begin{equation}
    N\to \infty\,,\quad |k|\to \infty\,,\quad e^2\to 0\,,\quad \kappa\equiv k/N\ {\rm fixed}\,,\quad \al\equiv e^2 N\ {\rm fixed}\,.
\end{equation}
In this limit we work with the canonically normalized gauge field, obtained by replacing $A_\mu\to eA_\mu$. Accordingly, $D_\mu=\nabla_\mu-ieA_\mu$, the propagator of section~\ref{sec:MCS} becomes $\langle A_\mu A_{\nu'}\rangle=G_{A,\mu\nu'}$, and the topological mass can also be written as $m=\kappa\alpha/(2\pi)$. So, up to boundary terms,
\begin{equation}\label{eq:SA}
S_{\rm MCS}=\int d^3x\sqrt{g}\left[\frac1{4}F_{\mu\nu}F^{\mu\nu}+\frac1{2\xi}(\nabla_\mu A^\mu)^2-\frac{i\kappa\alpha}{4\pi} \epsilon^{\mu\nu\rho}A_\mu\nabla_\nu A_\rho\right]\,.
\end{equation}

We first determine the phase diagram in AdS, comparing it with flat space, and then the boundary spectra in each phase at generic couplings.

\subsection{Phase diagram}
\label{subsec:scalar-ads-phase-diagram}
We follow the flat-space analysis of \cite{DiPietro:2023zqn}, adapting it to AdS$_3$. We introduce auxiliary fields $\sigma,\chi$ by inserting in the partition function the identity
\begin{equation}
    1= \int D\sigma\,\delta[|\Phi_i|^2/\sqrt{N}-\sigma]=\int D\sigma D\chi\, e^{-\int \frac{\chi}{\sqrt{N}} (|\Phi_i|^2-\sqrt{N}\si)}\,,
\end{equation}
where $\chi$ is integrated along the imaginary axis. We normalize the auxiliary fields so that their saddle values are of order $\sqrt N$ and their fluctuations have order-one connected two-point functions. This gives
\begin{equation}
    S_\Phi= \int d^3x\,\sqrt g\left[|D_\mu\Phi_i|^2+\mu^2|\Phi_i|^2+N U_{\rm int}\!\left(\frac{\si}{\sqrt{N}}\right) + \frac{\chi}{\sqrt{N}}(|\Phi_i|^2-\sqrt{N}\si)\right]\,,
\end{equation}
where we used that on-shell $\sqrt{N}\si=|\Phi_i|^2$. The action is now quadratic in the scalars. Introducing background fields $\phi_i$ and integrating out their quantum fluctuations, we obtain the 1PI effective action
\begin{equation}
\begin{aligned}
    \Gamma_\Phi^{\rm 1PI}[\phi_i]&=\int d^3x\,\sqrt g\left[|D_\mu\phi_i|^2+\mu^2|\phi_i|^2 +N U_{\rm int}\!\left(\frac{\si}{\sqrt{N}}\right) + \frac{\chi}{\sqrt{N}}(|\phi_i|^2-\sqrt{N}\si)\right]\\
    &\qquad + N\,\Tr\log\!\left(-D^2+\mu^2+\frac{\chi}{\sqrt{N}}\right)\,.
\end{aligned}
\end{equation}
At leading order in large $N$, gauge field fluctuations are subleading, since $e^2=\alpha/N$, and we may set $e=0$. Expanding around
\begin{equation}
    \phi_i = \sqrt{N}\phi_0\delta_{iN} + \di\phi_i\,,\qquad \si=\sqrt{N}\Sigma+\di\si\,,\qquad \chi=\sqrt{N}X+\di\chi\,,
\end{equation}
we find, in dimensional regularization,
\begin{equation}\label{eq:scalar-Veff-AdS}
    \frac1N V_{\rm eff}^{\rm AdS}(\phi_0,\Sigma,X)=\mu^2|\phi_0|^2+U_{\rm int}(\Sigma)+X(|\phi_0|^2-\Sigma)\mp\frac1{6\pi}(1+X+\mu^2)^{3/2}\,,
\end{equation}
where reality requires $X\geq -(1+\mu^2)$, and the upper and lower signs correspond to standard and alternate quantization. Varying $X$ gives $\Sigma=|\phi_0|^2+G_{\Delta_\phi}(x,x)=|\phi_0|^2\mp\frac1{4\pi}\sqrt{1+X+\mu^2}$, with $\Delta_\phi=1\pm \sqrt{1+X+\mu^2}$. The alternate-quantization potential is always higher than the standard-quantization one.

\begin{figure}[t]
  \centering
  \begin{subfigure}[t]{0.49\textwidth}
    \centering
    \begin{tikzpicture}
      \node[anchor=south west,inner sep=0] (plot) at (0,0)
        {\includegraphics[width=0.87\linewidth]{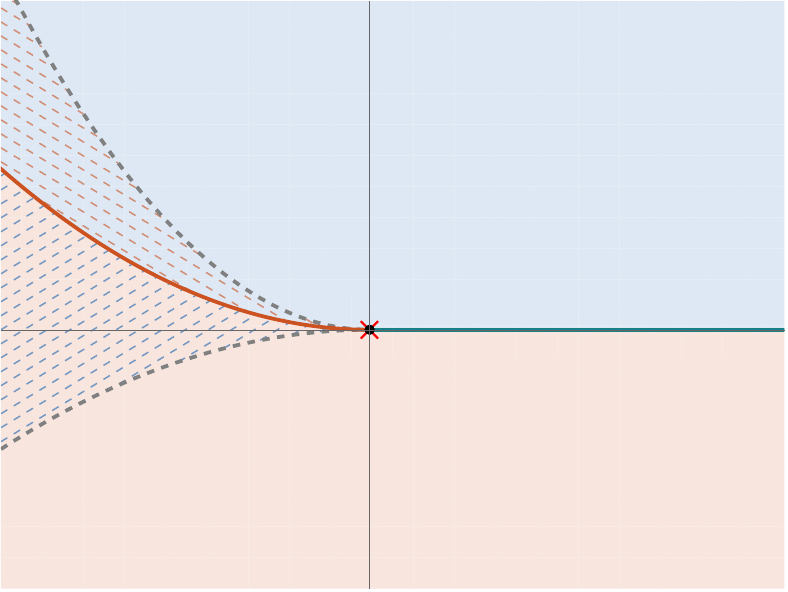}};
      \begin{scope}[x={(plot.south east)},y={(plot.north west)}]
        \node[anchor=south,inner sep=2pt] at (0.470,1) {$\mu^2$};
        \node[anchor=west,inner sep=2pt] at (1,0.440) {$\frac{\lambda}{4\pi}$};
        \node[text=phaseBlue, font=\small]
  at (0.72,0.83) {unHiggsed};
\node[text=phaseOrange, font=\small]
  at (0.72,0.12) {Higgsed};
      \end{scope}
    \end{tikzpicture}
    \caption{Flat space}
    \label{fig:flatphasediag}
  \end{subfigure}\hfill
  \begin{subfigure}[t]{0.49\textwidth}
  \centering
  \begin{tikzpicture}
    \node[anchor=south west,inner sep=0] (plot) at (0,0)
      {\includegraphics[width=0.87\linewidth]
        {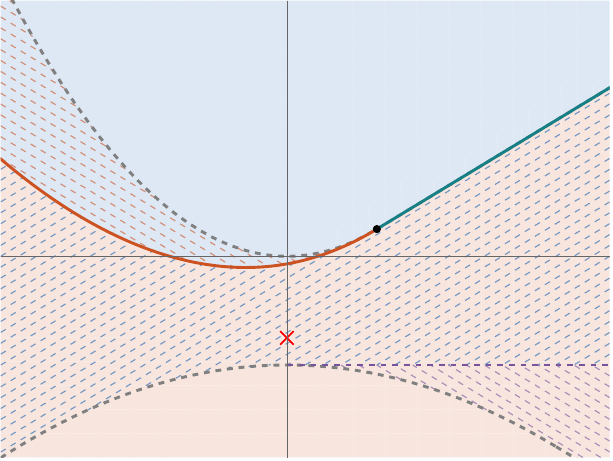}};
    \begin{scope}[x={(plot.south east)},y={(plot.north west)}]
      \node[anchor=south,inner sep=2pt]
        at (0.470,1) {$\mu^2$};
      \node[anchor=west,inner sep=2pt]
        at (1,0.440) {$\frac{\lambda}{4\pi}$};
      \node[text=phaseBlue,font=\small]
        at (0.72,0.83) {unHiggsed};
      \node[text=phaseOrange,font=\small]
        at (0.72,0.075) {Higgsed};
    \end{scope}
  \end{tikzpicture}
  \caption{AdS$_3$}
  \label{fig:adsphasediag}
\end{subfigure}
  \caption{Scalar phase diagrams at $h=4\pi^2$. Blue and orange denote unHiggsed and Higgsed minima. Background colors identify the lower-energy minimum; hatching marks an additional minimum, with purple reserved for alternate quantization. Gray dashed curves bound existence regions. In AdS, the purple dashed half-line marks $\Delta_\phi=1$. Orange curves mark equal vacuum energies, teal lines continuous Higgsing, black dots their junctions, and red crosses the bulk tricritical theories.}
  \label{fig:phasediags}
\end{figure}

Eliminating $\phi_0$ and $X$ gives two possible potentials, which we refer to as Higgsed and unHiggsed branches,
\begin{equation}
\begin{aligned}
\frac1N V_{\rm eff,H}^{\rm AdS}(\Sigma)&=\mu^2\Sigma+\frac\lambda2\Sigma^2+\frac h3\Sigma^3-\frac1{6\pi}\,,&&\Sigma\geq-\frac1{4\pi}\,,\\
\frac1N V_{\rm eff,uH}^{\rm AdS}(\Sigma)&=(\mu^2+1)\Sigma+\frac\lambda2\Sigma^2+\frac{h-16\pi^2}{3}\Sigma^3\,,&&\Sigma<\frac1{4\pi}\,.
\end{aligned}
\end{equation}
The unHiggsed branch includes both quantizations, with
\begin{equation}\label{eq:scalar-Delta-Sigma}
\Delta_\phi=1-4\pi\Sigma\,.
\end{equation}
Thus $\Sigma\leq0$ gives standard quantization and $0<\Sigma<1/(4\pi)$ alternate quantization, with the upper bound imposed by unitarity. The branches meet at $\Sigma=-1/(4\pi)$ and overlap for $-1/(4\pi)<\Sigma<1/(4\pi)$. At fixed $\Sigma$ in this interval, the Higgsed branch has lower potential, with equality only at $\Sigma=-1/(4\pi)$.

The unHiggsed and Higgsed branches are bounded from below for
\begin{equation}\label{eq:h-stability}
    h<16\pi^2\,,\qquad h>0\,,
\end{equation}
respectively. The upper bound is known as the Bardeen--Moshe--Bander point, where at $\lambda=\mu^2=0$ the large-$N$ flat-space unHiggsed potential develops a flat direction and scale invariance is spontaneously broken \cite{Bardeen:1983rv,Omid:2016jve}. On the unHiggsed branch, the stationary points are
\begin{equation}\label{eq:scalar-uH-saddles}
    \Sigma_{{\rm uH},\pm}^{\rm AdS}=\frac{-\lambda\pm\sqrt{\cal D}}{2(h-16\pi^2)}\,,\qquad {\cal D}\equiv\lambda^2+4(\mu^2+1)(16\pi^2-h)\,,
\end{equation}
which are real for ${\cal D}\geq0$ and unitary for $\Sigma<\frac1{4\pi}$, and represent a local minimum and a local maximum, respectively. In the following, $\Sigma_{\rm uH}^{\rm AdS}$ denotes either of them. In particular, for $\lambda>0$ the minimum continues smoothly into alternate quantization below $\mu^2=-1$, until it merges with the maximum or reaches $\Delta_\phi=0$.

The Higgsed minimum and its existence condition are
\begin{equation}
    \Sigma_{\rm H}^{\rm AdS}=\frac{\sqrt{\la^2-4\mu^2h}-\la}{2h}\,,\qquad \mu^2\leq\begin{cases}
        \frac{\lambda^2}{4h}\,, & \lambda\leq\frac{h}{2\pi}\,,\\
        \frac{\lambda}{4\pi}-\frac{h}{16\pi^2}\,, & \lambda>\frac{h}{2\pi}\,.
    \end{cases}
\end{equation}
At fixed bulk couplings, each unitary stationary point defines an AdS-invariant boundary condition for the matter fields. The Higgsed saddle breaks the global symmetry, and its images under the broken generators form a conformal manifold of symmetry-breaking boundary conditions, with the Goldstone modes giving the tilt operators \cite{Carmi:2018qzm}.

For comparison, removing the curvature shift, $1+X+\mu^2\to X+\mu^2$ in~\eqref{eq:scalar-Veff-AdS}, gives the flat-space potential \cite{DiPietro:2023zqn}
\begin{equation}
    \frac1N V_{\rm eff}(\Sigma)=\mu^2\Sigma + \frac\la 2\Sigma^2 + \frac13\left(h-16\pi^2\Theta(-\Sigma)\right)\Sigma^3\,,
\end{equation}
with minima $\Sigma_{\rm H}^{\rm flat}=\Sigma_{\rm H}^{\rm AdS}$, existing for $\mu^2\leq \frac{\la^2}{4h} \Theta(-\la)$, and
\begin{equation}
    \Sigma_{\rm uH}^{\rm flat}=\frac{\la-\sqrt{\la^2+4\mu^2(16\pi^2-h)}}{2(16\pi^2-h)}\,,
\end{equation}
existing for $\mu^2\geq -\frac{\la^2}{4(16\pi^2-h)}\Theta(-\la)$. For $\lambda<0$ both minima exist, and they are degenerate on the curve $\mu^2=\lambda^2 f(h)$, with
\begin{equation}
    f(h)=\frac{(\tau_0+1)(3\tau_0-1)}{16h\tau_0^2}\,,\qquad \tau_0=1-\frac{8\pi}{\sqrt{h}}\sin\left(\frac\pi 6+\frac13\arccos\frac{\sqrt{h}}{4\pi}\right)\,,
\end{equation}
where $\tau_0=\Sigma_{\rm H}^{\rm flat}/\Sigma_{\rm uH}^{\rm flat}$ on the degeneracy curve. The resulting phase diagram, figure~\ref{fig:flatphasediag}, contains an unHiggsed phase, which flows to $U(1)_k$ Chern--Simons theory, and a Higgsed phase, which flows to the $\mathbb{CP}^{N-1}$ model with a topological term. They are separated by the half-line $\mu^2=0$, $\la>0$, of second-order transitions, described by critical bosonic QED$_3$, and by the half-line $\mu^2=\la^2f(h)$, $\la<0$, of first-order transitions. The two half-lines meet at the tricritical point $\mu^2=\la=0$.

In AdS, figure~\ref{fig:adsphasediag}, the curvature has two effects. First, the line of second-order transitions moves away from the $\mu^2=0$ axis, to the half-line $\mu^2=\frac{\la}{4\pi}-\frac h{16\pi^2}$ starting from $(\la,\mu^2)= \left(\frac h{2\pi},\frac{h}{16\pi^2}\right)$. Correspondingly, the degeneracy curve no longer has a simple explicit form. In terms of $\tau=\Sigma_{\rm H}^{\rm AdS}/\Sigma_{\rm uH}^{\rm AdS}$, it reads
\begin{equation}
\lambda(\tau)= \frac{h}{4\pi}(1+\tau)\,\beta +\frac{4\pi}{1-\tau}\frac{1-\beta^2}{\beta}\,,\qquad \mu^2(\tau)=\frac{h}{16\pi^2}\tau\,\beta^2 +\frac{\tau}{1-\tau}(1-\beta^2)\,,
\end{equation}
where $\beta(\tau)=-4\pi\Sigma_{\rm uH}^{\rm AdS}\geq1$ is the root of $\bigl[\frac{h}{16\pi^2}(\tau-1)^3-3\tau+1\bigr]\beta^3+3(\tau+1)\beta-4=0$, and $\tau\leq1/\beta$. Second, now one has distinct boundary conditions to consider. Since distinct saddles define distinct boundary conditions of the same bulk theory, comparing their effective potentials compares the vacuum energies of these boundary conditions. On a rigid AdS background, vacuum degeneracy therefore does not necessarily imply a first-order transition. Moreover, while in flat space the higher-energy vacuum is metastable and decays through bubble nucleation, in AdS the curvature can hinder the expansion of the nucleating bubble or lead to a mixture of different phases in the bulk.

Features of the AdS effective potential, such as two saddles meeting, do not by themselves identify bulk conformal points. Conformality in AdS, which is Weyl-equivalent to conformality in flat \emph{half}-space, must hold at any finite $L$, whereas the flat-space critical behavior is recovered only as $L\to\infty$. Figure~\ref{fig:adsphasediag} illustrates this: the bulk tricritical theory lies at $(\lambda,\mu^2)=(0,-3/4)$, away from the transition lines. At this point the unHiggsed branch has a standard-quantization minimum and, for $h<12\pi^2$, a unitary alternate-quantization maximum. We therefore treat the critical points separately in sections~\ref{sec:scalar-critical} and~\ref{sec:scalar-tricritical}, by specifying the scalar boundary condition and analyzing the corresponding boundary spectrum.

\subsection{UnHiggsed phase}
\label{subsec:scalar-unhiggsed}
In the unHiggsed phase we expand around the saddle $\phi_0=0$, with $\Sigma=\Sigma_{\rm uH}^{\rm AdS}$ and $X_{\rm uH}\equiv U_{\rm int}'(\Sigma_{\rm uH}^{\rm AdS})$ determined by the gap equations. Here $\Sigma_{\rm uH}^{\rm AdS}$ can be either root in \eqref{eq:scalar-uH-saddles}. We introduce fluctuations around the nontrivial expectation values,
\begin{equation}
    \sigma(x)=\sqrt{N}\Sigma_{\rm uH}^{\rm AdS} + \delta\sigma(x)\,,\qquad \chi(x)=\sqrt{N}X_{\rm uH}+\delta\chi(x)\,,
\end{equation}
and we expand the effective action,
\begin{equation}
    \begin{aligned}
        S_{\rm eff,uH}&=S_A+\int d^3x\sqrt{g}\left[|D_\mu\phi_i|^2+(\mu^2+X_{\rm uH})|\phi_i|^2+\frac{1}{\sqrt{N}}\delta\chi |\phi_i|^2\right] \\
        &\qquad + S^{(2)}_{\si\chi,\rm uH}+S^{(2)}_{A,\rm uH}+\dots\,,
    \end{aligned}
\end{equation}
where the dots denote corrections that are subleading at large $N$. $S^{(2)}_{\si\chi,\rm uH}$ encodes quadratic fluctuations in the scalar sector,
\begin{equation}
    S^{(2)}_{\si\chi,\rm uH}=\frac12\int d^3x \sqrt{g_x}\int d^3y\sqrt{g_y}\begin{pmatrix}
            \delta\si(x)&\delta\chi(x)
        \end{pmatrix}\underbrace{\begin{pmatrix}
            \zeta_{\rm uH} \di(x,y)&-\di(x,y)\\
            -\di(x,y)&-B^{(0)}_{\Delta_\phi,0}(x,y)
        \end{pmatrix}}_{K_{\rm uH}(x,y)}\begin{pmatrix}
            \delta\si(y)\\\delta\chi(y)
        \end{pmatrix}\,,
\end{equation}
where we have introduced $\zeta_{\rm uH}\equiv U''_{\rm int}(\Sigma_{\rm uH}^{\rm AdS})=\lambda+2h\Sigma_{\rm uH}^{\rm AdS}$. $S^{(2)}_{A,\rm uH}$ instead encodes the terms quadratic in the gauge field that come from expanding the non-local term,
\begin{equation}\label{eq:S2AunHiggsed}
    S^{(2)}_{A,{\rm uH}}=\frac \al 2\int d^3x \sqrt{g_x}\int d^3y\sqrt{g_y} A^\mu(x) \Big[-B^{(1)}_{\Delta_\phi,0}(x,y)_{\mu\nu}+2 g_{\mu\nu}{\di(x,y)}G_{\Delta_\phi}(x,x)\Big]A^\nu(y)\,.
\end{equation}
The bubble functions are those defined in appendix~\ref{subsec:scalar-bubbles}. The second term in~\eqref{eq:S2AunHiggsed} is the contact contribution from the seagull term.

By \eqref{eq:scalar-Delta-Sigma}, the boundary dimension of the elementary scalar is $\Delta_\phi=1+\beta$, with $\beta\equiv-4\pi\Sigma_{\rm uH}^{\rm AdS}$, so that $\beta\geq0$ in standard and $-1<\beta<0$ in alternate quantization.

\paragraph{Scalar sector.}
From $S^{(2)}_{\sigma\chi,\rm uH}$ we read off the inverse propagator $K_{\rm uH}(x,y)$,
\begin{equation}
    K_{\rm uH}=\int_{\mathbb R} d\nu\,\widetilde K_{\rm uH}(\nu)\Omega_\nu\,,\qquad \widetilde K_{\rm uH}(\nu)=\begin{pmatrix}
        \zeta_{\rm uH} & -1\\
        -1 & -\widetilde B^{(0)}_{\Delta_\phi,0}(\nu)
    \end{pmatrix}\,,
\end{equation}
where $\widetilde B^{(0)}_{\Delta_\phi,0}(\nu)$ is the spectral function of the spin-zero scalar bubble. Inverting it gives
\begin{equation}\label{eq:KuH-inverse}
    \widetilde K_{\rm uH}^{-1}(\nu)=\frac{1}{1+\zeta_{\rm uH}\widetilde B^{(0)}_{\Delta_\phi,0}(\nu)}\begin{pmatrix}
        \widetilde B^{(0)}_{\Delta_\phi,0}(\nu) & -1\\
        -1 & -\zeta_{\rm uH}
    \end{pmatrix}\,.
\end{equation}
For $\zeta_{\rm uH}\neq0$, the scalar singlet boundary spectrum follows from the zeros of $1+\zeta_{\rm uH}\widetilde B^{(0)}_{\Delta_\phi,0}$.\footnote{For alternate quantization, the extra term proportional to $\Omega_{\nu_0}$ in the bubble, cf. appendix~\ref{subsec:scalar-bubbles}, drops out of the resummed correlators, since $(1+\zeta_{\rm uH}\widetilde B^{(0)}_{\Delta_\phi,0})^{-1}$ vanishes at $\nu=\pm\nu_0$.} Substituting the spectral function derived in appendix~\ref{subsec:scalar-bubbles} and setting $\nu=-i(\Delta-1)$, we obtain
\begin{equation}
0=\frac1{8\pi(\Delta-1)}\left[\psi\left(\Delta_\phi+\frac{\Delta}2-1\right)-\psi\left(\Delta_\phi-\frac{\Delta}2\right)\right] + \zeta_{\rm uH}^{-1}\,.
\end{equation}
At $\zeta_{\rm uH}=0$, the $\delta\sigma$ correlator reduces to the bubble $B^{(0)}_{\Delta_\phi,0}$. This is the generalized-free point, whose boundary spectrum is $\Delta_{n,0}=2\Delta_\phi+2n$, $n\in\ZZ_{\geq0}$.

The unHiggsed saddle is a local minimum of the effective potential for
\begin{equation}\label{eq:scalar-hom-min}
    \zeta_{\rm uH}+8\pi\beta\geq0\,.
\end{equation}
The lightest boundary singlet is irrelevant at a local minimum and relevant at a maximum, becoming marginal, $\Delta_{0,0}=2$, when this quantity vanishes. Bulk stability is weaker: it only excludes complex $\Delta_{0,0}$, i.e.\ it requires $1+\zeta_{\rm uH}\widetilde B^{(0)}_{\Delta_\phi,0}(0)\geq0$, or
\begin{equation}\label{eq:scalar-BF-stability}
    \zeta_{\rm uH}+\frac{8\pi}{\psi'\!\left(\beta+\frac12\right)}\geq0\,,
\end{equation}
with $\psi'$ the trigamma function. At ${\cal D}=0$ the minimum and the maximum merge with a marginal singlet and annihilate, as in a merger of conformal boundary conditions \cite{Kaplan:2009kr,Gorbenko:2018ncu,Lauria:2023uca}.

\begin{figure}[t]
  \centering
  \begin{tikzpicture}
    \definecolor{spectrumRed}{RGB}{204,78,26}
    \definecolor{spectrumPurple}{RGB}{122,81,158}
    \definecolor{spectrumTeal}{RGB}{20,128,137}
    \node[anchor=south west,inner sep=0] (plot) at (0,0)
      {\includegraphics[width=0.5\textwidth]{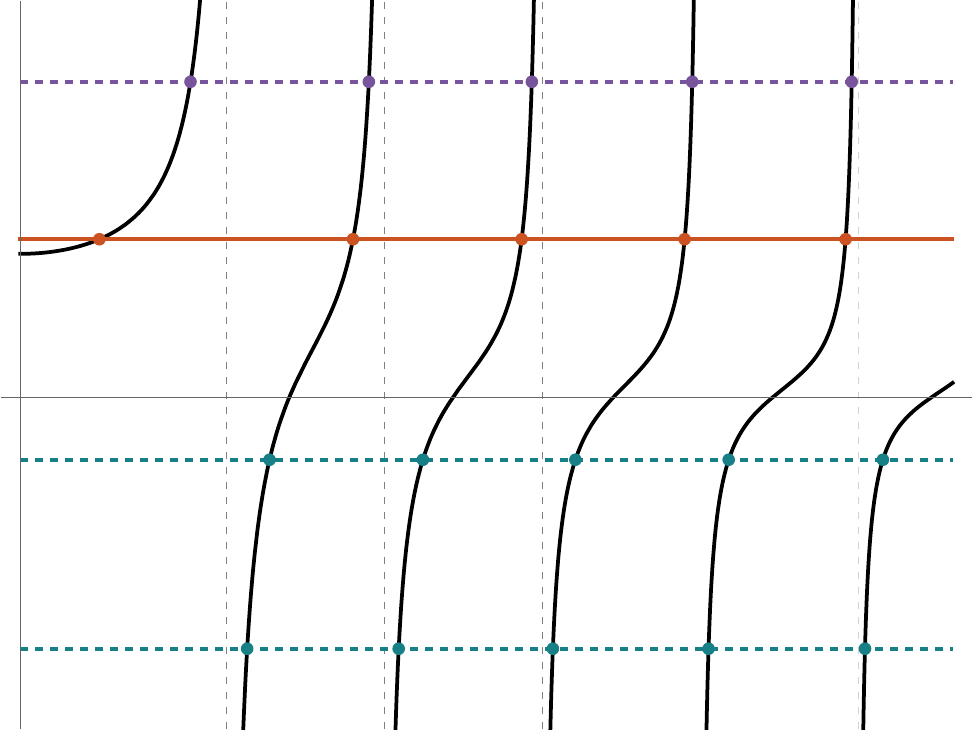}};
    \begin{scope}[x={(plot.south east)},y={(plot.north west)}]
      \node[anchor=south west,inner sep=2pt] at (0.020,1.015)
        {\small $\widetilde B^{(0)}_{\Delta_\phi,0}\!\left(-i(\Delta-1)\right)$};
      \node[anchor=west,inner sep=2pt] at (1,0.457) {\small $\Delta-1$};
      \node[text=spectrumRed,font=\small,anchor=north]
        at (0.115,0.625) {$\Delta=2$};
      \draw[spectrumRed,line width=0.4pt]
        (0.108,0.627) -- (0.102,0.664);
      \draw[->,spectrumPurple,line width=0.7pt]
        (0.953,0.735) -- (0.953,0.860)
        node[midway,right,font=\small] {$\zeta_{\rm uH}$};
      \draw[->,spectrumTeal,line width=0.7pt]
        (0.953,0.155) -- (0.953,0.325)
        node[midway,right,font=\small] {$\zeta_{\rm uH}$};
    \end{scope}
  \end{tikzpicture}
  \caption{Scalar singlet spectrum in the unHiggsed phase at fixed $\Delta_\phi$ in standard quantization. Intersections of the bubble (black) with $-1/\zeta_{\rm uH}$ give the boundary dimensions. Red marks $\zeta_{\rm uH}=-8\pi\beta$, where the lightest singlet has $\Delta=2$; purple and teal denote negative and positive $\zeta_{\rm uH}$. Arrows indicate increasing $\zeta_{\rm uH}$; vertical guides mark the generalized-free values $2\Delta_\phi+2n$.}
  \label{fig:plotscalarspin0symmetric}
\end{figure}

Figure~\ref{fig:plotscalarspin0symmetric} illustrates the pole condition at fixed $\Delta_\phi$. Positive $\zeta_{\rm uH}$ raises the lightest singlet above its generalized-free dimension, while negative $\zeta_{\rm uH}$ lowers it.

\paragraph{Vector sector.}
The gauge-field quadratic term~\eqref{eq:S2AunHiggsed} determines the self-energy through $S^{(2)}_{A,\rm uH}=-\frac12\int A\Pi_{A,\rm uH}^{\rm mat}A$, with the contact term ensuring transversality at coincident points. In spectral representation,
\begin{equation}
    \Pi_{A,\rm uH}^{\rm mat}=\alpha \int_{\mathbb R} d\nu\, \widetilde{B}_{\Delta_\phi,0}^{(1)}(\nu)\,\Omega^E_\nu = \alpha\int_{\mathbb R} d\nu\, \widetilde{B}_{\Delta_\phi,0}^{(1)}(\nu)\, \Omega^+_\nu\,,
\end{equation}
where we used that the spectral representation for the spin-one scalar bubble function $\widetilde{B}^{(1)}_{\Delta_\phi,0}(\nu)$ is even in $\nu$. Resumming bubble insertions, one obtains the spectral representation for the full large-$N$ photon propagator $G^{\rm full}_A(x,y)_{\mu\nu'}$, defined by $(G_A^{\rm full})^{-1}=G_{A}^{-1}-\Pi_A^{\rm mat}$,
\begin{equation}
    G^{\rm full}_{A,\rm uH}=\int_{\mathbb R} d\nu\, \frac{\Omega_\nu^+}{(\nu+i\varsigma 0)(\nu-im) - \alpha \widetilde{B}^{(1)}_{\Delta_\phi,0}(\nu)}+ G_A^\parallel\,,
\end{equation}
with $G_A^\parallel$ given in \eqref{eq:GApar}.

\begin{figure}[t]
\centering
\begin{subfigure}[t]{0.49\textwidth}
  \centering
  \begin{tikzpicture}
    \node[anchor=south west,inner sep=0] (plot) at (0,0)
      {\includegraphics[width=0.85\linewidth]{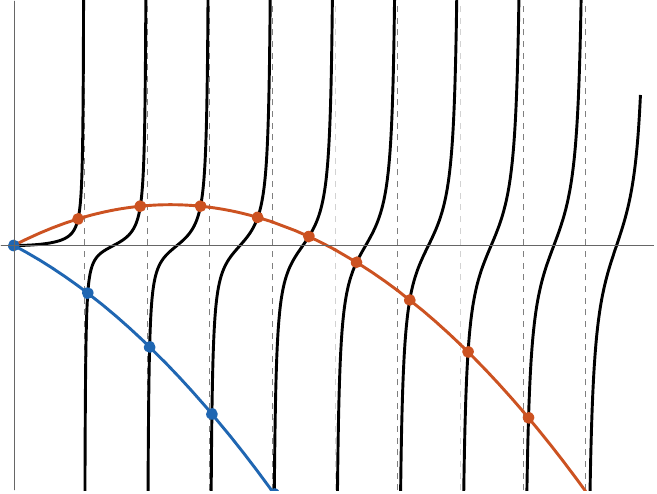}};
    \begin{scope}[x={(plot.south east)},y={(plot.north west)}]
      \node[anchor=south west,inner sep=2pt,font=\small] at (0.020,1.015)
        {$\widetilde B^{(1)}_{\Delta_\phi,0}\!\left(-i(\Delta-1)\right)$};
      \node[anchor=west,inner sep=2pt,font=\small] at (1,0.502) {$\Delta-1$};
    \end{scope}
  \end{tikzpicture}
  \caption{Spin-one spectrum}
\end{subfigure}\hfill
\begin{subfigure}[t]{0.49\textwidth}
  \centering
  \begin{tikzpicture}
    \node[anchor=south west,inner sep=0] (plot) at (0,0)
      {\includegraphics[width=0.85\linewidth]{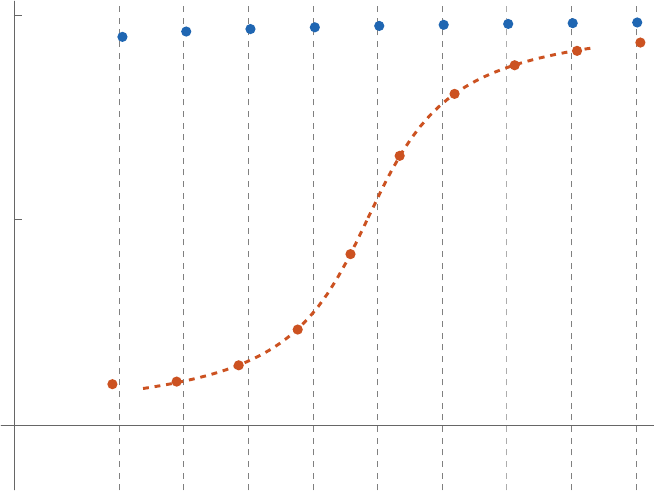}};
    \begin{scope}[x={(plot.south east)},y={(plot.north west)}]
      \node[anchor=south,inner sep=2pt,font=\small] at (0.021,1.015) {$\delta_1^s$};
      \node[anchor=west,inner sep=2pt,font=\small] at (1,0.141) {$\Delta$};
      \node[anchor=east,inner sep=2pt,font=\small] at (0.021,0.141) {$0$};
      \node[anchor=east,inner sep=2pt,font=\small] at (0.021,0.555) {$\pi/2$};
      \node[anchor=east,inner sep=2pt,font=\small] at (0.021,0.969) {$\pi$};
    \end{scope}
  \end{tikzpicture}
  \caption{Spectral phase shift}
\end{subfigure}
\par\smallskip
{\small
\begin{tikzpicture}[baseline=-0.5ex]
  \definecolor{helicityBlue}{RGB}{31,102,184}
  \definecolor{helicityOrange}{RGB}{204,78,26}
  \draw[helicityBlue,line width=1pt] (0,0)--(0.45,0);
  \node[anchor=west] at (0.5,0) {$s=+1$};
  \draw[helicityOrange,line width=1pt] (2.1,0)--(2.55,0);
  \node[anchor=west] at (2.6,0) {$s=-1$};
\end{tikzpicture}}
\captionsetup{singlelinecheck=false}
\caption{Spin-one spectrum and spectral phase shifts in the unHiggsed scalar phase, at $\Delta_\phi=9/8$, $\alpha=30$, and $\kappa=2\pi/3$. (a) Intersections of the bubble (black) with $-(\Delta-1)^2/\alpha-s\kappa(\Delta-1)/(2\pi)$ give the corresponding dimensions. (b) The $s=-1$ resonance and its Breit--Wigner fit~\eqref{eq:resonance-fit}, with $(\Delta_{\rm res},\Gamma_\Delta)\approx(10.96,3.84)$. Vertical guides mark generalized-free dimensions.}
\label{fig:plotscalarspin1symmetric}
\end{figure}

For $\kappa\neq0$, the $+i\varsigma0$ prescription assigns the pole at $\nu=0$ to the chiral current $J_i^\varsigma$ selected by the Dirichlet boundary condition, with $\Delta=1$ and helicity $s=\varsigma=\sgn\kappa$. The remaining spin-one operators correspond to the nonzero poles, which we parametrize as
\begin{equation}\label{eq:spin1-param}
    \nu=-is(\Delta-1)\,,\qquad s=\pm1\,,
\end{equation}
so that the positive and negative halves of the $i\nu$ axis correspond to helicity $+1$ and $-1$, respectively. The pole condition becomes
\begin{equation}
    0=\frac{(\Delta-1)^2}{\al}+\frac{s\kappa}{2\pi}(\Delta-1)+\widetilde B_{\Delta_\phi,0}^{(1)}\bigl(-is(\Delta-1)\bigr)\,.
\end{equation}
Figure~\ref{fig:plotscalarspin1symmetric}(a) illustrates this condition, showing the splitting between the two helicity sectors.

The massive Maxwell--Chern--Simons state is recovered by taking $\alpha\to0$ at fixed $m$. In this limit the matter polarization decouples and the nonzero pole approaches $\nu=im$, giving $\Delta=1+|m|$ with $s=-\varsigma$. At finite $\alpha$, its dimension is shifted by the matter polarization. As in the pure gauge theory, the limit $\kappa\to0$ is singular.

Turning on $\alpha>0$ also produces poles associated with spin-one double-trace operators, with different dimensions in the two helicity sectors. At weak coupling, mixing with the massive gauge mode shifts nearby two-particle levels downward below this mode and upward above it. This sign change occurs in the helicity sector $s=-\varsigma$ and is characteristic of a resonance \cite{Carmi:2018qzm}.

To characterize this behavior, we define the spectral phase shift
\begin{equation}\label{eq:phaseshiftgeneric}
    \delta_\ell^s(n)=\frac{\pi}{2}\left(\Delta_{n,\ell}^{s,\mathrm{GFF}}-\Delta_{n,\ell}^s\right)\,,\qquad n\geq0\,.
\end{equation}
Here $\Delta_{n,\ell}^{s,\mathrm{GFF}}$ denotes the generalized-free two-particle dimension, and the interacting levels are ordered by increasing dimension within each helicity sector. For scalar exchange the helicity label is absent. In the flat-space limit, an appropriate average over nearby levels determines the scattering phase shift \cite{Paulos:2016fap}. A narrow resonance produces a rapid variation of the scattering phase shift of order $\pi$. We characterize its AdS counterpart by fitting the discrete spectral phase shifts to the Breit--Wigner form
\begin{equation}\label{eq:resonance-fit}
    \delta_\ell^s(n)\simeq\frac{\pi}{2}+\arctan\!\left[\frac{2\bigl(\Delta_{n,\ell}^s-\Delta_{\mathrm{res}}\bigr)}{\Gamma_\Delta}\right]\,,
\end{equation}
where $\Delta_{\mathrm{res}}$ and $\Gamma_\Delta$ specify the resonance center and width in dimension space. In the present case, $\ell=1$ and $\Delta_{n,1}^{s,\mathrm{GFF}}=2\Delta_\phi+2n+1$, with $\Delta_{0,1}^{\varsigma}=1$ for the protected current. Figure~\ref{fig:plotscalarspin1symmetric}(b) shows the resonant variation associated with the massive gauge mode in the $s=-\varsigma$ sector, which is absent in the $s=\varsigma$ sector.

\subsection{Higgsed phase}
\label{subsec:scalar-higgsed}
In the Higgsed phase we expand around the saddle $|\phi_0|^2=v^2$, with $\Sigma=\Sigma_{\rm H}^{\rm AdS}=v^2-\frac1{4\pi}$ and $X_{\rm H}=-\mu^2$. We parametrize the fluctuations around this saddle as
\begin{equation}
    \begin{aligned}
        \phi_a(x)&=\pi_a(x)\,,\quad a=1,\dots,N-1\,, &\qquad
        \phi_N(x)&=\left(\sqrt{N}v+\frac{\rho(x)}{\sqrt{2}}\right)e^{\frac{i\theta(x)}{\sqrt{2N}v}}\,,\\
        \si(x)&=\sqrt{N}\Sigma_{\rm H}^{\rm AdS}+\delta\si(x)\,, &
        \chi(x)&=\sqrt{N}X_{\rm H}+\delta\chi(x)\,.
    \end{aligned}
\end{equation}
We impose standard quantization on the scalar fluctuations around this background. We will also find it convenient to modify the gauge-fixing term,
\begin{equation}
    S_{\rm gf}'=\frac1{2\xi}\int d^3 x\,\sqrt{g}\,\left(\nabla_\mu A^\mu - \sqrt{2\alpha}\xi v \theta\right)^2\,,
\end{equation}
to avoid kinetic mixing between $A_\mu$ and $\theta$. With this choice, the effective action expanded near the symmetry-breaking vacuum reads
\begin{equation}
    \begin{aligned}
        S_{\rm eff,H}&=S_A +\int d^3 x\,\sqrt{g}\,\left[|D_\mu \pi_a|^2 + \frac12(\partial_\mu \theta)^2+ \alpha \xi v^2\theta^2 + \frac1{\sqrt{N}}\delta\chi|\pi_a|^2\right]\\
        &\qquad + S^{(2)}_{\rho\si\chi,\rm H} + S^{(2)}_{A,\rm H} + \dots\,,
    \end{aligned}
\end{equation}
where now $S^{(2)}_{\rho\si\chi,\rm H}$ encodes fluctuations quadratic in the scalars $\rho$, $\delta\si$, $\delta\chi$,
\begin{equation}
\begin{aligned}
        S^{(2)}_{\rho\si\chi,\rm H}&=\frac12 \int d^3 x\sqrt{g_x}\int d^3y\sqrt{g_y}\\
        &\quad\times\begin{pmatrix}
        \rho(x)&\delta\si(x)&\delta\chi(x)
    \end{pmatrix}\underbrace{\begin{pmatrix}
        -\nabla^2_x\di(x,y) &0&\sqrt{2}v\di(x,y)\\
        0&\zeta_{\rm H}\di(x,y)&-\di(x,y)\\
        \sqrt{2}v\di(x,y)&-\di(x,y)&-B_{2,0}^{(0)}(x,y)
    \end{pmatrix}}_{K_{\rm H}(x,y)}\begin{pmatrix}
        \rho(y)\\\delta\si(y)\\\delta\chi(y)
    \end{pmatrix}\,,
\end{aligned}
\end{equation}
and $S^{(2)}_{A,\rm H}$ encodes the terms quadratic in the gauge field,
\begin{equation}
    S^{(2)}_{A,\rm H}=\frac \al 2\int d^3x \sqrt{g_x}\int d^3y\sqrt{g_y} A^\mu(x) \Big[-B^{(1)}_{2,0}(x,y)_{\mu\nu}+2 g_{\mu\nu}{\di(x,y)}\left(v^2+G_{2}(x,x)\right)\Big]A^\nu(y)\,.
\end{equation}
Compared with $S^{(2)}_{A,\rm uH}$, this also contains the Higgs mass term for the photon $M_W^2\equiv 2\alpha v^2$, and has massless $\pi$ scalars $\Delta_\pi=2$ running in the loops. Dots denote fluctuations which are subleading at large $N$.

\paragraph{Scalar sector.}
Going to spectral representation, the quadratic kernel is
\begin{equation}
    \widetilde K_{\rm H}(\nu)=\begin{pmatrix}
        \nu^2+1 & 0 & \sqrt{2}v\\
        0 & \zeta_{\rm H} & -1\\
        \sqrt{2}v & -1 & -\widetilde B^{(0)}_{2,0}(\nu)
    \end{pmatrix}\,,\qquad \zeta_{\rm H}\equiv\lambda+2h\Sigma_{\rm H}^{\rm AdS}\,,
\end{equation}
with $\zeta_{\rm H}\geq0$ on the Higgsed branch. The radial mode $\rho$ is more conveniently traded for its linearized equation-of-motion operator,
\begin{equation}
    \mathscr E_\rho\equiv-\nabla^2\rho+\sqrt{2}v\,\delta\chi\,,
\end{equation}
whose correlators with $\delta\sigma$ and $\delta\chi$ are contact terms. In the basis $\bigl(\mathscr E_\rho,\delta\sigma,\delta\chi\bigr)$, the propagator is block-diagonal,
\begin{equation}
    \widetilde K_{\rm H,\mathscr E}^{-1}(\nu)=\begin{pmatrix}
        \nu^2+1 & 0\\
        0 & \widetilde K_{\rm H}^{-1}\big|_{\sigma\chi}(\nu)
    \end{pmatrix}\,,
\end{equation}
where the first entry is a contact term, and the $\sigma\chi$ block is~\eqref{eq:KuH-inverse} with $\zeta_{\rm uH}\to\zeta_{\rm H}$ and $\widetilde B^{(0)}_{\Delta_\phi,0}\to\widetilde B_{\rm H}^{(0)}$, with
\begin{equation}\label{eq:Higgsed-kernels}
\widetilde B_{\rm H}^{(0)}(\nu)\equiv\widetilde B^{(0)}_{2,0}(\nu)+\frac{2v^2}{\nu^2+1}\,.
\end{equation}

\begin{figure}[t]
\centering
\begin{subfigure}[t]{0.49\textwidth}
  \centering
  \begin{tikzpicture}
    \node[anchor=south west,inner sep=0] (plot) at (0,0)
      {\includegraphics[width=0.85\linewidth]{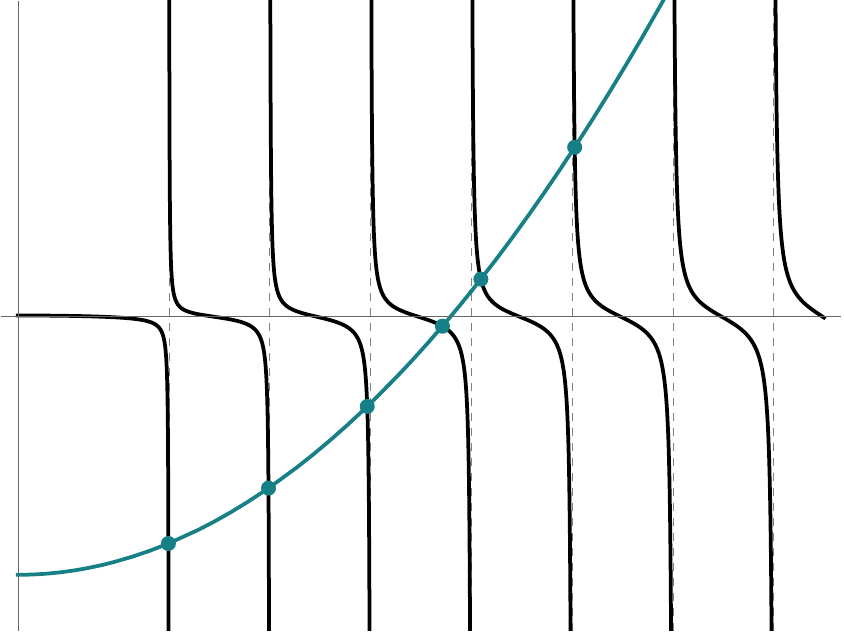}};
    \begin{scope}[x={(plot.south east)},y={(plot.north west)}]
      \node[anchor=south west,inner sep=2pt,font=\small] at (0.020,1.015)
        {$f_{\mathrm H}\!\left(-i(\Delta-1)\right)$};
      \node[anchor=west,inner sep=2pt,font=\small] at (1,0.502) {$\Delta-1$};
    \end{scope}
  \end{tikzpicture}
  \caption{Scalar singlet spectrum}
\end{subfigure}\hfill
\begin{subfigure}[t]{0.49\textwidth}
  \centering
  \begin{tikzpicture}
    \node[anchor=south west,inner sep=0] (plot) at (0,0)
      {\includegraphics[width=0.85\linewidth]{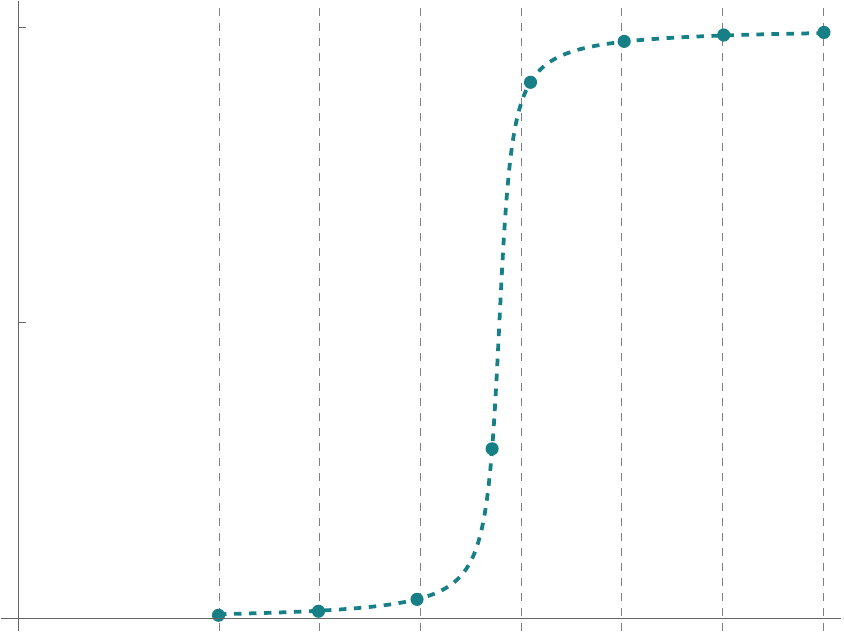}};
    \begin{scope}[x={(plot.south east)},y={(plot.north west)}]
      \node[anchor=south,inner sep=2pt,font=\small] at (0.021,1.015) {$\delta_0$};
      \node[anchor=west,inner sep=2pt,font=\small] at (1,0.027) {$\Delta$};
      \node[anchor=east,inner sep=2pt,font=\small] at (0.021,0.027) {$0$};
      \node[anchor=east,inner sep=2pt,font=\small] at (0.021,0.494) {$\pi/2$};
      \node[anchor=east,inner sep=2pt,font=\small] at (0.021,0.958) {$\pi$};
    \end{scope}
  \end{tikzpicture}
  \caption{Spectral phase shift}
\end{subfigure}
\captionsetup{singlelinecheck=false}
\caption{Scalar singlet spectrum and spectral phase shift in the Higgsed phase, at $\zeta_{\rm H}=3$ and $m_\rho^2=73$. (a) Intersections of $f_{\rm H}(\nu)$ (black) with $-2v^2-(\nu^2+1)/\zeta_{\rm H}$ (teal), at $\nu=-i(\Delta-1)$, give the boundary dimensions. (b) The radial-mode resonance, fitted by~\eqref{eq:resonance-fit} with $(\Delta_{\rm res},\Gamma_\Delta)\approx(9.58,0.37)$. Vertical guides mark generalized-free values $\Delta=4+2n$.}
\label{fig:plotscalarspin0higgsed}
\end{figure}

For $\zeta_{\rm H}>0$, to avoid spurious singularities, we write the pole condition as
\begin{equation}
    0=f_{\rm H}(\nu)+2v^2+\frac{\nu^2+1}{\zeta_{\rm H}}\,,\qquad f_{\rm H}(\nu)\equiv(\nu^2+1)\widetilde B^{(0)}_{2,0}(\nu)\,.
\end{equation}
Figure~\ref{fig:plotscalarspin0higgsed}(a) illustrates the pole condition: the scalar singlet dimensions are given by the intersections of $f_{\rm H}(\nu)$ with $-2v^2-(\nu^2+1)/\zeta_{\rm H}$, evaluated at $\nu=-i(\Delta-1)$. The limit $v^2\to\infty$ at fixed $\zeta_{\rm H}$ gives the generalized-free pion double-trace spectrum. At $\zeta_{\rm H}=0$, the $\delta\sigma$ correlator is $\widetilde B_{\rm H}^{(0)}$, whose pole at $\nu^2+1=0$ corresponds to a scalar of dimension $\Delta=2$.

A different limit is sending $v^2\to \infty$ at fixed $m^2_\rho\equiv 2v^2\zeta_{\rm H}$, for which one finds
\begin{equation}
    \nu^2+1+m^2_\rho=0\implies \Delta_\rho=1+\sqrt{1+m^2_\rho}\,,
\end{equation}
corresponding to the radial mode $\rho$ with mass squared $m^2_\rho$. For small positive $\zeta_{\rm H}$, the spectrum also contains levels close to the pion double-trace dimensions $2\Delta_\pi+2n=4+2n$, which mix with the radial mode as $\zeta_{\rm H}$ grows. The spectral phase shift \eqref{eq:phaseshiftgeneric}, with the radial mode included in the ordered spectrum, then varies by order $\pi$ across the radial level, see figure~\ref{fig:plotscalarspin0higgsed}(b).

\begin{figure}[t]
\centering
\begin{subfigure}[t]{0.49\textwidth}
  \centering
  \begin{tikzpicture}
    \node[anchor=south west,inner sep=0] (plot) at (0,0)
      {\includegraphics[width=0.85\linewidth]{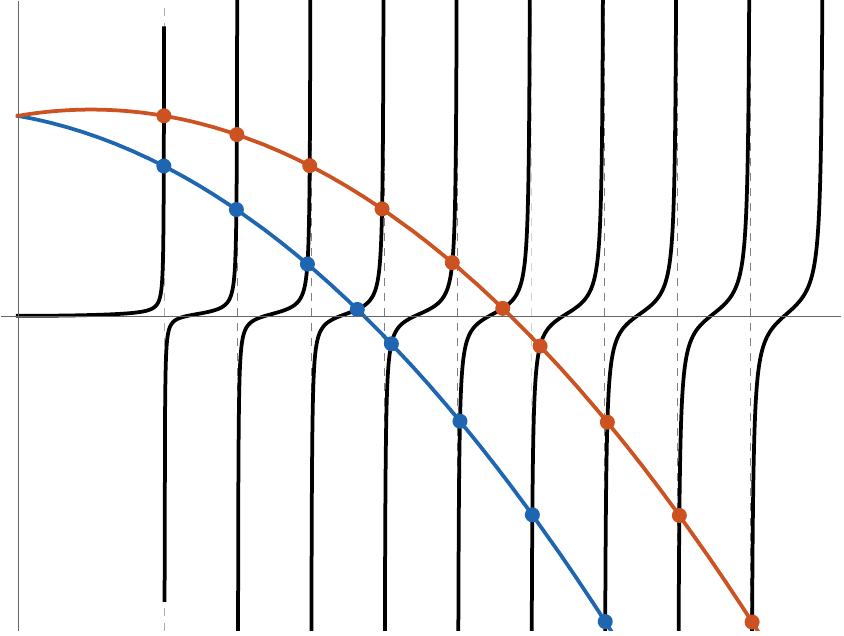}};
    \begin{scope}[x={(plot.south east)},y={(plot.north west)}]
      \node[anchor=south west,inner sep=2pt,font=\small] at (0.020,1.015)
        {$\widetilde B^{(1)}_{2,0}\!\left(-i(\Delta-1)\right)$};
      \node[anchor=west,inner sep=2pt,font=\small] at (1,0.502) {$\Delta-1$};
    \end{scope}
  \end{tikzpicture}
  \caption{Spin-one spectrum}
\end{subfigure}\hfill
\begin{subfigure}[t]{0.49\textwidth}
  \centering
  \begin{tikzpicture}
    \node[anchor=south west,inner sep=0] (plot) at (0,0)
      {\includegraphics[width=0.85\linewidth]{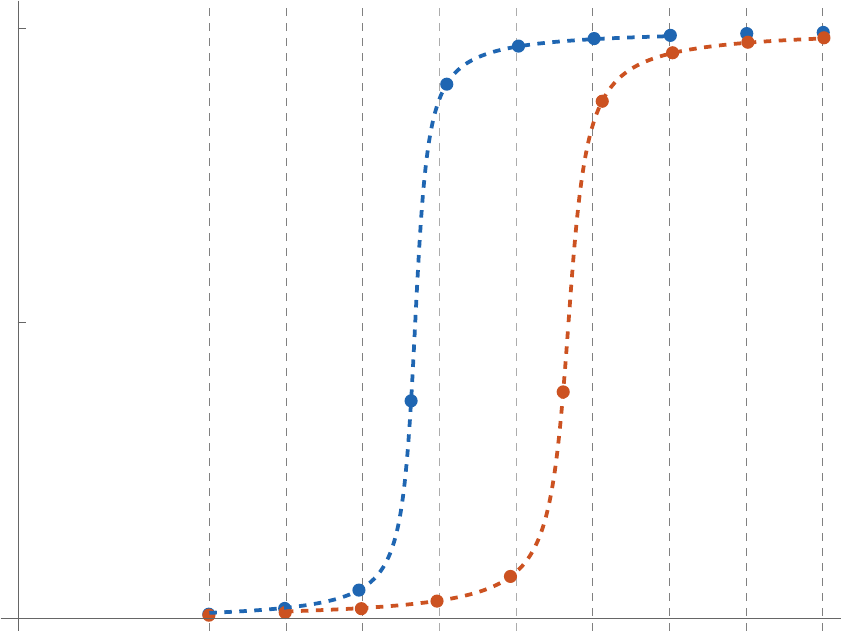}};
    \begin{scope}[x={(plot.south east)},y={(plot.north west)}]
      \node[anchor=south,inner sep=2pt,font=\small] at (0.021,1.015) {$\delta_1^s$};
      \node[anchor=west,inner sep=2pt,font=\small] at (1,0.027) {$\Delta$};
      \node[anchor=east,inner sep=2pt,font=\small] at (0.021,0.027) {$0$};
      \node[anchor=east,inner sep=2pt,font=\small] at (0.021,0.493) {$\pi/2$};
      \node[anchor=east,inner sep=2pt,font=\small] at (0.021,0.958) {$\pi$};
    \end{scope}
  \end{tikzpicture}
  \caption{Spectral phase shift}
\end{subfigure}
\par\smallskip
{\small
\begin{tikzpicture}[baseline=-0.5ex]
  \definecolor{helicityBlue}{RGB}{31,102,184}
  \definecolor{helicityOrange}{RGB}{204,78,26}
  \draw[helicityBlue,line width=1pt] (0,0)--(0.45,0);
  \node[anchor=west] at (0.5,0) {$s=+1$};
  \draw[helicityOrange,line width=1pt] (2.1,0)--(2.55,0);
  \node[anchor=west] at (2.6,0) {$s=-1$};
\end{tikzpicture}}

\captionsetup{singlelinecheck=false}
\caption{Spin-one spectra and spectral phase shifts in the Higgsed scalar phase, at $\alpha=10$, $m=4$, and $M_W^2=127$. (a) Intersections of the bubble (black) with $-(\Delta-1)^2/\alpha-s\kappa(\Delta-1)/(2\pi)+2v^2$ give the boundary dimensions. (b) Resonances in both helicities, with Breit--Wigner fits~\eqref{eq:resonance-fit} at $(\Delta_{\rm res}^+,\Gamma_\Delta^+)\approx(10.37,0.49)$ and $(\Delta_{\rm res}^-,\Gamma_\Delta^-)\approx(14.37,0.72)$. Vertical guides mark the generalized-free values $\Delta=5+2n$.}
\label{fig:plotscalarspin1higgsed}
\end{figure}

\paragraph{Vector sector.}
The self-energy read from $S^{(2)}_{A,\rm H}$ has the spectral representation
\begin{equation}
    \Pi_{A,\rm H}^{\rm mat}=\alpha \int_{\mathbb R}d\nu\,\widetilde B_{\rm H}^{(1)}(\nu)\Omega_\nu^E-2\alpha v^2 \nabla \nabla'G_2\,,
\end{equation}
where we have introduced
\begin{equation}
\widetilde B_{\rm H}^{(1)}(\nu)\equiv\widetilde B^{(1)}_{2,0}(\nu)-2v^2\,.
\end{equation}
Resumming these insertions gives
\begin{equation}
    G^{\rm full}_{A,\rm H}=\int_{\mathbb R} d\nu\,\frac{\Omega_\nu^+}{\nu(\nu-im)-\alpha\widetilde B_{\rm H}^{(1)}(\nu)}+G_{A,\rm H}^\parallel\,,
\end{equation}
where we recall that $M_W^2=2\alpha v^2$, and the longitudinal, gauge-dependent part is
\begin{equation}
    G_{A,\rm H}^{\parallel}=\xi \int_{\mathbb R}d\nu\, \frac{\nabla \nabla' \Omega_\nu}{(\nu^2+1)(\nu^2+1+\xi M_W^2)}\,.
\end{equation}
The Higgs term makes the transverse spectral function regular at $\nu=0$, so no $+i\varsigma0$ prescription is needed and importantly there is no protected current. The spin-one spectrum of flavor singlets exchanged in the $\langle \pi^*\pi\pi^*\pi\rangle$ four-point function can be inferred from the condition
\begin{equation}
    \widetilde B_{\rm H}^{(1)}(\nu)=\frac{\nu^2}{\alpha}-\frac{i\kappa\nu}{2\pi}\,.
\end{equation}
For $\alpha\to 0$ at fixed $m$ and $M_W^2$, it reduces to $\nu^2-im\nu+M_W^2=0$, whose roots correspond to two spin-one operators of helicity $\pm1$ and dimensions $\Delta_\pm=1+ \frac12\left(\sqrt{m^2+4M_W^2}\mp m\right)$. Higgsing then replaces the protected current $J_i^\varsigma$ and the single massive vector $V_i^{-\varsigma}$ by two massive states $W_i^\pm$. At finite coupling, each mixes with pion double-trace operators of the same helicity. The corresponding spectral phase shifts \eqref{eq:phaseshiftgeneric}, with $\Delta_{n,1}^{s,\mathrm{GFF}}=2\Delta_\pi+2n+1$, display a variation of order $\pi$ across the levels associated with $W_i^\pm$, see figure~\ref{fig:plotscalarspin1higgsed}.

The disappearance of the protected current has a natural interpretation in terms of symmetry breaking. In the bulk, the condensate Higgses the $U(1)$ gauge field, and its phase $\theta$ supplies the longitudinal mode. At the boundary, the condensate breaks the $U(1)$ generated by $J_i^\varsigma$, and the would-be tilt operator dual to $\theta$ recombines with the current into a long spin-one multiplet.

\section{Fermionic matter}
\label{sec:fermion}
We now turn to Maxwell--Chern--Simons theory coupled to fermionic matter,
\begin{equation}
    S_{\rm fMCS}=S_{\rm MCS}+ S_\Psi\,,
\end{equation}
with
\begin{equation}
    S_\Psi=\int d^3x\sqrt{g}\left[\frac12\bar\Psi_i \overset{\leftrightarrow}{\slashed D}\Psi^i + \bar\Psi_iM^i{}_j\Psi^j + \frac{g}{2N}(\bar\Psi_i\Psi^i)^2\right]\,.
\end{equation}
Here, $\Psi^i$, $i=1,\dots,N$, are Dirac fermions, $M^i{}_j=M\delta^i{}_j$ is a flavor-symmetric mass matrix, $\overset{\leftrightarrow}{\slashed D}= \Gamma^\mu\overset{\rightarrow}D_\mu-\overset{\leftarrow}D_\mu\Gamma^\mu$, with
\begin{equation}
    \overset{\rightarrow}D_\mu \Psi^i=\overset{\rightarrow}{\nabla}_\mu \Psi^i -iA_\mu \Psi^i\,,\qquad \bar\Psi_i\overset{\leftarrow}{D}_\mu=\bar\Psi_i\overset{\leftarrow}{\nabla}_\mu + i \bar\Psi_iA_\mu\,.
\end{equation}
Here $\Gamma_\mu$ are AdS$_3$ Dirac gamma matrices, satisfying
\begin{equation}\label{eq:gammaalgebra}
    \Gamma_\mu\Gamma_\nu = g_{\mu\nu} \mathbf{1} - i \epsilon_{\mu\nu\rho} \Gamma^\rho\,,\qquad {\rm tr}\,\mathbf{1}=2\,.
\end{equation}

We must specify boundary conditions for the fermions. For a single Dirac fermion of mass $M$ in AdS$_3$, a solution of the free Dirac equation, upon expansion near the boundary, has the general behavior
\begin{equation}
    \Psi(x,z)\sim z^{\Delta_+}\hat\psi_+(x)+z^{\Delta_-}\hat\psi_-(x)\,,
\end{equation}
with $\Delta_\pm=1\pm M$, where $\hat\psi_\pm$ are boundary Weyl spinors. We label the two choices by $\eta=\pm1$,
\begin{equation}
    \left.\gamma_0\Psi\right|_\partial =-\eta\left.\Psi\right|_\partial\,,\qquad \left.{\bar\Psi}\gamma_0\right|_\partial =\left.\eta\,{\bar\Psi}\right|_\partial\,.
\end{equation}
Here $\gamma_0$ is the gamma matrix along the radial direction. The corresponding boundary dimension is $\Delta_\psi=\Delta_\eta=1+\eta M$. For a nonzero mass, standard quantization corresponds to $\eta=\operatorname{sgn}(M)$, while alternate quantization corresponds to $\eta=-\operatorname{sgn}(M)$, and is consistent with unitarity, $\Delta_\psi>\frac12$, for $|M|<\frac12$ \cite{Giombi:2021cnr}. At zero mass, $\eta$ continues to distinguish the two boundary conditions. More generally, in a theory with $N$ Dirac fermions we can work with flavor-mixing boundary conditions,
\begin{align}
 \left.\gamma_0\Psi^i\right|_\partial = \left.-H^i{}_j\Psi^j\right|_\partial\,,\qquad \left.\bar{\Psi}_i \gamma_0\right|_\partial = \left.\bar{\Psi}_j H^j{}_i\right|_\partial\,,
\label{eq:fermbc}
\end{align}
with $H^2=\mathbf{1}$, $H^\dagger =H$. In the following we will fix $H^i{}_j= \eta\, \delta^i{}_j$. In particular, in this case the full theory has global symmetry $G=PSU(N)\times U(1)_{\rm top}$, like its scalar counterpart.

With these choices, the bulk-to-bulk fermion propagator, satisfying
\begin{equation}
    (\slashed{\nabla}+M)S=\delta \mathbf{1}\,,
\end{equation}
can be written as \cite{Giombi:2021cnr}
\begin{equation}
\begin{aligned}
    S^i{}_j(x,x')&\equiv\left\langle\Psi^i(x)\bar\Psi_j(x')\right\rangle_\eta=\delta^i{}_j S_\eta\,,\\
S_\eta &=\left[\eta\, A_{1}(u)+A_{2}(u)\, n_\mu \Gamma^\mu\right]\Lambda(x,x')\,,
\end{aligned}
\end{equation}
where $n_\mu$ ($n_{\nu'}$) is the unit vector at $x$ ($x'$) generating the geodesic connecting the two points,
\begin{equation}
n_\mu = \frac{\nabla_\mu u}{\sqrt{u(u+2)}}\,, \qquad n_\mu n^\mu =1\,.
\end{equation}
We denote by $g_{\mu\nu'}(x,x')$ and $\Lambda(x,x')$ the vector and spinor parallel transports along the geodesic. They satisfy
\begin{equation}
        g_{\mu\nu'}=-\nabla_\mu \nabla_{\nu'}u + \frac1{u+2}\nabla_\mu u\nabla_{\nu'}u\,,\qquad g_\mu{}^{\nu'}n_{\nu'}=-n_{\mu}\,,
\end{equation}
and
\begin{equation}\label{eq:spinortransport}
    \Lambda(x',x)=\Lambda^{-1}(x,x')\,,\qquad \Lambda \,\Gamma_{\nu'}\,\Lambda^{-1}=\Gamma_\mu\, g^{\mu}{}_{\nu'}\,.
\end{equation}
The two functions appearing in the propagator are
\begin{equation}
    \begin{aligned}
A_1(u)&=\frac{e^{-(\Delta_\psi-1)r}}{8\pi\sinh r}\left[2(\Delta_\psi-1)+\tanh\frac r2\right]\,,\\
A_2(u)&=\frac{e^{-(\Delta_\psi-1)r}}{8\pi\sinh r}\left[2(\Delta_\psi-1)+\coth\frac r2\right]\,.
\end{aligned}
\end{equation}
These expressions hold in both quantizations.

\subsection{Phase diagram}\label{subsec:fermion-phase-diagram}
Again, we take the large-$N$ limit with $\kappa=k/N$ and $\alpha=e^2N$ held fixed, and we use canonically normalized gauge fields. To capture the large-$N$ physics, unlike in the scalar case, it suffices to introduce a single Hubbard--Stratonovich field $\sigma$, normalized so that its saddle value is of order $\sqrt N$ and its fluctuation has an order-one propagator. After integrating out the quantum fluctuations of the fermions, we obtain the following one-loop effective action for the background fields,
\begin{equation}
    \Gamma_\Psi^{\rm 1PI}[\psi,\sigma]={}\int d^3x\sqrt{g}\left[\frac12\bar\psi_i\overset{\leftrightarrow}{\slashed D}\psi^i+\left(M+\frac{\sigma}{\sqrt N}\right)\bar\psi_i\psi^i-\frac{1}{2g}\sigma^2\right]-N\tr\log\!\left(\slashed D+M+\frac{\sigma}{\sqrt N}\right)\,.
\end{equation}
The auxiliary field $\sigma$ is a scalar of classical dimension one, for which we may add a potential,
\begin{equation}
    S_\sigma=\int d^3x\sqrt{g}\left[\sqrt N\,m_{\si, \rm b}^2\sigma+\frac {f_{\rm b}}2\sigma^2+\frac{y}{3\sqrt N}\sigma^3\right]\,.
\end{equation}

Expanding around a homogeneous configuration, $\sigma=\sqrt N\,\Sigma+\delta\sigma$, and trading $\Sigma$ for the effective fermion mass $\calM\equiv M+\Sigma$, the unrenormalized effective potential is, up to a field-independent constant,
\begin{equation}
    \frac1N V_{\rm F,eff,b}^{\rm AdS}(\calM)=\mu_{\rm F,b}^2\calM+\frac {\lambda_{\rm F,b}}2\calM^2+\frac y3\calM^3-\frac1{{\rm vol}({\rm AdS}_3)}\tr\log(\slashed D+\calM)\,,
\end{equation}
with
\begin{equation}
    \lambda_{\rm F,b}\equiv f_{\rm b}-\frac1g-2yM\,,\qquad \mu_{\rm F,b}^2\equiv m_{\si,\rm b}^2-\left(f_{\rm b}-\frac1g\right)M+yM^2\,.
\end{equation}
We evaluate the fermion determinant at fixed boundary condition $\eta$, for either sign of $\calM$, with $\eta\calM>-\frac12$ by unitarity. The corresponding gap equation involves the fermion trace $\tr\frac1{\slashed D+\calM}$, which is divergent and requires regularization. We adopt the parity-preserving Pauli--Villars prescription of \cite{Carmi:2018qzm}: assuming $N$ even, we introduce $N$ commuting Dirac fermions, half with mass $\calM+M_{\rm PV}$ and half with mass $\calM-M_{\rm PV}$. In the limit $M_{\rm PV}\to\infty$, one finds
\begin{equation}
    -\frac1{{\rm vol}({\rm AdS}_3)}\tr\frac1{\slashed D+\calM}\longrightarrow\frac{\eta}{2\pi}\left(\calM^2-\frac14\right)-\frac{M_{\rm PV}}{\pi}\calM\,.
\end{equation}
The divergence is absorbed into the renormalized coupling
\begin{equation}
    \lambda_{\rm F}=\lambda_{\rm F,b}-\frac{M_{\rm PV}}{\pi}\,,
\end{equation}
while $\mu_{\rm F}^2\equiv\mu_{\rm F,b}^2$ is finite. The renormalized effective potential and gap equation are
\begin{equation}\label{eq:fermion-effective-potential}
    \frac1N V_{{\rm F,eff},\eta}^{\rm AdS}(\calM)=\mu_{\rm F}^2\calM+\frac{\lambda_{\rm F}}2\calM^2+\frac y3\calM^3+\frac{\eta\,\calM}{6\pi}\left(\calM^2-\frac34\right)\,,
\end{equation}
and
\begin{equation}
    0=\mu_{\rm F}^2+\lambda_{\rm F}\calM-\frac{\eta}{8\pi}+\left(y+\frac{\eta}{2\pi}\right)\calM^2\,.
\end{equation}
For each boundary condition, the potential is bounded from below on the unitary domain $\eta\calM>-\frac12$ for
\begin{equation}\label{eq:y-stability}
    \eta y>-\frac1{2\pi}\,,
\end{equation}
whereas flat-space stability requires $|y|<\frac1{2\pi}$ \cite{DiPietro:2023zqn}.

At fixed $\eta$, the stationary points of \eqref{eq:fermion-effective-potential} are
\begin{equation}\label{eq:fermion-saddles}
    \calM_{\eta,\pm}=\frac{-\lambda_{\rm F}\pm\sqrt{\mathcal D_\eta}}{2\left(y+\frac{\eta}{2\pi}\right)}\,,\qquad \mathcal D_\eta=\lambda_{\rm F}^2-4\left(y+\frac{\eta}{2\pi}\right)\left(\mu_{\rm F}^2-\frac{\eta}{8\pi}\right)\,,
\end{equation}
with $\calM_{\eta,+}$ a local minimum and $\calM_{\eta,-}$ a local maximum. Saddles with $\eta\calM\geq0$ are in standard quantization and those with $-\frac12<\eta\calM<0$ in alternate quantization. In particular, for $\lambda_{\rm F}>0$ the minimum continues smoothly into alternate quantization for $\eta\mu_{\rm F}^2>\frac1{8\pi}$, until it merges with the maximum or reaches $\Delta_\psi=\frac12$.

In the overlap of the two existence regions, we compare the vacuum energies of the two minima. For $y\neq0$, the equal-energy curve is given parametrically by $\tau\equiv\calM_{+,+}/\calM_{-,+}<0$,
\begin{equation}
\begin{aligned}
    \calM_{-,+}^2(\tau)&=-\frac{3(\tau+1)}{2\left[\tau^3-3\tau^2-3\tau+1+2\pi y(\tau-1)^3\right]}\,,\\
    \lambda_{\rm F}(\tau)&=\frac{\frac1{4\pi}+\left[y-\frac1{2\pi}-\left(y+\frac1{2\pi}\right)\tau^2\right]\calM_{-,+}^2(\tau)}{(\tau-1)\calM_{-,+}(\tau)}\,,\\
    \mu_{\rm F}^2(\tau)&=\frac1{8\pi}-\tau\calM_{-,+}(\tau)\lambda_{\rm F}(\tau)-\left(y+\frac1{2\pi}\right)\tau^2\calM_{-,+}^2(\tau)\,,
\end{aligned}
\end{equation}
with $\calM_{-,+}<0$ and $\tau$ restricted to values for which both minima are real and in standard quantization. For $y=0$, the curve reduces to $\mu_{\rm F}^2=0$.

\begin{figure}[t]
\centering
\definecolor{massBlue}{RGB}{31,102,184}
\definecolor{massOrange}{RGB}{204,78,26}
\begin{subfigure}[t]{0.49\textwidth}
  \centering
  \begin{tikzpicture}
    \node[anchor=south west,inner sep=0] (plot) at (0,0)
      {\includegraphics[width=0.87\linewidth]{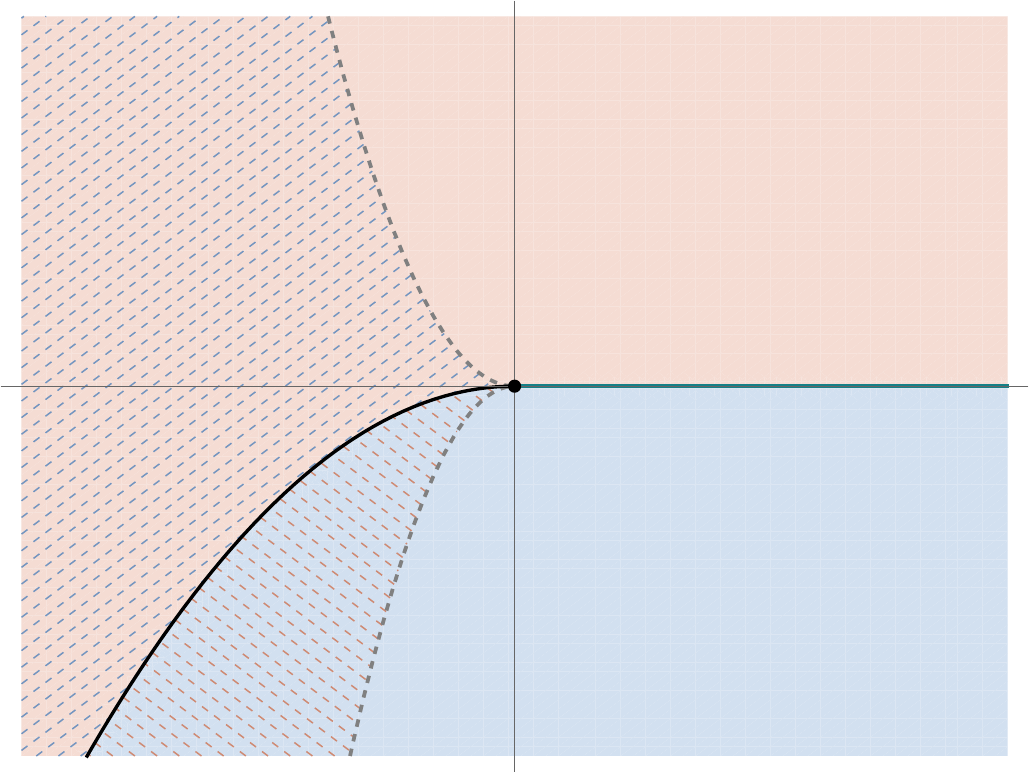}};
    \begin{scope}[x={(plot.south east)},y={(plot.north west)}]
      \node[anchor=south,inner sep=2pt,font=\small] at (0.500,1.010) {$\mu_{\mathrm F}^2$};
      \node[anchor=west,inner sep=2pt,font=\small] at (1,0.502) {$\lambda_{\mathrm F}$};
      \node[text=massOrange,font=\small] at (0.75,0.81) {$\calM<0$};
      \node[text=massBlue,font=\small] at (0.75,0.19) {$\calM>0$};
    \end{scope}
  \end{tikzpicture}
  \caption{Flat space}
  \label{fig:fermion-phase-flat}
\end{subfigure}
\hfill
\begin{subfigure}[t]{0.49\textwidth}
  \centering
  \begin{tikzpicture}
    \node[anchor=south west,inner sep=0] (plot) at (0,0)
      {\includegraphics[width=0.87\linewidth]{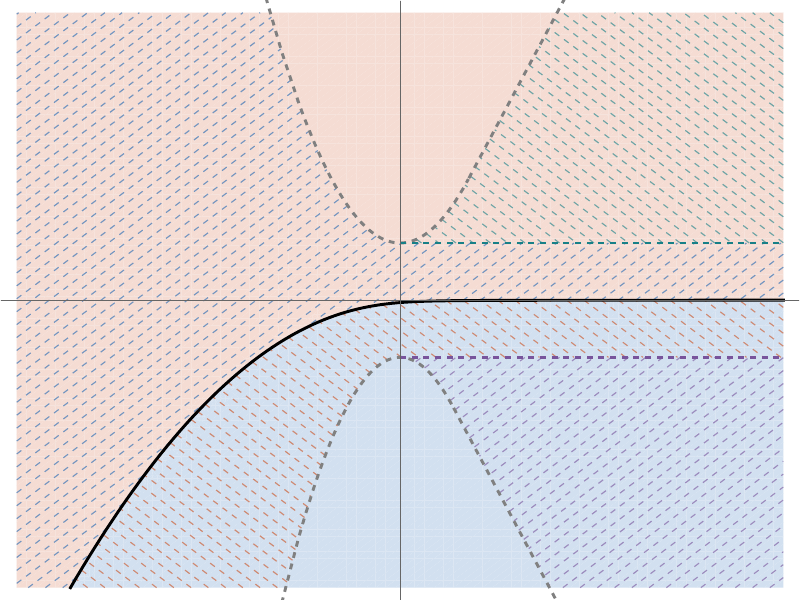}};
    \begin{scope}[x={(plot.south east)},y={(plot.north west)}]
      \node[anchor=south,inner sep=2pt,font=\small]
        at (0.498,1.010) {$\mu_{\mathrm F}^2$};
      \node[anchor=west,inner sep=2pt,font=\small]
        at (1,0.502) {$\lambda_{\mathrm F}$};
      \node[text=massOrange,font=\small]
        at (0.57,0.86) {$\eta=-$};
      \node[text=massBlue,font=\small]
        at (0.57,0.14) {$\eta=+$};
    \end{scope}
  \end{tikzpicture}
  \caption{AdS$_3$}
  \label{fig:fermion-phase-ads}
\end{subfigure}
\caption{Fermionic phase diagrams at $y=1/(16\pi)$. Blue and orange denote positive- and negative-mass minima in flat space, and $\eta=+$ and $\eta=-$ minima in AdS. Background colors identify the lower-energy minimum; hatching marks an additional minimum. In AdS, teal and purple hatching denote alternate quantization for $\eta=+$ and $\eta=-$, and horizontal dashed half-lines mark the standard/alternate crossovers. Gray dashed curves bound existence regions; black curves mark equal vacuum energies. In flat space, the teal half-line marks the continuous transition at $\mu_{\rm F}^2=0$, $\lambda_{\rm F}>0$.}
\label{fig:fermion-phase-diagram}
\end{figure}

For comparison, setting $\eta=\sgn\calM$ in \eqref{eq:fermion-effective-potential} and removing the curvature contribution $-|\calM|/(8\pi)$ gives the flat-space phase diagram shown in figure~\ref{fig:fermion-phase-flat}. The first-order transition line at $\lambda_{\rm F}<0$ ends at the Gross--Neveu point $(\lambda_{\rm F},\mu_{\rm F}^2)=(0,0)$. For $\lambda_{\rm F}>0$, the mass vanishes continuously at $\mu_{\rm F}^2=0$, where the infrared theory is the regular fermion CFT.

The AdS phase diagram is shown in figure~\ref{fig:fermion-phase-ads}. At fixed $y$ and $\lambda_{\rm F}$, the $\eta=+$ minimum has lower energy below the degeneracy value of $\mu_{\rm F}^2$, while the $\eta=-$ minimum has lower energy above it. For finite couplings, both minima remain massive on the degeneracy line, with the same caveat concerning first-order transitions as in the scalar case. As for scalars, alternate minima are never the lowest.

Integrating out the massive fermions shifts the Chern--Simons level by $\eta N/2$ \cite{Redlich:1983dv}, so the effective levels of the two saddles differ by $N$.

\subsection{Gapped phases}
\label{subsec:fermion-gapped}
We now study the fluctuations around a saddle $\calM_\eta$ of $V^{\rm AdS}_{{\rm F,eff},\eta}$, either $\calM_{\eta,+}$ or $\calM_{\eta,-}$ in \eqref{eq:fermion-saddles}, with $\eta$ the boundary condition of section~\ref{subsec:fermion-phase-diagram}. The boundary fermion has dimension $\Delta_\psi=\Delta_\eta\equiv1+\eta\calM_\eta$, with $\Delta_\eta\geq1$ in standard and $\frac12<\Delta_\eta<1$ in alternate quantization. Expanding $\sigma=\sqrt N\,\Sigma_\eta+\delta\sigma$, with $\Sigma_\eta=\calM_\eta-M$, and integrating out the fermions, the quadratic effective action for the singlet fluctuations takes the form
\begin{align}
    S_{\eta}^{(2)}&=S_A+\frac12\int d^3x\sqrt{g_x}\int d^3y\sqrt{g_y}\,\delta\sigma(x)\left[\zeta_\eta^{\rm F}\delta(x,y)-B_{\Delta_\eta,1/2}^{(0)}(x,y)\right]\delta\sigma(y)\notag\\
    &-\frac{\alpha}{2}\int d^3x\sqrt{g_x}\int d^3y\sqrt{g_y}\,A^\mu(x)B_{\Delta_\eta,1/2}^{(1)}(x,y)_{\mu\nu}A^\nu(y) + \dots \,,
\end{align}
where $\zeta_\eta^{\rm F}\equiv\lambda_{\rm F}+2y\calM_\eta$, and $B_{\Delta_\eta,1/2}^{(0)}$ and $B_{\Delta_\eta,1/2}^{(1)}$ are the single-fermion bubbles derived in appendix~\ref{subsec:fermion-bubbles}. The parity-even parts depend on the saddle through $\Delta_\eta$, while the parity-odd spin-one contribution also carries the sign $\eta$.

\paragraph{Scalar sector.}
The scalar singlet boundary spectrum is determined by
\begin{equation}
    \zeta_\eta^{\rm F}-\widetilde B_{\Delta_\eta,1/2}^{(0)}\bigl(-i(\Delta-1)\bigr)=0\,.
\end{equation}
At the saddles $\calM_{\eta,\pm}$, the second derivative of the effective potential is
\begin{equation}\label{eq:fermion-hom-min}
    \frac1N\,\partial_\calM^2V^{\rm AdS}_{{\rm F,eff},\eta}\Big|_{\calM_{\eta,\pm}}=\zeta_\eta^{\rm F}+\frac{\Delta_\eta-1}{\pi}=\pm\sqrt{\mathcal D_\eta}\,.
\end{equation}
As for scalars, the lightest singlet is irrelevant at a local minimum and relevant at a maximum, becoming marginal when this quantity vanishes. Bulk stability only excludes complex $\Delta_{0,0}$, i.e.\ it requires $\zeta_\eta^{\rm F}\geq\widetilde B^{(0)}_{\Delta_\eta,1/2}(0)$, or
\begin{equation}\label{eq:fermion-BF-stability}
    \zeta_\eta^{\rm F}+\frac{(\Delta_\eta-1)^2}{2\pi}\psi'(\Delta_\eta)+\frac{2\Delta_\eta-1}{4\pi}\geq0\,.
\end{equation}
At $\mathcal D_\eta=0$ the two saddles merge and annihilate.

For a local minimum, the lightest scalar dimension satisfies $2\leq\Delta_{0,0}<2\Delta_\eta+1$, while the higher levels obey $2\Delta_\eta+2n-1<\Delta_{n,0}<2\Delta_\eta+2n+1$, $n\geq1$. As $\zeta_\eta^{\rm F}$ is increased, this state moves toward the first pole of the bubble, while each higher state remains between two consecutive poles. In the limit $\zeta_\eta^{\rm F}\to+\infty$, the spectrum approaches the generalized-free scalar double-trace tower, $\Delta_{n,0}\to2\Delta_\eta+2n+1$.

\paragraph{Vector sector.}
The induced Chern--Simons level shift found in~\eqref{eq:CSlevelshift} motivates introducing
\begin{equation}
    \kappa_{\rm eff}\equiv\kappa+\frac{\eta}{2}\,,\qquad m_{\rm eff}\equiv\frac{\alpha\kappa_{\rm eff}}{2\pi}=m+\frac{\alpha\eta}{4\pi}\,,\qquad \varsigma_{\rm eff}\equiv\sgn(\kappa_{\rm eff})\,.
\end{equation}
Resumming the fermion-bubble insertions, the gauge propagator can then be written as
\begin{equation}
    G_{A,\eta}^{\rm full}=\int_{\mathbb R} d\nu\,\frac{\Omega_\nu^+}{(\nu+i\varsigma_{\rm eff}0)(\nu-i m_{\rm eff})-\alpha\left[\widetilde B_{\Delta_\eta,1/2}^{(1,+)}(\nu)-\frac{i\eta\nu}{4\pi}\right]}+G_A^\parallel\,.
\end{equation}
The quantity in square brackets starts at $O(\nu^2)$, so near the origin the propagator is that of Maxwell--Chern--Simons theory with $m\to m_{\rm eff}$. In each phase, we choose the Dirichlet boundary polarization according to the sign of the effective Chern--Simons level. The contour prescription is then fixed by $\varsigma_{\rm eff}$ and, as in section~\ref{subsec:scalar-unhiggsed}, for $\kappa_{\rm eff}\neq0$ the pole at $\nu=0$ gives the protected current $J_i^{\varsigma_{\rm eff}}$, with $\Delta^{\varsigma_{\rm eff}}_{0,1}=1$. The locus $\kappa_{\rm eff}=0$ is the same singular value as $\kappa=0$ in the pure gauge theory.

The remaining spin-one spectrum follows from the nonzero poles. Using \eqref{eq:spin1-param}, one finds
\begin{equation}\label{eq:fermion_vector_spectrum}
    \widetilde B_{\Delta_\eta,1/2}^{(1,+)}\bigl(-is(\Delta-1)\bigr)-\frac{s\eta}{4\pi}(\Delta-1)=-\frac{(\Delta-1)^2}{\alpha}-\frac{s\kappa_{\rm eff}}{2\pi}(\Delta-1)\,.
\end{equation}

\begin{figure}[t]
\centering
\definecolor{helicityBlue}{rgb}{0.12,0.40,0.72}
\definecolor{helicityOrange}{rgb}{0.90,0.40,0.12}
\begin{subfigure}[t]{0.49\textwidth}
  \centering
  \begin{tikzpicture}
    \node[anchor=south west,inner sep=0] (plot) at (0,0)
      {\includegraphics[width=0.91\linewidth]{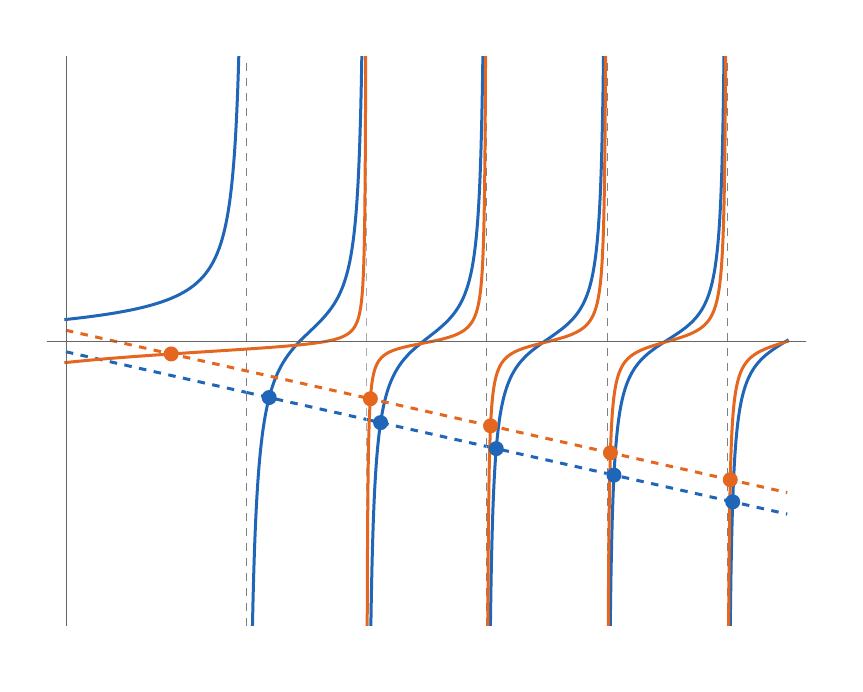}};
    \begin{scope}[x={(plot.south east)},y={(plot.north west)}]
      \node[anchor=south west,inner sep=1pt,font=\small] at (0.077,0.925)
        {$\frac{1}{\Delta-1}\widetilde B^{(1,+)}_{2,1/2}(-is(\Delta-1))$};
      \node[anchor=west,inner sep=1pt,font=\small] at (0.94,0.506)
        {$\Delta-1$};
    \end{scope}
  \end{tikzpicture}
  \caption{$\eta=+1$}
\end{subfigure}
\hfill
\begin{subfigure}[t]{0.49\textwidth}
  \centering
  \begin{tikzpicture}
    \node[anchor=south west,inner sep=0] (plot) at (0,0)
      {\includegraphics[width=0.91\linewidth]{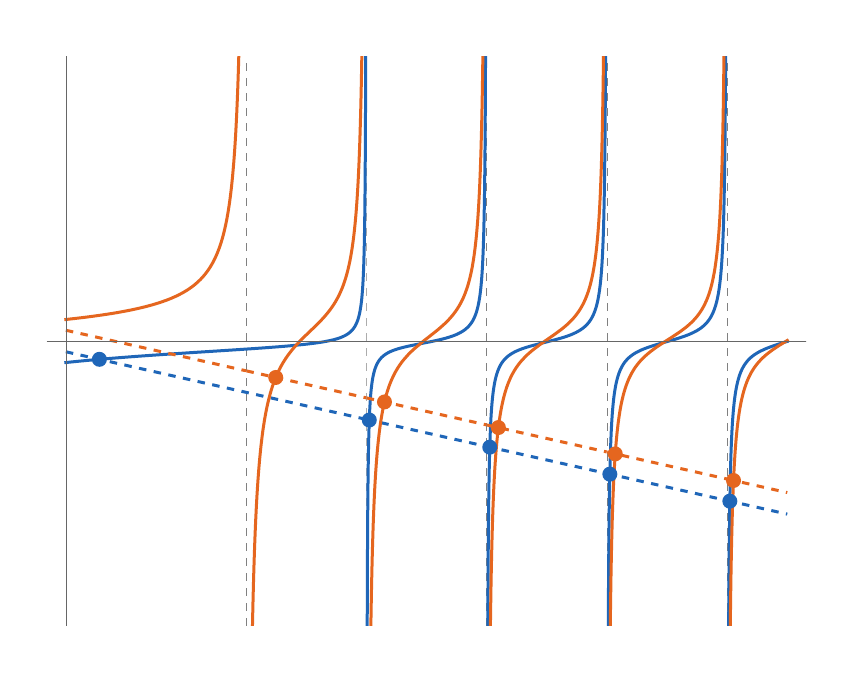}};
    \begin{scope}[x={(plot.south east)},y={(plot.north west)}]
      \node[anchor=south west,inner sep=1pt,font=\small] at (0.077,0.925)
        {$\frac{1}{\Delta-1}\widetilde B^{(1,+)}_{2,1/2}(-is(\Delta-1))$};
      \node[anchor=west,inner sep=1pt,font=\small] at (0.94,0.506)
        {$\Delta-1$};
    \end{scope}
  \end{tikzpicture}
  \caption{$\eta=-1$}
\end{subfigure}
\par\smallskip
\begin{tikzpicture}[baseline=-0.5ex]
  \draw[helicityBlue,line width=1pt] (0,0)--(0.45,0);
  \node[anchor=west,font=\small] at (0.5,0) {$s=+1$};
  \draw[helicityOrange,line width=1pt] (2.1,0)--(2.55,0);
  \node[anchor=west,font=\small] at (2.6,0) {$s=-1$};
\end{tikzpicture}
\captionsetup{singlelinecheck=false}
\caption{Spin-one spectra at $|\calM|=1$, $\alpha=20$, and $\kappa=1/4$, with $\eta=\operatorname{sgn}\calM$. Intersections of the helicity-plus bubble divided by $\Delta-1$ (solid) with $-(\Delta-1)/\alpha-s\kappa/(2\pi)$ (dashed) give the unprotected dimensions. The protected current has helicity $\operatorname{sgn}(\kappa+\eta/2)$. Vertical guides mark generalized-free values $\Delta=4+2n$; the lowest matter level occurs only for $s=\eta$.}
\label{fig:fermion-gapped-vector-spectrum}
\end{figure}

The state associated with the massive Maxwell--Chern--Simons mode is identified by taking $\alpha\to 0$ with $m_{\rm eff}$ fixed, which reproduces a state with $\Delta=1+|m_{\rm eff}|$ and helicity $-\varsigma_{\rm eff}$. At finite $\alpha$, this state is shifted by the fermion polarization and mixes with the two-fermion singlets of the same helicity. When this mode enters the two-fermion spectrum, its mixing with the two-particle levels produces the same resonance pattern discussed in the scalar theory, with a spectral phase shift varying by order $\pi$ across the resonance. Figure~\ref{fig:fermion-gapped-vector-spectrum} shows the spin-one spectra for the two boundary conditions. Unlike the Higgsed scalar phase, the gapped fermion phases thus retain the protected chiral current, alongside the massive Maxwell--Chern--Simons mode and the two-fermion towers.

\section{Bulk conformal points at large $N$}
\label{sec:conformal-points}
We now take the bulk conformal limits, sending $\alpha\to+\infty$ at fixed $\kappa\neq0$, together with the appropriate limit of the matter couplings. This also sends $|m|\to \infty$, so that the massive Maxwell--Chern--Simons mode decouples. We treat the critical and tricritical scalar theories with ordinary, special, and normal boundary conditions in sections~\ref{sec:scalar-critical} and~\ref{sec:scalar-tricritical}, and the regular fermion and Gross--Neveu theories in section~\ref{sec:fermion-conformal}. In each case, bulk data follow from the bulk-channel expansion of the scalar-singlet and topological-current correlators, as detailed in appendix~\ref{app:bulk-channel}. Scalar boundary data follow from the scalar-singlet correlators, while the spin-one boundary spectra are extracted from the dressed gauge propagators of sections~\ref{subsec:scalar-unhiggsed}, \ref{subsec:scalar-higgsed}, and~\ref{subsec:fermion-gapped}. All results hold at strict large $N$.

In order to distinguish bulk and boundary conformal quantities, in this section we add a hat to boundary objects while leaving bulk objects unhatted. We normalize the two-point function of a bulk scalar primary $\mathcal O$ of dimension $\Delta_{\mathcal O}$ in the theory on $\mathbb R^{3}$ as
\begin{equation}
    \langle \mathcal O(x)\mathcal O(0)\rangle_{\mathbb R^3}=\frac{C_{\mathcal O}}{|x|^{2\Delta_{\mathcal O}}}\,,
\end{equation}
while in the presence of the boundary we define its one-point coefficient by
\begin{equation}
    \langle\mathcal O(x,z)\rangle_{\mathbb R^2\times\mathbb R_+}=\frac{a_{\mathcal O}}{z^{\Delta_{\mathcal O}}}\,,\qquad \langle\mathcal O\rangle_{\rm AdS}=a_{\mathcal O}\,.
\end{equation}
For scalar primaries, the bulk OPE takes the schematic form
\begin{equation}
    \mathcal O_1(x)\mathcal O_2(0)\sim\sum_{\mathcal O_3}C_{\mathcal O_1\mathcal O_2}{}^{\mathcal O_3}\,|x|^{\Delta_{\mathcal O_3}-\Delta_{\mathcal O_1}-\Delta_{\mathcal O_2}}\bigl[\mathcal O_3(0)+\text{descendants}\bigr]\,,
\end{equation}
where $C_{\mathcal O_1\mathcal O_2}{}^{\mathcal O_3}$ denotes the bulk OPE coefficient. In particular, $C_{\mathcal O}=C_{\mathcal O\mathcal O}{}^{\mathbf 1}$. Similarly, we write the bulk-to-boundary OPE as
\begin{equation}
    \mathcal O(x,z)\sim\sum_{\hat{\mathcal O}}b_{\mathcal O}{}^{\hat{\mathcal O}}\,z^{\hat\Delta_{\hat{\mathcal O}}-\Delta_{\mathcal O}}\,\hat{\mathcal O}(x)+\ldots\,,
\end{equation}
where $b_{\mathcal O}{}^{\hat{\mathcal O}}$ is the bulk-to-boundary OPE coefficient. In particular, $b_{\mathcal O}{}^{\hat{\mathbf 1}} = a_{\mathcal O}$.

We normalize boundary operators as in section~\ref{subsec:boundary-limits},
\begin{equation}
    \langle \hat {\cal O}(x)\hat{\cal O}(0)\rangle_{\mathbb R^2} =\frac{\hat C_{\hat {\cal O}}}{|x|^{2\hat{\Delta}_{\hat {\cal O}}}}\,,\qquad \hat C_{\hat{\mathcal O}}=\mathcal C_{\hat\Delta_{\hat{\mathcal O}},0}=\frac1{2\pi}\,,
\end{equation}
independently of $\hat\Delta_{\hat{\mathcal O}}$. With this normalization, the bulk-to-boundary OPE coefficients follow from the boundary-channel poles of the spectral function $\widetilde G_{\mathcal O\mathcal O}(\nu)$ of the bulk two-point function $\langle \mathcal O\mathcal O\rangle$:
\begin{equation}\label{eq:scalar-bope-residue}
    \left(b_{\mathcal O}{}^{\hat{\mathcal O}}\right)^2=\frac{\hat\Delta_{\hat{\cal O}}-1}{\pi}\,\varrho_{\mathcal O\hat{\mathcal O}}\,,\qquad \varrho_{\mathcal O\hat{\mathcal O}}\equiv-2\pi i\,\underset{\nu=-i(\hat\Delta_{\hat{\cal O}}-1)}{\operatorname{Res}}\,\widetilde G_{\mathcal O\mathcal O}(\nu)\,.
\end{equation}
If several boundary operators of the same dimension appear in the bulk-to-boundary OPE of $\mathcal O$, the left-hand side is replaced by the sum of their squared coefficients.

Taking the conformal limit of the dressed gauge propagator and a curl at each endpoint, the two tensor structures of the topological-current correlator have spectral functions
\begin{equation}\label{eq:Gtop-general}
\widetilde G_E^{\rm top}(\nu)=\frac{\nu^2}{4\pi^2N}\,\frac{\mathcal B_E(\nu)}{\mathcal B_E(\nu)^2+\mathcal B_O(\nu)^2}\,,\qquad \widetilde G_O^{\rm top}(\nu)=-\frac{i\nu^2}{4\pi^2N}\,\frac{\mathcal B_O(\nu)}{\mathcal B_E(\nu)^2+\mathcal B_O(\nu)^2}\,,
\end{equation}
where
\begin{equation}
\mathcal B_E(\nu)=\begin{cases}
\widetilde B^{(1)}_{\hat\Delta_\phi,0}(\nu),&\text{unHiggsed scalars},\\
\widetilde B_{\rm H}^{(1)}(\nu),&\text{Higgsed scalars},\\
\widetilde B^{(1,E)}_{\hat\Delta_\psi,1/2}(\nu),&\text{fermions},
\end{cases}\quad \mathcal B_O(\nu)=\begin{cases}
\frac{\kappa\nu}{2\pi},&\text{scalars},\\[3pt]
\frac{\kappa\nu}{2\pi}-i\widetilde B^{(1,O)}_{\hat\Delta_\psi,1/2}(\nu),&\text{fermions}.
\end{cases}
\end{equation}

The spin-one boundary spectra follow instead from the poles of the dressed gauge propagator, and are shown as functions of $\kappa$ in figures~\ref{fig:scalar-conformal-spin1-spectrum} and~\ref{fig:fermion-conformal-spin1-spectrum}. When the effective Chern--Simons level is nonzero and the gauge symmetry is unbroken, the pole at $\nu=0$ gives the protected current, with $\hat\Delta^{\varsigma}_{0,1}=1$ for scalars and $\hat\Delta^{\varsigma_{\rm eff}}_{0,1}=1$ for fermions; Higgsing removes it. The remaining poles form two helicity towers. As $|\kappa|\to\infty$, the unprotected levels approach generalized-free double-trace dimensions, with $O(1/|\kappa|)$ corrections of opposite sign in the two helicity sectors. In the scalar theories, the two helicities become degenerate as $\kappa\to0$. Finally, the bulk data extracted with different boundary conditions agree, as required, once their different one-point coefficients are taken into account.

\subsection{Critical scalar QED$_3$}
\label{sec:scalar-critical}
We consider the critical scalar theory, obtained by taking $\lambda\to +\infty$ at fixed $\mu^2$, $h$, and $\kappa$, coupled to the $\alpha\to +\infty$ limit of Maxwell--Chern--Simons theory. The unHiggsed and Higgsed saddles realize ordinary (symmetry-preserving) and normal (symmetry-breaking) boundary conditions, respectively, for the scalars. There is no special boundary condition in this case, as the maximum $\Sigma_{{\rm uH},-}^{\rm AdS}$ leaves the unitarity window.

The propagating scalar singlet is represented by the $\chi$ field. To identify the corresponding bulk primary operator, we separate the conformal curvature coupling from the mass term. On unit-radius AdS$_3$, the conformal scalar Laplacian is $-\nabla^2+R/8=-\nabla^2-3/4$, whereas $\mu^2$ in our action includes both the mass and curvature contributions. We then write
\begin{equation}
    -\nabla^2+\mu^2+\frac{\chi}{\sqrt N}=-\nabla^2-\frac34+\frac{\boldsymbol{\chi}}{\sqrt N}\,,\qquad \boldsymbol{\chi}\equiv\chi+\sqrt N\left(\mu^2+\frac34\right)\mathbf1\,.
\end{equation}
The operator $\boldsymbol{\chi}$ is the Weyl transform of the flat-space critical scalar primary. Evaluating its expectation value on the saddle $\chi=\sqrt N X+\delta\chi$ gives, at leading large $N$,
\begin{equation}\label{eq:critical-scalar-onepoint}
    a_\chi=\langle\boldsymbol{\chi}\rangle_{\rm AdS}=\sqrt N\left(\mu^2+X+\frac34\right)\,.
\end{equation}

\subsubsection{Ordinary boundary conditions}
\label{sec:ordinary-bc}
On the symmetry-preserving branch, taking $\lambda\to+\infty$ at fixed $\mu^2\geq-1$ gives
\begin{equation*}
\Sigma_{\rm uH}^{\rm AdS}\to0\,,\qquad \mu^2+X_{\rm uH}\to-1\,,\qquad \zeta_{\rm uH}\to+\infty\,,\qquad \hat\Delta_\phi\to1\,.
\end{equation*}
This saddle realizes the ordinary boundary condition of the critical scalar theory, with one-point coefficient $a_\chi^{\rm ord}=-\sqrt{N}/4$. The scalar singlet propagator becomes
\begin{equation}\label{eq:KinvuHcritchichi}
    \bigl[\widetilde K_{\rm uH,crit}^{-1}(\nu)\bigr]_{\chi\chi}=-\bigl(\widetilde B^{(0)}_{1,0}(\nu)\bigr)^{-1}=-8\nu\coth\frac{\pi\nu}{2}\,.
\end{equation}
Its short-distance behavior fixes the bulk dimension and two-point normalization, cf. appendix~\ref{subapp:extracting-bulk-data},
\begin{equation}\label{eq:crit-bulk-data}
\Delta_\chi^{{\rm crit}}=2\,,\qquad C_\chi^{{\rm crit}}=\frac{8}{\pi^2}\,.
\end{equation}

The poles and residues of~\eqref{eq:KinvuHcritchichi} determine the scalar boundary data,
\begin{equation}
    \hat\Delta_{\hat\chi_n}=3+2n\,,\qquad \varrho_{\chi \hat \chi_n} =64(n+1)\,,\qquad n\in\mathbb Z_{\geq0}\,.
\end{equation}
The lowest state has $\hat\Delta_{\hat\chi_0}=3$ and is identified with the displacement operator.

We next consider the topological current correlator. The scalar spin-one bubble is
\begin{equation}
    \widetilde B^{(1)}_{1,0}(\nu)=-\frac{\nu^3}{16(\nu^2+1)}\coth\frac{\pi\nu}{2}\,.
\end{equation}
The bulk-channel expansion of the topological current correlator~\eqref{eq:Gtop-general} determines its normalization,
\begin{equation}\label{eq:CJtop}
    C_{J^{\rm top}}(\kappa)=\frac{8}{\pi^4N}\,\frac{1}{1+\left(\frac{8\kappa}{\pi}\right)^2}\,,
\end{equation}
which at $\kappa=0$ agrees with the large-$N$ QED$_3$ result of \cite{Chester:2016wrc,Giombi:2016fct}, 
as well as the bulk OPE coefficients associated with the two $J^{\rm top}J^{\rm top}\chi$ tensor structures,
\begin{equation}
    C^{E,{\rm crit}}_{J^{\rm top}J^{\rm top}}{}^{\chi}=-\frac{2C_{J^{\rm top}}(\kappa)}{\sqrt{N}}\,\frac{1-\left(\frac{8\kappa}{\pi}\right)^2}{1+\left(\frac{8\kappa}{\pi}\right)^2}\,,\qquad C^{O,{\rm crit}}_{J^{\rm top}J^{\rm top}}{}^{\chi}=-\frac{8i\kappa\,C_{J^{\rm top}}(\kappa)}{\sqrt{N}}\,\frac{1}{1+\left(\frac{8\kappa}{\pi}\right)^2}\,.
\end{equation}

Taking the conformal limit of the dressed gauge propagator in section~\ref{subsec:scalar-unhiggsed}, its nonzero poles give the spin-one boundary dimensions through~\eqref{eq:spin1-param},
\begin{equation}
    0=\frac{s\kappa(\hat\Delta-1)}{2\pi}+\frac{(\hat\Delta-1)^3}{16\bigl[1-(\hat\Delta-1)^2\bigr]}\cot\frac{\pi(\hat\Delta-1)}{2}\,,\qquad s=\pm1\,.
\end{equation}
As $\kappa\to0$, the lightest state of helicity $-\varsigma$ approaches dimension one, while the higher levels become degenerate in the two helicities,
\begin{equation}
\hat\Delta^{-\varsigma}_{0,1}=1+4|\kappa|+O(|\kappa|^3)\,,\qquad \hat\Delta^s_{n,1}\to2n+2\,,\quad n\geq1\,,\ s=\pm1\,.
\end{equation}
At large $|\kappa|$, the unprotected levels approach
\begin{equation}
\hat\Delta^\varsigma_{n,1}\to2n+1\,,\quad n\geq1\,,\qquad \hat\Delta^{-\varsigma}_{n,1}\to2n+3\,,\quad n\geq0\,.
\end{equation}

\subsubsection{Normal boundary conditions}
On the Higgsed branch, the same critical limit gives
\begin{equation}
    \Sigma_{\rm H}^{\rm AdS}\longrightarrow0\,,\qquad v^2\longrightarrow\frac{1}{4\pi}\,,\qquad \zeta_{\rm H}\longrightarrow+\infty\,,
\end{equation}
while $\mu^2+X_{\rm H}=0$ gives $a_\chi^{\rm norm}=3\sqrt N/4=-3a_\chi^{\rm ord}$. The condensate breaks the global $PSU(N)$ symmetry, and the $N-1$ complex Goldstone fields $\pi_a$ give rise to boundary tilt operators of dimension $\hat\Delta_\pi=2$.

The scalar singlet propagator is
\begin{equation}
\bigl[\widetilde K_{\rm H,crit}^{-1}(\nu)\bigr]_{\chi\chi}=-\bigl[\widetilde B_{\rm H}^{(0)}(\nu)\bigr]^{-1}=-8\nu\coth\frac{\pi\nu}{2}\,.
\end{equation}
This coincides with the ordinary critical propagator, so the scalar singlet bulk normalization and boundary data are unchanged.

In the spin-one sector, the condensate contributes to the transverse inverse gauge propagator. The relevant combination is
\begin{equation}
    \widetilde B_{\rm H}^{(1)}(\nu)=-\frac{\nu(\nu^2+4)}{16(\nu^2+1)}\coth\frac{\pi\nu}{2}\,.
\end{equation}
The bulk OPE coefficients for the topological current agree with those of the ordinary boundary condition in both tensor structures.

In the dressed gauge propagator of section~\ref{subsec:scalar-higgsed}, the Higgs term removes the pole at $\nu=0$, so there is no protected current. In the conformal limit, the remaining poles satisfy
\begin{equation}
    0=\frac{s\kappa(\hat\Delta-1)}{2\pi}-\frac{(\hat\Delta-1)\bigl[4-(\hat\Delta-1)^2\bigr]}{16\bigl[1-(\hat\Delta-1)^2\bigr]}\cot\frac{\pi(\hat\Delta-1)}{2}\,.
\end{equation}
As $\kappa\to0$, no state approaches dimension one; the two helicities become degenerate at
\begin{equation}
\hat\Delta^s_{n,1}\to2n+4\,,\qquad n\geq0\,.
\end{equation}
At large $|\kappa|$, the lightest state has helicity $\varsigma$ and approaches the would-be conserved current, while the higher levels approach the pion double-trace dimensions:
\begin{equation}
\hat\Delta^\varsigma_{0,1}=1+\frac1{|\kappa|}+O(|\kappa|^{-3})\,,\qquad \hat\Delta^\varsigma_{n,1}\to2n+3\,,\ n\geq1\,,\qquad \hat\Delta^{-\varsigma}_{n,1}\to2n+5\,,\ n\geq0\,.
\end{equation}

Figure~\ref{fig:scalar-conformal-spin1-spectrum}(a) compares the spin-one spectra for ordinary and normal boundary conditions.

\definecolor{helicityBlue}{rgb}{0.12,0.40,0.70}
\definecolor{helicityOrange}{rgb}{0.80,0.32,0.13}

\begin{figure}[t]
  \centering
  \begin{subfigure}[t]{0.49\textwidth}
    \centering
    \begin{tikzpicture}
      \node[anchor=south west,inner sep=0] (plot) at (0,0)
        {\includegraphics[width=0.88\linewidth]{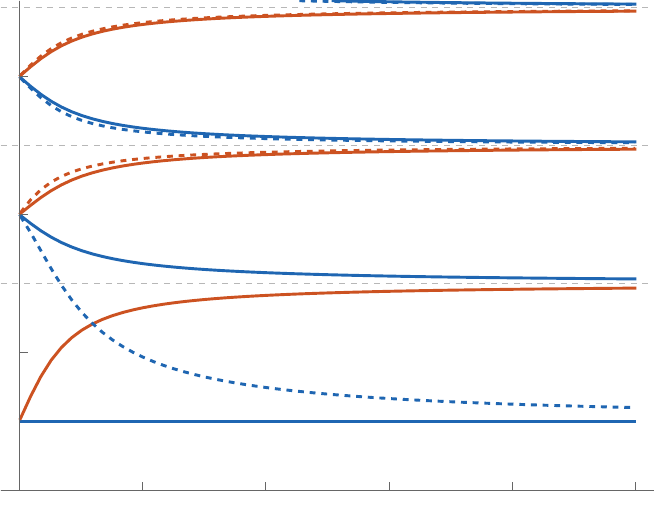}};
      \begin{scope}[x={(plot.south east)},y={(plot.north west)}]
        \node[anchor=south,inner sep=2pt]
          at (0.02927,1.005) {$\hat\Delta$};
        \node[anchor=west,inner sep=2pt]
          at (1.005,0.05870) {$\kappa$};
        \foreach \k in {0,1,2,3,4,5} {
          \node[anchor=north,inner sep=2pt,font=\small]
            at ({0.02927+0.18766*\k},0.05870) {$\k$};
        }
        \foreach \d in {2,4,6} {
          \node[anchor=east,inner sep=2pt,font=\small]
            at (0.02927,{0.05870+0.13238*\d}) {$\d$};
        }
      \end{scope}
    \end{tikzpicture}
    \caption{Critical}
  \end{subfigure}\hfill
  \begin{subfigure}[t]{0.49\textwidth}
    \centering
    \begin{tikzpicture}
      \node[anchor=south west,inner sep=0] (plot) at (0,0)
        {\includegraphics[width=0.88\linewidth]{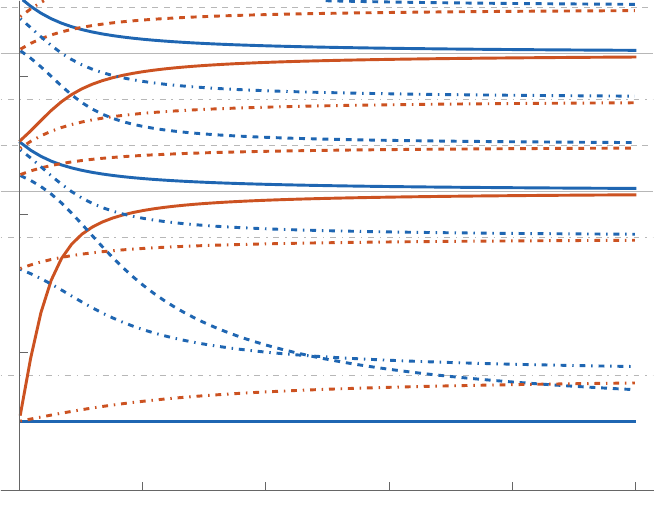}};
      \begin{scope}[x={(plot.south east)},y={(plot.north west)}]
        \node[anchor=south,inner sep=2pt]
          at (0.02927,1.005) {$\hat\Delta$};
        \node[anchor=west,inner sep=2pt]
          at (1.005,0.05870) {$\kappa$};
        \foreach \k in {0,1,2,3,4,5} {
          \node[anchor=north,inner sep=2pt,font=\small]
            at ({0.02927+0.18766*\k},0.05870) {$\k$};
        }
        \foreach \d in {2,4,6} {
          \node[anchor=east,inner sep=2pt,font=\small]
            at (0.02927,{0.05870+0.13238*\d}) {$\d$};
        }
      \end{scope}
    \end{tikzpicture}
    \caption{Tricritical, $h=7\pi^2$}
  \end{subfigure}

  \par\smallskip
  \begin{tikzpicture}[baseline=-0.5ex]
    \draw[helicityBlue,line width=1pt]
      (0,0)--(0.4,0);
    \node[anchor=west,font=\small]
      at (0.45,0) {$s=+1$};

    \draw[helicityOrange,line width=1pt]
      (1.8,0)--(2.2,0);
    \node[anchor=west,font=\small]
      at (2.25,0) {$s=-1$};

    \draw[black,line width=1pt]
      (3.6,0)--(4.0,0);
    \node[anchor=west,font=\small]
      at (4.05,0) {ordinary};

    \draw[black,line width=1pt,
      dash pattern=on 2.8pt off 2.8pt]
      (5.9,0)--(6.3,0);
    \node[anchor=west,font=\small]
      at (6.35,0) {normal};

    \draw[black,line width=1pt,
      dash pattern=on 0.7pt off 2.8pt on 2.8pt off 2.8pt]
      (8.0,0)--(8.5,0);
    \node[anchor=west,font=\small]
      at (8.55,0) {special};
  \end{tikzpicture}

  \captionsetup{singlelinecheck=false}
  \caption{Spin-one boundary spectra of (a) the critical scalar theory and (b) the tricritical theory at $h=7\pi^2$, for $\kappa>0$. Colors label helicity; solid, dashed, and dot-dashed curves denote ordinary, normal, and special boundary conditions. Ordinary and special spectra share the protected $s=+1$ current at $\hat\Delta=1$. Gray lines mark generalized-free dimensions: $3+2n$ (ordinary) and $5+2n$ (normal) in (a); $13/3+2n$ (ordinary), $5+2n$ (normal), and $5/3+2n$ (special) in (b), with $n\geq0$.}
  \label{fig:scalar-conformal-spin1-spectrum}
\end{figure}

\subsection{Tricritical scalar QED$_3$}
\label{sec:scalar-tricritical}
We next consider the scalar theory at $\mu^2=-\frac34$ and $\lambda=0$, taking $\alpha\to+\infty$ with $h$ and $\kappa$ fixed. The scalar mass takes its conformally coupled value, while the sextic interaction is marginal at leading large $N$. We discuss separately the ordinary, special, and normal boundary conditions, realized by the unHiggsed minimum for $h<16\pi^2$, the unHiggsed maximum for $0<h<12\pi^2$, and the Higgsed minimum for $h>0$, respectively. Without the gauge field, the same theory on flat half-space was studied at large $N$ in \cite{Herzog:2020lel}, including the stability of its boundary phases and the bulk-channel decomposition of the current two-point function.

The scalar singlet primary is represented by the renormalized bilinear $|\Phi|^2/\sqrt N$. The auxiliary-field constraint identifies this operator with the field $\sigma$, so that $\boldsymbol{\sigma}=\sigma$. Evaluating the constraint on a homogeneous saddle gives
\begin{equation}\label{eq:tricritical-scalar-onepoint}
    a_\sigma=\langle\boldsymbol{\sigma}\rangle_{\rm AdS}=\frac{\langle|\Phi|^2\rangle_{\rm ren}}{\sqrt N}=\sqrt N\,\Sigma^{\rm AdS}\,.
\end{equation}
Here $\Sigma^{\rm AdS}$ is the value of $\Sigma$ on the relevant saddle.

The ordinary and special boundary conditions correspond to the two unHiggsed saddles. Writing $\beta=\hat\Delta_\phi-1$, their common saddle relations are
\begin{equation}\label{eq:tricritical-saddle}
\Sigma_{\rm uH}^{\rm AdS}=-\frac{\beta}{4\pi}\,,\qquad X_{\rm uH}=\frac{h\beta^2}{16\pi^2}\,,\qquad \zeta_{\rm uH}=-\frac{h\beta}{2\pi}\,,\qquad \mu^2+X_{\rm uH}=\beta^2-1\,,
\end{equation}
with
\begin{equation}\label{eq:scalar-special-beta}
\beta=\beta_\pm\equiv\pm\frac{2\pi}{\sqrt{16\pi^2-h}}\,,\qquad a_\sigma=-\frac{\sqrt N\,\beta}{4\pi}\,.
\end{equation}
The upper and lower signs give the ordinary and special branches, respectively. Their scalar singlet correlator is
\begin{equation}\label{eq:tricritical-common-correlator}
\widetilde G_{\sigma\sigma}(\nu)=\frac{\widetilde B^{(0)}_{1+\beta,0}(\nu)}{1+\zeta_{\rm uH}\widetilde B^{(0)}_{1+\beta,0}(\nu)}\,.
\end{equation}
The pole at $\hat\Delta=3$, which corresponds to the boundary displacement operator, is common to both branches, as can be checked explicitly.

\subsubsection{Ordinary boundary conditions}
\label{sec:scalar-tricritical-ordinary}
The ordinary branch has $\beta=\beta_+$ and is a local minimum, since $\zeta_{\rm uH}+8\pi\beta=\sqrt{16\pi^2-h}>0$.

The bulk-channel expansion, detailed in appendix~\ref{subapp:extracting-bulk-data}, gives
\begin{equation}\label{eq:tri-bulk-data}
\Delta_\sigma^{\rm tri}=1\,,\qquad C_\sigma^{\rm tri}=\frac1{16\pi^2}\,,\qquad C^{\rm tri}_{\sigma\sigma}{}^\sigma=\frac1{2\pi\sqrt N}\left(1-\frac h{64}\right)\,.
\end{equation}
This agrees with the leading flat-space result for the scalar singlet three-point function in the sextic vector model \cite{Goykhman:2020ffn}.

For $h\neq0$, the scalar boundary dimensions obey
\begin{equation}\label{eq:tricritical-pole-condition}
1+\zeta_{\rm uH}\widetilde B^{(0)}_{1+\beta,0}\!\left(-i(\hat\Delta-1)\right)=0\,.
\end{equation}
The displacement is the lowest pole, $\hat\Delta_{\hat\sigma_0}=3$. The higher roots are $\hat\Delta_{\hat\sigma_n}\in (2\hat\Delta_\phi+2n-2,2\hat\Delta_\phi+2n)$, $n\geq 1$, and their bulk-to-boundary OPE coefficients can be computed by residues. For $h\to0$, the boundary data simplify to
\begin{equation}
    \hat\Delta_{\hat\sigma_n}=3+2n\,,\qquad \varrho_{\sigma \hat\sigma_n}=\frac{1}{4(n+1)}\,,\qquad n\geq0\,.
\end{equation}

The topological-current correlator follows from~\eqref{eq:Gtop-general}, using the scalar bubble~\eqref{eq:B1-scalar} at $\hat\Delta_\phi=1+\beta$. Its normalization is~\eqref{eq:CJtop}, as in the critical theory. Dividing the scalar-exchange coefficients by $a_\sigma^{\rm ord}$, we obtain
\begin{equation}\label{eq:tricritical-CEO}
    C^{E,{\rm tri}}_{J^{\rm top}J^{\rm top}}{}^{\sigma}=-\frac{4\pi\,C_{J^{\rm top}}(\kappa)}{\sqrt{N}}\frac{1-\left(\frac{8\kappa}{\pi}\right)^2}{1+\left(\frac{8\kappa}{\pi}\right)^2}\,,\qquad C^{O,{\rm tri}}_{J^{\rm top}J^{\rm top}}{}^{\sigma}=-\frac{256i\kappa\,C_{J^{\rm top}}(\kappa)}{\pi\sqrt{N}}\frac{1}{1+\left(\frac{8\kappa}{\pi}\right)^2}\,.
\end{equation}
Both coefficients are independent of $h$ at this order.

The nonzero poles of the dressed gauge propagator in section~\ref{subsec:scalar-unhiggsed} give, in the conformal limit,
\begin{equation}\label{eq:tricritical-vector-spectrum}
0=\frac{s\kappa(\hat\Delta-1)}{2\pi}+\widetilde B^{(1)}_{1+\beta,0}\!\left(-is(\hat\Delta-1)\right)\,,\qquad \hat\Delta>1\,.
\end{equation}
As $\kappa\to0$, the lightest state of helicity $-\varsigma$ approaches dimension one,
\begin{equation}\label{eq:tricritical-small-kappa}
    \hat\Delta^{-\varsigma}_{0,1}=1+\frac{4|\kappa|}{2\beta^2\psi'(1+\beta)-2\beta+1}+O(|\kappa|^3)\,,
\end{equation}
with $\psi'$ the trigamma function, while the remaining levels split linearly around their degenerate $\kappa=0$ values. At large $|\kappa|$, the unprotected levels approach
\begin{equation}
\hat\Delta^\varsigma_{n,1}\to2\hat\Delta_\phi+2n-1\,,\ n\geq1\,,\qquad \hat\Delta^{-\varsigma}_{n,1}\to2\hat\Delta_\phi+2n+1\,,\ n\geq0\,.
\end{equation}

\subsubsection{Special boundary conditions}
\label{sec:scalar-special}
The special branch has $\beta=\beta_-$ in~\eqref{eq:scalar-special-beta}. For $h<0$ this saddle violates \eqref{eq:scalar-BF-stability}, and for $h>12\pi^2$ it violates unitarity. For $0<h<12\pi^2$, $0<\hat\Delta_\phi<\frac12$ and $a_\sigma^{\rm sp}=-a_\sigma^{\rm ord}$. Since $\zeta_{\rm uH}+8\pi\beta=-\sqrt{16\pi^2-h}<0$, the lightest singlet is relevant. Without the gauge field, the ordinary and special saddles reduce to the two disordered phases of \cite{Herzog:2020lel}.

The scalar singlet correlator and its bulk data are those of section~\ref{sec:scalar-tricritical-ordinary}, with $\beta$ given by \eqref{eq:scalar-special-beta}. The boundary dimensions obey \eqref{eq:tricritical-pole-condition}. The lowest root, $\hat\Delta_{\hat\sigma_0}\in(2-2\hat\Delta_\phi,2)$, is relevant. The next root is the shared displacement pole, $\hat\Delta_{\hat\sigma_1}=3$. The higher roots are $\hat\Delta_{\hat\sigma_n}\in(2\hat\Delta_\phi+2n,2\hat\Delta_\phi+2n+2)$, $n\geq2$. For $h\to0$, $\hat\Delta_\phi\to\frac12$ and $\hat\Delta_{\hat\sigma_0}\to1$.

For the topological current, the scalar-exchange coefficients change sign together with $a_\sigma$, so that \eqref{eq:CJtop} and \eqref{eq:tricritical-CEO} are unchanged. The asymptotic spin-one spectra of section~\ref{sec:scalar-tricritical-ordinary}, including \eqref{eq:tricritical-small-kappa}, hold with $\beta$ given by \eqref{eq:scalar-special-beta}. In particular, at large $|\kappa|$ the lightest unprotected states of both helicities approach $2\hat\Delta_\phi+1<2$.

\subsubsection{Normal boundary conditions}
The Higgsed saddle at the same bulk tuning is
\begin{equation}
    \zeta_{\rm H}=\sqrt{3h}\,,\qquad \Sigma_{\rm H}^{\rm AdS}=\frac{3}{2\zeta_{\rm H}}\,,\qquad v^2=\frac{1}{4\pi}+\frac{3}{2\zeta_{\rm H}}\,,\qquad X_{\rm H}=\frac34\,.
\end{equation}
The condensate breaks the global $PSU(N)$ symmetry. As in the normal critical theory, the $N-1$ complex Goldstone fields give rise to boundary tilt operators of dimension two.

The scalar singlet correlator is
\begin{equation}
    \widetilde G_{\sigma\sigma}(\nu)=\frac{\widetilde B_{\rm H}^{(0)}(\nu)}{1+\zeta_{\rm H}\widetilde B_{\rm H}^{(0)}(\nu)}=\frac{\frac{\nu^2+1}{8\nu}\tanh\frac{\pi\nu}{2}+\frac{3}{\zeta_{\rm H}}}{\nu^2+4+\frac{\zeta_{\rm H}(\nu^2+1)}{8\nu}\tanh\frac{\pi\nu}{2}}\,.
\end{equation}
Its bulk-channel expansion reproduces the bulk data of section~\ref{sec:scalar-tricritical-ordinary}, while \eqref{eq:tricritical-scalar-onepoint} gives $a_\sigma^{\rm norm}=3\sqrt N/(2\zeta_{\rm H})$.

The scalar boundary dimensions obey
\begin{equation}
    \frac{1-(\hat\Delta-1)^2}{8(\hat\Delta-1)}\tan\frac{\pi(\hat\Delta-1)}{2}+\frac{4-(\hat\Delta-1)^2}{\zeta_{\rm H}}=0\,.
\end{equation}
This equation contains the exact solution $\hat\Delta=3$, identified with the displacement operator, together with an $h$-dependent tower. Figure~\ref{fig:tricritical-scalar-singlets} compares the low-lying scalar-singlet spectra for ordinary, normal, and special boundary conditions; only the special branch contains a relevant singlet.

In the spin-one sector, the combination entering the transverse inverse gauge propagator is
\begin{equation}
    \widetilde B_{\rm H}^{(1)}(\nu)=-\frac{\nu(\nu^2+4)}{16(\nu^2+1)}\coth\frac{\pi\nu}{2}-\frac{3}{\zeta_{\rm H}}\,.
\end{equation}
The topological current normalization obtained from~\eqref{eq:Gtop-general} is again~\eqref{eq:CJtop}, and we reproduce both OPE coefficients obtained with ordinary boundary conditions, once we account for the different one-point coefficients.

The dressed gauge propagator in section~\ref{subsec:scalar-higgsed} has no protected-current pole. Its conformal limit gives the spin-one boundary spectrum through
\begin{equation}\label{eq:tricritical-normal-vector-spectrum}
0=\frac{s\kappa(\hat\Delta-1)}{2\pi}+\widetilde B_{\rm H}^{(1)}\!\left(-is(\hat\Delta-1)\right)\,,\qquad \hat\Delta>1\,.
\end{equation}
At $\kappa=0$, the two helicities are degenerate, with
\begin{equation}
\hat\Delta^s_{n,1}=\hat\Delta_{n,1}^{(0)}(\zeta_{\rm H})\in(2n+4,\,2n+5)\,,\qquad n\geq0\,.
\end{equation}
At large $|\kappa|$ and fixed $h>0$, the lightest state again approaches dimension one,
\begin{equation}
\hat\Delta^\varsigma_{0,1}=1+\frac{1+6\pi/\zeta_{\rm H}}{|\kappa|}+O(|\kappa|^{-3})\,,
\end{equation}
while the higher levels approach those of the normal critical theory.

Figure~\ref{fig:scalar-conformal-spin1-spectrum}(b) compares the ordinary, special, and normal spectra at $h=7\pi^2$.

\subsection{Conformal fermionic theories}
\label{sec:fermion-conformal}
We now consider the regular fermion theory and the Gross--Neveu critical family, retaining the flavor-preserving boundary conditions labeled by $\eta=\pm1$. The Gross--Neveu family admits in addition an alternate-quantization boundary condition for each $\eta$ with $\eta y>0$. Figure~\ref{fig:fermion-conformal-spin1-spectrum} compares their spin-one spectra. In both theories, the large-$\nu$ expansion of the current correlator depends on $\kappa$ rather than $\kappa_{\rm eff}$. This is expected: the large-$\nu$ expansion probes short distances and determines bulk data, which cannot depend on the boundary condition, whereas the shift $\kappa_{\rm eff}=\kappa+\eta/2$ arises from the infrared, small-$\nu$ behavior of the parity-odd polarization.

\definecolor{singletBlue}{rgb}{0.12,0.40,0.70}
\definecolor{singletOrange}{rgb}{0.80,0.32,0.13}
\definecolor{singletPurple}{rgb}{0.47,0.33,0.61}

\begin{figure}[t]
  \centering
  \begin{subfigure}[t]{0.49\textwidth}
    \centering
    \begin{tikzpicture}
      \node[anchor=south west,inner sep=0] (plot) at (0,0)
        {\includegraphics[width=0.84\linewidth]{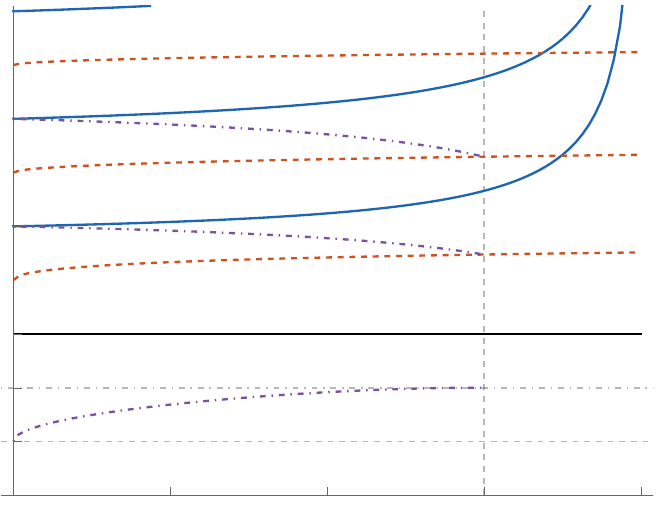}};
      \begin{scope}[x={(plot.south east)},y={(plot.north west)}]
        \node[anchor=south,inner sep=2pt]
          at (0.02045,1.005) {$\hat\Delta$};
        \node[anchor=west,inner sep=2pt]
          at (1.005,0.05669) {$\frac{h}{\pi^2}$};
        \foreach \h in {0,4,8,12,16} {
          \node[anchor=north,inner sep=2pt,font=\small]
            at ({0.02045+0.059746*\h},0.05669) {$\h$};
        }
        \foreach \d in {1,2,3,5,7,9} {
          \node[anchor=east,inner sep=2pt,font=\small]
            at (0.02045,{0.05669+0.102419*\d}) {$\d$};
        }
      \end{scope}
    \end{tikzpicture}

    \par\smallskip
    \begin{tikzpicture}[baseline=-0.5ex]
      \draw[singletBlue,line width=1pt]
        (0,0)--(0.5,0);
      \node[anchor=west,font=\small]
        at (0.6,0) {ordinary};

      \draw[singletOrange,line width=1pt,
        dash pattern=on 2.8pt off 2.8pt]
        (0,-0.4)--(0.5,-0.4);
      \node[anchor=west,font=\small]
        at (0.6,-0.4) {normal};

      \draw[singletPurple,line width=1pt,
        dash pattern=on 0.7pt off 2.8pt on 2.8pt off 2.8pt]
        (0,-0.8)--(0.5,-0.8);
      \node[anchor=west,font=\small]
        at (0.6,-0.8) {special};
    \end{tikzpicture}
    \caption{Tricritical scalar}
    \label{fig:tricritical-scalar-singlets}
  \end{subfigure}\hfill
  \begin{subfigure}[t]{0.49\textwidth}
    \centering
    \begin{tikzpicture}
      \node[anchor=south west,inner sep=0] (plot) at (0,0)
        {\includegraphics[width=0.84\linewidth]{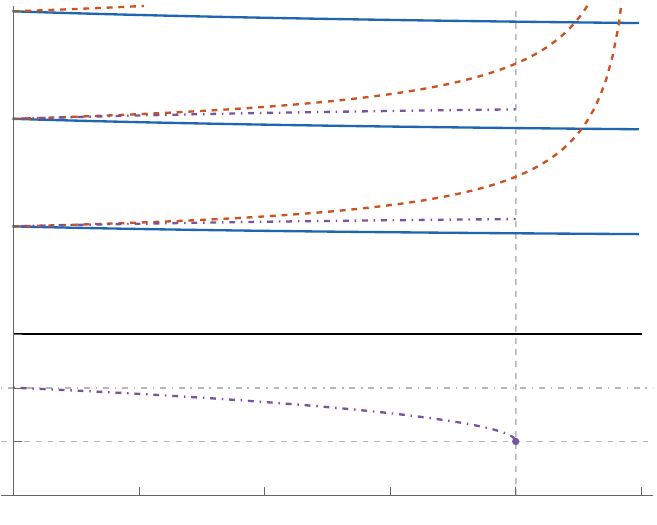}};
      \begin{scope}[x={(plot.south east)},y={(plot.north west)}]
        \node[anchor=south,inner sep=2pt]
          at (0.02045,1.005) {$\hat\Delta$};
        \node[anchor=west,inner sep=2pt]
          at (1.005,0.05669) {$2\pi y$};
        \foreach \t/\lab in
          {0/0,0.2/0.2,0.4/0.4,0.6/0.6,0.8/0.8,1/1} {
          \node[anchor=north,inner sep=2pt,font=\small]
            at ({0.02045+0.95593*\t},0.05669) {$\lab$};
        }
        \foreach \d in {1,2,3,5,7,9} {
          \node[anchor=east,inner sep=2pt,font=\small]
            at (0.02045,{0.05669+0.102419*\d}) {$\d$};
        }
      \end{scope}
    \end{tikzpicture}

    \par\smallskip
    \begin{tikzpicture}[baseline=-0.5ex]
      \draw[singletBlue,line width=1pt]
        (0,0)--(0.5,0);
      \node[anchor=west,font=\small]
        at (0.6,0) {standard $\eta=+$};

      \draw[singletOrange,line width=1pt,
        dash pattern=on 2.8pt off 2.8pt]
        (0,-0.4)--(0.5,-0.4);
      \node[anchor=west,font=\small]
        at (0.6,-0.4) {standard $\eta=-$};

      \draw[singletPurple,line width=1pt,
        dash pattern=on 0.7pt off 2.8pt on 2.8pt off 2.8pt]
        (0,-0.8)--(0.5,-0.8);
      \node[anchor=west,font=\small]
        at (0.6,-0.8) {alternate $\eta=+$};
    \end{tikzpicture}
    \caption{Gross--Neveu}
    \label{fig:gn-scalar-singlets}
  \end{subfigure}

  \captionsetup{singlelinecheck=false}
  \caption{Low-lying scalar-singlet boundary spectra in (a) the tricritical scalar theory and (b) the Gross--Neveu theory for $y\geq0$. The shared displacement operator at $\hat\Delta=3$ is shown in black; horizontal gray guides mark BF saturation ($\hat\Delta=1$) and marginality ($\hat\Delta=2$). Vertical guides mark the special-BC unitarity endpoint $h=12\pi^2$ and the alternate GN BF threshold $2\pi y\simeq0.80067$, where the lowest singlet reaches $\hat\Delta=1$.}
  \label{fig:scalar-singlet-spectra}
\end{figure}

\definecolor{helicityBlue}{rgb}{0.12,0.40,0.70}
\definecolor{helicityOrange}{rgb}{0.80,0.32,0.13}

\begin{figure}[t]
  \centering
  \begin{subfigure}[t]{0.49\textwidth}
    \centering
    \begin{tikzpicture}
      \node[anchor=south west,inner sep=0] (plot) at (0,0)
        {\includegraphics[width=0.88\linewidth]{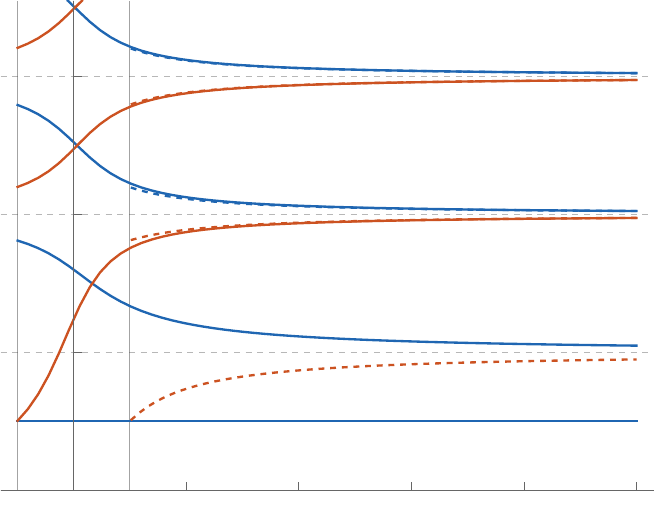}};
      \begin{scope}[x={(plot.south east)},y={(plot.north west)}]
        \node[anchor=south,inner sep=2pt]
          at (0.11182,1.005) {$\hat\Delta$};
        \node[anchor=west,inner sep=2pt]
          at (1.005,0.05870) {$\kappa$};
        \foreach \k in {0,1,2,3,4,5} {
          \node[anchor=north,inner sep=2pt,font=\small]
            at ({0.11182+0.17163*\k},0.05870) {$\k$};
        }
        \foreach \d in {2,4,6} {
          \node[anchor=east,inner sep=2pt,font=\small]
            at (-0.01,{0.05870+0.13238*\d}) {$\d$};
        }
      \end{scope}
    \end{tikzpicture}
    \caption{Regular fermion}
  \end{subfigure}\hfill
  \begin{subfigure}[t]{0.49\textwidth}
    \centering
    \begin{tikzpicture}
      \node[anchor=south west,inner sep=0] (plot) at (0,0)
        {\includegraphics[width=0.88\linewidth]{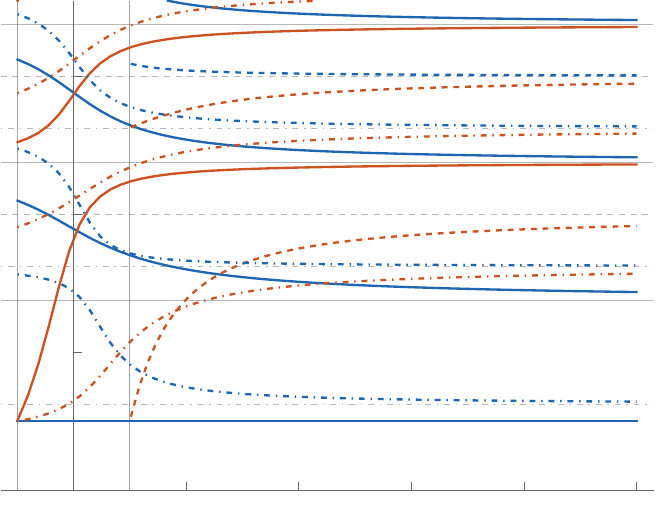}};
      \begin{scope}[x={(plot.south east)},y={(plot.north west)}]
        \node[anchor=south,inner sep=2pt]
          at (0.11182,1.005) {$\hat\Delta$};
        \node[anchor=west,inner sep=2pt]
          at (1.005,0.05870) {$\kappa$};
        \foreach \k in {0,1,2,3,4,5} {
          \node[anchor=north,inner sep=2pt,font=\small]
            at ({0.11182+0.17163*\k},0.05870) {$\k$};
        }
        \foreach \d in {2,4,6} {
          \node[anchor=east,inner sep=2pt,font=\small]
            at (-0.01,{0.05870+0.13238*\d}) {$\d$};
        }
      \end{scope}
    \end{tikzpicture}
    \caption{Gross--Neveu, $y=3/(8\pi)$}
  \end{subfigure}

  \par\smallskip
  \begin{tikzpicture}[baseline=-0.5ex]
    \draw[helicityBlue,line width=1pt] (0,0)--(0.4,0);
    \node[anchor=west,font=\small] at (0.45,0) {$s=+1$};
    \draw[helicityOrange,line width=1pt] (1.8,0)--(2.2,0);
    \node[anchor=west,font=\small] at (2.25,0) {$s=-1$};
  \end{tikzpicture}

  \par\smallskip
  \begin{tikzpicture}[baseline=-0.5ex]
    \draw[black,line width=1pt] (0,0)--(0.5,0);
    \node[anchor=west,font=\small]
      at (0.55,0) {standard $\eta=+$};

    \draw[black,line width=1pt,
      dash pattern=on 2.8pt off 2.8pt]
      (3.6,0)--(4.1,0);
    \node[anchor=west,font=\small]
      at (4.15,0) {standard $\eta=-$};

    \draw[black,line width=1pt,
      dash pattern=on 0.7pt off 2.8pt on 2.8pt off 2.8pt]
      (7.2,0)--(7.7,0);
    \node[anchor=west,font=\small]
      at (7.75,0) {alternate $\eta=+$};
  \end{tikzpicture}

  \captionsetup{singlelinecheck=false}
  \caption{Spin-one boundary spectra of (a) the regular fermion theory and (b) the Gross--Neveu theory at $y=3/(8\pi)$. Colors label helicity and line styles label boundary conditions. Each branch is shown for $\kappa+\eta/2>0$ and includes the protected $s=+1$ current at $\hat\Delta=1$. Vertical gray lines mark $\kappa=\pm1/2$. Horizontal guides mark generalized-free dimensions: $2+2n$ in (a); $2+2/\sqrt7+2n$, $4+2n$, and $2-2/\sqrt7+2n$ for standard $\eta=+$, standard $\eta=-$, and alternate $\eta=+$ in (b), respectively, with $n\geq0$. The lowest generalized-free level of each branch occurs only for $s=\eta$.}
  \label{fig:fermion-conformal-spin1-spectrum}
\end{figure}

\subsubsection{Regular fermion}
\label{sec:regular-fermion}
The regular fermion theory is obtained by taking $\lambda_{\rm F}\to+\infty$ and $\alpha\to+\infty$ at fixed $\mu_{\rm F}^2$, $y$, and $\kappa$. On a branch compatible with the chosen boundary condition, in standard quantization, the gap equation gives
\begin{equation*}
\calM_\eta\to0\,,\qquad \zeta_\eta^{\rm F}\to+\infty\,,\qquad \hat\Delta_\psi\to1\,.
\end{equation*}
As noted in section~\ref{sec:fermion}, the two choices of $\eta$ remain distinct at vanishing mass. As in the critical scalar case, alternate quantization gives no further unitary boundary condition.

The auxiliary fluctuation $\delta\sigma$ decouples in this limit. We denote by $\boldsymbol{\chi}$ the scalar primary represented at leading large $N$ by the renormalized bilinear $\bar\Psi_i\Psi^i/\sqrt N$.

Differentiating the determinant contribution in~\eqref{eq:fermion-effective-potential} and taking $\calM\to0$ at fixed $\eta$ gives
\begin{equation}
    a_\chi^{\rm reg}=\langle\boldsymbol{\chi}\rangle_{\rm AdS}=-\frac{\eta\sqrt N}{8\pi}\,.
\end{equation}
The connected correlator of $\boldsymbol{\chi}$ follows from the auxiliary-field equation, after rescaling the $\delta\sigma$ propagator and subtracting a contact term,
\begin{equation}
    \widetilde G_{\chi\chi}(\nu)=\lim_{\zeta_\eta^{\rm F}\to\infty}\left[\frac{(\zeta_\eta^{\rm F})^2}{\zeta_\eta^{\rm F}-\widetilde B^{(0)}_{1,1/2}(\nu)}-\zeta_\eta^{\rm F}\right]=-\frac{\nu}{8}\coth\frac{\pi\nu}{2}\,.
\end{equation}
The short-distance singularity fixes, cf. appendix~\ref{subapp:extracting-bulk-data},
\begin{equation}\label{eq:reg-bulk-data}
\Delta_\chi^{{\rm reg}}=2\,,\qquad C_\chi^{{\rm reg}}=\frac{1}{8\pi^2}\,.
\end{equation}
The boundary poles and residues give
\begin{equation}
    \hat\Delta_{\hat\chi_n}=3+2n\,,\qquad \varrho_{\chi \hat\chi_n}=n+1\,,\qquad n\geq0\,.
\end{equation}
The lowest state is the displacement operator.

We next consider the topological current. The fermion polarization simplifies to
\begin{equation}
    \widetilde B^{(1,E)}_{1,1/2}(\nu)=-\frac{\nu}{16}\tanh\frac{\pi\nu}{2}\,,\qquad \widetilde B^{(1,O)}_{1,1/2}(\nu)=\frac{i\eta\nu}{4\pi(\nu^2+1)}\,.
\end{equation}
The odd component remains nonzero because the boundary condition breaks parity. The identity contribution to the topological-current correlator reproduces the normalization~\eqref{eq:CJtop}. The dimension-two singlet contributes to both tensor structures. Using the extracted value of $a_\chi^{\rm reg}$, the bulk-channel expansion gives
\begin{equation}
    C^{E,{\rm reg}}_{J^{\rm top}J^{\rm top}}{}^\chi=\frac{256\kappa\,C_{J^{\rm top}}(\kappa)}{\pi\sqrt N}\frac{1}{1+\left(\frac{8\kappa}{\pi}\right)^2}\,,\qquad C^{O,{\rm reg}}_{J^{\rm top}J^{\rm top}}{}^\chi=-\frac{4i\pi\,C_{J^{\rm top}}(\kappa)}{\sqrt N}\frac{1-\left(\frac{8\kappa}{\pi}\right)^2}{1+\left(\frac{8\kappa}{\pi}\right)^2}\,.
\end{equation}
At $\kappa=0$ the $\boldsymbol{\chi}$ exchange survives only in the parity-odd structure.

The nonzero gauge-propagator poles in~\eqref{eq:fermion_vector_spectrum} give, in the conformal limit,
\begin{equation}
    0=\frac{\hat\Delta-1}{16}\tan\frac{\pi(\hat\Delta-1)}{2}+\frac{s\kappa_{\rm eff}(\hat\Delta-1)}{2\pi}+\frac{s\eta(\hat\Delta-1)^3}{4\pi\bigl[1-(\hat\Delta-1)^2\bigr]}\,.
\end{equation}
At large $|\kappa|$, the unprotected levels in the two helicity sectors approach
\begin{equation}
\hat\Delta^{\varsigma_{\rm eff}}_{n,1}\to 2n+1-\varsigma_{\rm eff}\eta\,,\ n\geq1\,,\qquad \hat\Delta^{-\varsigma_{\rm eff}}_{n,1}\to 2n+3+\varsigma_{\rm eff}\eta\,,\ n\geq0\,.
\end{equation}
The sector of helicity $\varsigma_{\rm eff}$ also contains the protected current, $\hat\Delta^{\varsigma_{\rm eff}}_{0,1}=1$. The lowest unprotected state of helicity $\eta$ therefore has $n=1$ if $\eta=\varsigma_{\rm eff}$ and $n=0$ otherwise. It approaches dimension two and corresponds to the flavor-singlet boundary bilinear $:\widehat{\bar\psi}_i\gamma_a\widehat\psi^{\,i}:$. The boundary Weyl projection removes its opposite-helicity component, whose tower then starts at dimension four. This unpaired lowest matter level is visible in figure~\ref{fig:fermion-conformal-spin1-spectrum}(a).

\subsubsection{Gross--Neveu critical family}
\label{sec:gn-critical}
We now tune $\mu_{\rm F}^2=\lambda_{\rm F}=0$ and take $\alpha\to+\infty$ at fixed $y$ and $\kappa$. At leading order in $1/N$, the cubic interaction is marginal and parametrizes a family of bulk conformal theories. For each boundary condition $\eta=\pm1$, the gap equation has the roots $\calM_{\eta,\pm}=\pm\eta/(2\sqrt{1+2\pi\eta y})$. The standard boundary conditions correspond to the minima, on which
\begin{equation}
    \calM_\eta=\calM_{\eta,+}=\frac{\eta}{2\sqrt{1+2\pi\eta y}}\,,\qquad \hat\Delta_\psi=\hat\Delta_\eta\equiv1+\eta\calM_\eta\,,\qquad \zeta_\eta^{\rm F}=2y\calM_\eta\,.
\end{equation}
At $y=0$, in particular, $\calM_\eta=\frac{\eta}{2}$, $\hat\Delta_\psi=\frac32$, and $\zeta_\eta^{\rm F}=0$.
In the following, the scalar primary $\boldsymbol{\sigma}$ is the mass field appearing in the Dirac operator. 

\paragraph{Standard boundary conditions.}
The expectation value of $\boldsymbol{\sigma}$ is
\begin{equation}
    a_\sigma^{\rm GN}=\langle\boldsymbol{\sigma}\rangle_{\rm AdS}=\sqrt N(M+\Sigma_\eta)=\sqrt N\,\calM_\eta=\frac{\eta\sqrt N}{2\sqrt{1+2\pi\eta y}}\,.
\end{equation}
Its connected two-point function is the auxiliary fluctuation propagator,
\begin{equation}
    \widetilde G_{\sigma\sigma}(\nu)=\frac{1}{\zeta_\eta^{\rm F}-\widetilde B^{(0)}_{\hat\Delta_\eta,1/2}(\nu)}\,.
\end{equation}
The bulk-channel expansion, detailed in appendix~\ref{subapp:extracting-bulk-data}, gives
\begin{equation}\label{eq:GN-bulk-data}
\Delta_\sigma^{\rm GN}=1\,,\qquad C_\sigma^{\rm GN}=\frac4{\pi^2}\,,\qquad C^{\rm GN}_{\sigma\sigma}{}^\sigma=-\frac{32y}{\pi\sqrt N}\,.
\end{equation}
As in flat space, where the fermion triangle vanishes, the leading three-point coefficient arises solely from the cubic vertex: it vanishes at $y=0$, while the one-point function remains nonzero.

The scalar boundary spectrum is determined by
\begin{equation}
    \zeta_\eta^{\rm F}-\widetilde B^{(0)}_{\hat\Delta_\eta,1/2}\!\left(-i(\hat\Delta-1)\right)=0\,.
\end{equation}
For any $y$, the displacement gives the exact solution $\hat\Delta=3$. At $y=0$, in particular,
\begin{equation}
    \hat\Delta_{\hat\sigma_n}=3+2n\,,\qquad \varrho_{\sigma \hat\sigma_n}=\frac{64(n+1)}{(2n+1)(2n+3)}\,,\qquad n\geq0\,.
\end{equation}

In the vector sector, the bulk-channel expansion of the topological-current correlator reproduces~\eqref{eq:CJtop}, independently of $y$. Using $a_\sigma^{\rm GN}$, the first subleading terms give
\begin{equation}
    C^{E,{\rm GN}}_{J^{\rm top}J^{\rm top}}{}^\sigma=-\frac{8\kappa\,C_{J^{\rm top}}(\kappa)}{\sqrt N}\frac{1}{1+\left(\frac{8\kappa}{\pi}\right)^2}\,,\qquad C^{O,{\rm GN}}_{J^{\rm top}J^{\rm top}}{}^\sigma=\frac{2i\,C_{J^{\rm top}}(\kappa)}{\sqrt N}\frac{1-\left(\frac{8\kappa}{\pi}\right)^2}{1+\left(\frac{8\kappa}{\pi}\right)^2}\,.
\end{equation}
Both coefficients are independent of $y$ and $\eta$ at this order.

Taking the conformal limit of the gauge-propagator pole condition~\eqref{eq:fermion_vector_spectrum}, the unprotected spin-one dimensions satisfy
\begin{equation}\label{eq:GN-vector-spectrum}
0=\frac{s\kappa(\hat\Delta-1)}{2\pi}+\widetilde B^{(1,+)}_{\hat\Delta_\eta,1/2}\!\left(-is(\hat\Delta-1)\right)\,,\qquad \hat\Delta>1\,.
\end{equation}
At large $|\kappa|$ and fixed $y$, the unprotected levels in the two helicity sectors approach the boundary fermion bilinear dimensions
\begin{equation}
\hat\Delta^{\varsigma_{\rm eff}}_{n,1}\to 2\hat\Delta_\eta+2n-1-\varsigma_{\rm eff}\eta\,,\ n\geq1\,,\qquad \hat\Delta^{-\varsigma_{\rm eff}}_{n,1}\to 2\hat\Delta_\eta+2n+1+\varsigma_{\rm eff}\eta\,,\ n\geq0\,.
\end{equation}
The protected current remains at $\hat\Delta^{\varsigma_{\rm eff}}_{0,1}=1$. As in the regular theory, the lowest unprotected state approaches $2\hat\Delta_\eta$ and occurs only for helicity $\eta$, owing to the boundary Weyl projection.

Figure~\ref{fig:fermion-conformal-spin1-spectrum}(b) illustrates this structure at $y=3/(8\pi)$, where the three boundary conditions have different generalized-free limiting dimensions.

\paragraph{Alternate boundary conditions.} For $\eta y>0$, the second root is unitary and defines an alternate boundary condition, with
\begin{equation}\label{eq:GN-alt-saddle}
    \calM_{\eta,-}=-\frac{\eta}{2\sqrt{1+2\pi|y|}}\,,\qquad \hat\Delta_\psi=1-\frac1{2\sqrt{1+2\pi|y|}}\,,\qquad \zeta_\eta^{\rm F}=-\frac{|y|}{\sqrt{1+2\pi|y|}}\,,
\end{equation}
so that $\frac12<\hat\Delta_\psi<1$ and $a_\sigma^{\rm alt}=-a_\sigma^{\rm GN}$. This saddle is a local maximum, and its lightest singlet is relevant. It satisfies \eqref{eq:fermion-BF-stability} up to $|y|\approx 0.1274$, where the lightest singlet reaches $\hat\Delta=1$.

The scalar singlet correlator, its bulk data, and the boundary-spectrum condition are those of the standard case with $\calM_\eta\to\calM_{\eta,-}$. In particular, the displacement operator again sits at $\hat\Delta=3$, since both roots satisfy the gap equation. For $y\to0$, $\hat\Delta_\psi\to\frac12$ and the lightest singlet approaches $\hat\Delta=2$. Figure~\ref{fig:gn-scalar-singlets} compares the standard and alternate scalar-singlet spectra, including the approach of the lowest alternate level to the BF bound.

In the vector sector, we evaluate the topological-current correlator~\eqref{eq:Gtop-general} and the gauge-propagator pole condition~\eqref{eq:GN-vector-spectrum} at the alternate value of $\hat\Delta_\psi$. The scalar-exchange coefficients change sign together with $a_\sigma$, so that $C^{E,{\rm GN}}_{J^{\rm top}J^{\rm top}}{}^\sigma$ and $C^{O,{\rm GN}}_{J^{\rm top}J^{\rm top}}{}^\sigma$ are unchanged. The same ordered asymptotic spectrum applies with $\hat\Delta_\eta\to\hat\Delta_\psi$. The lowest unprotected state of helicity $\eta$ has $n=1$ if $\eta=\varsigma_{\rm eff}$ and $n=0$ otherwise, and approaches $2\hat\Delta_\psi$.

\section{Outlook}\label{sec:outlook}
We have studied Maxwell--Chern--Simons theory coupled to $N$ scalars or Dirac fermions on AdS$_3$ at large $N$, with Dirichlet boundary conditions for the gauge field. When the matter does not condense and the effective Chern--Simons level is nonzero, the boundary chiral current remains protected, while Higgsing lifts it. The other spin-one operators form two helicity towers, whose dimensions depend on the couplings and display resonances associated with the massive gauge modes. At the bulk conformal points, the bulk data extracted with different boundary conditions agree, as required, while the boundary spectra distinguish them.

A natural question concerns going beyond the strict large-$N$ limit. At order $1/N$, the couplings $h$ and $y$ acquire nonzero beta functions, lifting the continuous $N=\infty$ families \cite{DiPietro:2020fya, DiPietro:2023zqn},
\begin{align}\label{eq:betahgeneric}
\beta_h(h,\kappa)&=\frac{1}{\pi^2N}\bigg[-\frac{h^3}{256}+\frac{3h^2}{4}+\frac{128\pi^2(\pi^2-128\kappa^2)}{(\pi^2+64\kappa^2)^2}\,h-\frac{16384\pi^4(\pi^2-192\kappa^2)}{(\pi^2+64\kappa^2)^3}\bigg]\,,\\
\beta_y(y,\kappa)&=\frac{32}{3\pi^2N}\bigg[-96y^3+\frac{3(4096\kappa^4+640\pi^2\kappa^2-3\pi^4)}{(\pi^2+64\kappa^2)^2}\,y+\frac{12\pi^3(320\kappa^2-3\pi^2)\kappa}{(\pi^2+64\kappa^2)^3}\bigg]\,.
\end{align}
On the branch connected to tricritical scalar QED$_3$, the sextic deformation is irrelevant \cite{DiPietro:2023zqn}, and $h_*(\kappa)$ decreases from $32(\sqrt{17}-1)$ at $\kappa=0$ to zero at $|\kappa|=\kappa_c=\pi/(8\sqrt3)$. All three boundary conditions exist for $|\kappa|<\kappa_c$, and the normal saddle, $v^2=\frac1{4\pi}+\frac12\sqrt{3/h_*(\kappa)}$, diverges as $|\kappa|\to\kappa_c^-$. Beyond $\kappa_c$, $h_*<0$ and only the ordinary boundary condition survives, until the branch ends at a fixed-point merger at $|\kappa|\approx 0.229$. The largest real root of $\beta_h$ is a UV fixed point, which tends to $h=192$ as $|\kappa|\to\infty$ \cite{Pisarski:1982vz}, beyond the stability bound~\eqref{eq:h-stability} \cite{Bardeen:1983rv}, and in the ungauged model is robust to $1/N$ corrections \cite{Kvedaraite:2025lgi}. It starts from $h=256$ at $\kappa=0$ and falls below $16\pi^2$ for $0.518\lesssim|\kappa|\lesssim1.082$, where it admits ordinary and normal boundary conditions. For fermions, the branch on which $y$ is irrelevant appears at a fixed-point merger at $|\kappa|\approx 0.273$ and extends to $|\kappa|\to\infty$, where it approaches the ungauged Gross--Neveu fixed point. Along it, both standard boundary conditions exist, while the alternate one exists for $\eta=\operatorname{sgn}y_*$ and degenerates where $y_*$ changes sign, at $|\kappa|=\pi\sqrt{3/320}$. However, the existence of a bulk fixed point is necessary but not sufficient for having a BCFT. It would be interesting to determine which of the conformal boundary conditions we discussed admit an extension at finite $N$ and $k$.

A further direction is to extend our analysis to operators charged under $U(1)_{\rm top}$. These include monopole operators, which insert magnetic flux $\int_{S^2}F=2\pi q$ and, at nonzero Chern--Simons level, require matter dressing to be gauge invariant. Their bulk spectrum has been studied at large $N$ in both scalar and fermionic Chern--Simons-matter theories \cite{Chester:2017vdh}. Since our boundary conditions preserve $U(1)_{\rm top}$, a scalar dressed monopole admits the boundary OPE
\begin{equation}
    \mathcal M_q(x,z)\sim\sum_a b_{q,a}\,z^{\widehat\Delta_{q,a}-\Delta_q}\widehat{\mathcal M}_{q,a}(x)+\dots\,,
\end{equation}
where all exchanged operators carry charge $q$, excluding the identity for $q\neq0$. Computing $\widehat\Delta_{q,a}$ and $b_{q,a}$ would extend the BCFT data obtained here to the charged sector and allow us to investigate how this sector distinguishes ordinary and normal boundary conditions.

The charged sector would also allow us to investigate particle-vortex duality and three-dimensional bosonization, which relate monopoles to matter operators in dual descriptions \cite{Karch:2016sxi,Seiberg:2016gmd}. Quantitative evidence exists from bulk monopole dimensions: large-$N$ results extrapolated to a single flavor have been compared with charged operators of the $O(2)$ Wilson--Fisher CFT and of a free Dirac fermion \cite{Chester:2022wur}. Extending these comparisons to the boundary dimensions and bulk-to-boundary OPE coefficients would provide tests sensitive to the boundary condition. This requires identifying the
 dual theories and boundary conditions, including any boundary degrees of freedom needed for anomaly matching \cite{Aitken:2017nfd}. Since AdS-invariant deformations retain boundary conformal data, such comparisons could also be made away from criticality, along matched RG flows \cite{Minwalla:2015sca}, provided the curvature couplings and boundary interactions are matched as well.

Extending our analysis to non-Abelian $SU(N_c)_k$ Chern--Simons-matter theories would bring anyonic statistics into play. For boundary conditions supporting an $\widehat{\mathfrak{su}}(N_c)_k$ current algebra \cite{Witten:1988hf,Elitzur:1989nr}, correlators of charged boundary operators decompose into affine blocks whose braiding encodes that of the bulk anyons. In the flat-space limit, this structure could reproduce the modified crossing relations of Chern--Simons-matter S-matrices~\cite{Jain:2014nza,Mehta:2022lgq} and provide a framework to bootstrap them. Matching AdS boundary data across level-rank dual descriptions would further test non-Abelian bosonization away from criticality \cite{Hsin:2016blu}.

\section*{Acknowledgments}
We thank Ankur, Lorenzo Di Pietro, and Shota Komatsu for discussions. The work of the authors is partially supported by the Israel Science Foundation (ISF) Grant No. 1487/21, by the MOST NSF/BSF Physics Grant No. 2022726, and by the BSF Grant No. 2024187.

\appendix
\section{Harmonic functions in AdS}
\label{app:harmonic-functions}
In this appendix we collect the scalar and spin-one harmonic functions used in the spectral representations throughout the paper. We first review their definition in Euclidean AdS$_{d+1}$, with unit AdS radius. We then specialize to AdS$_3$, where the transverse spin-one sector admits an additional parity-odd structure. We follow \cite{Costa:2014kfa} in generic $d$ and \cite{Bhattacharya:2025udq} in $d=2$, with slightly different conventions with respect to the latter.

\subsection{Scalar harmonics}
\label{subapp:scalar-harmonics}
The scalar harmonic function with spectral parameter $\nu\in\mathbb R$ can be defined as
\begin{equation}
    \Omega_\nu(X_1,X_2)=\frac{i\nu}{2\pi}\left[G_{\frac d2+i\nu}(X_1,X_2)-G_{\frac d2-i\nu}(X_1,X_2)\right]\,,
\end{equation}
where $G_\Delta$ denotes the scalar bulk-to-bulk propagator of dimension $\Delta$. It is even under $\nu\to-\nu$ and satisfies
\begin{equation}
    -\nabla_1^2\Omega_\nu=\left(\nu^2+\frac {d^2}4\right)\Omega_\nu\,.
\end{equation}

In terms of the chordal distance $u=-1-X_1\cdot X_2$,
\begin{equation}
    \Omega_\nu(u)=C_\nu^{(0)}\,{}_2F_1\!\left(\frac d2+i\nu,\frac d2-i\nu;\frac d2+\frac12;-\frac u2\right)\,,
\end{equation}
where
\begin{equation}
    C_\nu^{(0)}\equiv\Omega_\nu(0)=\frac{\nu\sinh(\pi\nu)\Gamma(\frac d2)\Gamma(\frac d2+i\nu)\Gamma(\frac d2-i\nu)}{4\pi^{\frac d2+2}\Gamma(d)}\,.
\end{equation}
With this normalization, the scalar harmonics obey the convolution identity
\begin{equation}
    \int dY\,\Omega_\nu(X_1,Y)\Omega_{\nu'}(Y,X_2)=\frac{\delta(\nu-\nu')+\delta(\nu+\nu')}{2}\Omega_\nu(X_1,X_2)\,,
\end{equation}
and the completeness relation
\begin{equation}
    \int_{\mathbb R}d\nu\,\Omega_\nu(X_1,X_2)=\delta(X_1,X_2)\,.
\end{equation}

Accordingly, throughout the paper we write the spectral representation of a scalar bilocal function as
\begin{equation}
    F(X_1,X_2)=\int_{\mathbb R}d\nu\,\widetilde F(\nu)\,\Omega_\nu(X_1,X_2)\,,
\end{equation}
and refer to $\widetilde F(\nu)$ as its spectral function. In particular, the scalar bulk-to-bulk propagator admits the spectral representation
\begin{equation}\label{eq:G-spectral}
    G_\Delta(X_1,X_2)=\int_{\mathbb R}d\nu\,\frac{\Omega_\nu(X_1,X_2)}{\nu^2+(\Delta-\frac d2)^2}\,.
\end{equation}
This representation holds for $\Delta>\frac d2$, i.e.\ in standard quantization. For $\Delta<\frac d2$, that is, in alternate quantization, the right-hand side gives $G_{d-\Delta}$, and instead \cite{Giombi:2017hpr}
\begin{equation}\label{eq:G-spectral-alt}
    G_\Delta(X_1,X_2)=\int_{\mathbb R}d\nu\,\frac{\Omega_\nu(X_1,X_2)}{\nu^2+(\Delta-\frac d2)^2}-\frac{2\pi}{\frac d2-\Delta}\,\Omega_{i(\frac d2-\Delta)}(X_1,X_2)\,.
\end{equation}

\subsection{Transverse spin-one harmonics}
\label{subapp:transverse-spin-one-harmonics}
The transverse spin-one harmonic in AdS$_{d+1}$ is defined by \cite{Costa:2014kfa}
\begin{equation}
    \Omega_\nu^{(1)}(X_1,X_2;W_1,W_2)=\frac{i\nu}{2\pi}\left[\Pi_{\frac d2+i\nu}(X_1,X_2;W_1,W_2)-\Pi_{\frac d2-i\nu}(X_1,X_2;W_1,W_2)\right]\,,
\end{equation}
where $\Pi_\Delta$ is the massive Proca propagator. It is even under $\nu\to-\nu$, transverse at both points, and satisfies
\begin{equation}\label{eq:spin1-harmonic-eom}
\left(\nabla_1^2+\frac{d^2}4+\nu^2+1\right)\Omega_\nu^{(1)}=0\,,\qquad {\rm Div}\,\Omega_\nu^{(1)}=0\,.
\end{equation}

The harmonic can be decomposed in the invariant tensor basis as
\begin{equation}
    \Omega_\nu^{(1)}(X_1,X_2;W_1,W_2)=(W_1\cdot W_2)\,\omega_0^\nu(u)+(W_1\cdot X_2)(W_2\cdot X_1)\,\omega_1^\nu(u)\,.
\end{equation}
Defining
\begin{equation}
    \mathsf F_\nu^{(a)}(u)={}_2F_1\!\left(\frac d2+a+i\nu,\frac d2+a-i\nu;\frac d2+a+\frac12;-\frac u2\right)\,,\qquad a=0,1\,,
\end{equation}
one has
\begin{equation}
    \omega_0^\nu(u)=\frac{C_\nu^{(1)}}{d}\left[\mathsf F_\nu^{(0)}(u)-\frac{u+1}{d+1}\mathsf F_\nu^{(1)}(u)\right]\,,
\end{equation}
\begin{equation}
    \omega_1^\nu(u)=\frac{C_\nu^{(1)}}{d(d+1)u(u+2)}\left[(d+1)(u+1)\mathsf F_\nu^{(0)}(u)-\bigl(u(u+2)+d+1\bigr)\mathsf F_\nu^{(1)}(u)\right]\,,
\end{equation}
with
\begin{equation}
    C_\nu^{(1)}=\frac{d\,\nu(d^2+4\nu^2)\sinh(\pi\nu)\Gamma(\frac d2-1+i\nu)\Gamma(\frac d2-1-i\nu)}{(4\pi)^{\frac{d+3}{2}}\Gamma\!\left(\frac{d+1}{2}\right)}\,.
\end{equation}

The transverse harmonics do not by themselves form a complete basis for spin-one fields: the longitudinal harmonics $(W_1\cdot\nabla_1)(W_2\cdot\nabla_2)\Omega_\nu/(\nu^2+\frac{d^2}{4})$ complete them, and integrate to $Q$ below. It is useful to define
\begin{equation}
    Q(X_1,X_2;W_1,W_2)\equiv(W_1\cdot\nabla_1)(W_2\cdot\nabla_2)G_d(X_1,X_2)\,,\quad I(X_1,X_2;W_1,W_2)\equiv(W_1\cdot W_2)\delta(X_1,X_2)\,,
\end{equation}
so that the transverse completeness relation reads simply
\begin{equation}
    \int_{\mathbb R}d\nu\,\Omega_\nu^{(1)}=I-Q\,.
\end{equation}

We also define the convolution of two spin-one bitensors in embedding space as
\begin{equation}
    (F*H)(X_1,X_2;W_1,W_2)=\int dY\,F(X_1,Y;W_1,K_Y)\,H(Y,X_2;W_Y,W_2)\,,
\end{equation}
where $K_Y$ contracts the intermediate polarization. In generic AdS$_{d+1}$, it reads
\begin{equation}
    K_Y^A=\left(\frac{d-1}{2}+W_Y\cdot\partial_{W_Y}\right)\left[\partial_{W_Y}^A+Y^A(Y\cdot\partial_{W_Y})\right]-\frac12W_Y^A\left[\partial_{W_Y}^{\,2}+(Y\cdot\partial_{W_Y})^2\right]\,.
\end{equation}
Consequently, one has
\begin{equation}\label{eq:spin1-convolution}
    \Omega_{\nu}^{(1)}*\Omega_{\nu'}^{(1)} =\frac{d-1}{2}\,\frac{\delta(\nu-\nu')+\delta(\nu+\nu')}{2}\,\Omega_\nu^{(1)}\,.
\end{equation}

\subsection{Parity-odd and helicity harmonics in AdS$_3$}
\label{subapp:parity-odd-harmonics}
We now specialize to $d=2$. In three bulk dimensions the curl maps transverse vectors into transverse vectors. We denote the transverse parity-even harmonic by
\begin{equation}\label{eq:OmegaE-Proca}
\Omega_\nu^E\equiv\Omega_\nu^{(1)}=\frac{i\nu}{2\pi}\left(\Pi_{1+i\nu}-\Pi_{1-i\nu}\right)\,,
\end{equation}
which satisfies \eqref{eq:spin1-harmonic-eom} at $d=2$.

We define the parity-odd harmonic by
\begin{equation}\label{eq:OmegaO-Proca}
\Omega_\nu^O=\frac1\nu\,\mathcal C\Omega_\nu^E=\frac{i}{2\pi}\,\mathcal C\left(\Pi_{1+i\nu}-\Pi_{1-i\nu}\right)\,.
\end{equation}
Since $\Omega_\nu^E$ is even in $\nu$, $\Omega_\nu^O$ is odd. Despite appearances, it is regular at $\nu=0$, with $\Omega_\nu^O \sim O(\nu)$. The two harmonics satisfy
\begin{equation}
    \mathcal C\Omega_\nu^E=\nu\Omega_\nu^O\,,\qquad \mathcal C\Omega_\nu^O=\nu\Omega_\nu^E\,.
\end{equation}
Consequently, the combinations
\begin{equation}
    \Omega_\nu^\pm=\Omega_\nu^E\pm\Omega_\nu^O
\end{equation}
diagonalize the curl,
\begin{equation}\label{eq:helicity-harmonics}
\mathcal C\Omega_\nu^\pm=\pm\nu\Omega_\nu^\pm\,,\qquad \Omega_{-\nu}^\pm=\Omega_\nu^\mp\,.
\end{equation}
The sign $\pm$ is the helicity of section~\ref{subsec:boundary-conditions}, namely the sign of the spin of the corresponding $SO(3,1)$ principal series representation.

The parity-odd harmonic has the following structure,
\begin{equation}
    \Omega_\nu^O(X_1,X_2;W_1,W_2)=\omega_2^\nu(u)\,\epsilon(X_1,W_1,X_2,W_2)\,,
\end{equation}
with
\begin{equation}
    \omega_2^\nu(u)=-\frac{\nu(\nu^2+1)}{12\pi^2}{}_2F_1\!\left(2+i\nu,2-i\nu;\frac52;-\frac u2\right)\,.
\end{equation}

Accordingly, a spin-one bitensor in AdS$_3$ admits the spectral representation
\begin{equation}
\begin{aligned}
    F(X_1,X_2;W_1,W_2)&=\int_{\mathbb R}d\nu\,\left[\widetilde F_E(\nu)\,\Omega_\nu^E(X_1,X_2;W_1,W_2)+\widetilde F_O(\nu)\,\Omega_\nu^O(X_1,X_2;W_1,W_2)\right]\\
    &\qquad + \int_{\mathbb R}d\nu\, \widetilde{F}_L(\nu) \,(W_1\cdot \nabla_1)(W_2\cdot \nabla_2)\,\Omega_\nu(X_1,X_2)\,.
\end{aligned}
\end{equation}
We use the labels $E$, $O$, and $L$ for the corresponding spectral functions.

\paragraph{Identities.}
For $d=2$, \eqref{eq:spin1-convolution} gives $\Omega_\nu^E*\Omega_{\nu'}^E=\frac14\left[\delta(\nu-\nu')+\delta(\nu+\nu')\right]\Omega_\nu^E$, and the curl identities then give
\begin{equation}
\begin{aligned}
    \Omega_\nu^E*\Omega_{\nu'}^O&=\frac{\delta(\nu-\nu')-\delta(\nu+\nu')}{4}\,\Omega_\nu^O\,,\\
    \Omega_\nu^O*\Omega_{\nu'}^E&=\frac{\delta(\nu-\nu')+\delta(\nu+\nu')}{4}\,\Omega_\nu^O\,,\\
    \Omega_\nu^O*\Omega_{\nu'}^O&=\frac{\delta(\nu-\nu')-\delta(\nu+\nu')}{4}\,\Omega_\nu^E\,.
\end{aligned}
\end{equation}
Equivalently, in the helicity basis,
\begin{equation}
\begin{aligned}
    \Omega_\nu^\pm*\Omega_{\nu'}^\pm&=\frac12\delta(\nu-\nu')\,\Omega_\nu^\pm\,,\\
    \Omega_\nu^\pm*\Omega_{\nu'}^\mp&=\frac12\delta(\nu+\nu')\,\Omega_\nu^\pm\,.
\end{aligned}
\end{equation}
Finally, the completeness relations generalize to
\begin{equation}\label{eq:completeness-AdS3}
\int_{\mathbb R}d\nu\,\Omega_\nu^E= \int_{\mathbb R}d\nu\,\Omega_\nu^\pm=I-Q\,,\qquad \int_{\mathbb R}d\nu\,\Omega_\nu^O=0\,.
\end{equation}

\paragraph{Contact terms.} We finally specify how local contact terms enter the spectral representation. For scalar two-point functions, an AdS-invariant contact term $P(-\nabla^2)\delta$ has spectral coefficient $P(\nu^2+d^2/4)$, with $P$ a polynomial. For vector two-point functions in AdS$_3$, we write
\begin{equation}
\delta F=\int_{\mathbb R}d\nu\left[\delta\widetilde F_E(\nu)\Omega_\nu^E+\delta\widetilde F_O(\nu)\Omega_\nu^O+\delta\widetilde F_L(\nu)\nabla\nabla'\Omega_\nu\right]\,.
\end{equation}
The general AdS-invariant local contact term has coefficients
\begin{equation}
\begin{aligned}
\delta\widetilde F_E(\nu)&=a_0+\nu^2p(\nu^2)\,,\\
\delta\widetilde F_O(\nu)&=\nu q(\nu^2)\,,\\
\delta\widetilde F_L(\nu)&=\frac{a_0}{\nu^2+1}+r(\nu^2+1)\,,
\end{aligned}
\end{equation}
where $a_0$ is a constant and $p,q,r$ are polynomials. In particular, taking $p$ and $q$ to be constant gives the Maxwell and Chern--Simons contact terms, respectively.

\section{Matter bubbles}
\label{app:matter-bubbles}
\subsection{Scalar bubbles}
\label{subsec:scalar-bubbles}
We collect the scalar bubbles that will enter the large-$N$ quadratic effective actions. In the following, we denote by $B_{\Delta,J}^{(\ell)}$ the spin-$\ell$ bubble built from spin-$J$ bulk matter fields dual to boundary operators of dimension $\Delta$. Let $\Phi$ be a single complex scalar of boundary dimension $\Delta_\phi$, with bulk-to-bulk propagator $G_{\Delta_\phi}(u)$.

\paragraph{Scalar bubble.} The spin-zero bubble is simply the connected two-point function of the singlet bilinear,
\begin{equation}
    B_{\Delta_\phi,0}^{(0)}(x,x')\equiv\langle\Phi^\dagger\Phi(x)\,\Phi^\dagger\Phi(x')\rangle_{\rm c}=G_{\Delta_\phi}(u)^2\,.
\end{equation}
In AdS$_3$ the square of a propagator decomposes as
\begin{equation}
    G_{\Delta_\phi}(u)^2=\sum_{n=0}^{\infty}a_n(\Delta_\phi)\,G_{2\Delta_\phi+2n}(u)\,,\qquad a_n(\Delta_\phi)=\frac1{2\pi}\,,
\end{equation}
and therefore
\begin{equation}
\begin{aligned}
    \widetilde B_{\Delta_\phi,0}^{(0)}(\nu)&=\frac{1}{2\pi}\sum_{n=0}^{\infty}\frac{1}{\nu^2+(2\Delta_\phi+2n-1)^2}\\
    &=\frac{i}{8\pi\nu}\left[\psi\left(\Delta_\phi-\frac{1+i\nu}{2}\right)-\psi\left(\Delta_\phi-\frac{1-i\nu}{2}\right)\right]\,,
\end{aligned}
\end{equation}
in agreement with \cite{Carmi:2018qzm}. For $\Delta_\phi<\frac12$, which occurs only in alternate quantization, the $n=0$ term is represented as in \eqref{eq:G-spectral-alt}, and $B^{(0)}_{\Delta_\phi,0}$ acquires an extra term proportional to $\Omega_{\nu_0}$, with $\nu_0=i(1-2\Delta_\phi)$.

\paragraph{Spin-one bubble.} The spin-one bubble is the two-point function of the matter current
\begin{equation}
    j_\mu^{\rm mat}=i\,\Phi^\dagger\overset{\leftrightarrow}{\nabla}_\mu\Phi\,.
\end{equation}
At separated points, performing Wick contractions gives
\begin{equation}
\begin{aligned}
    B_{\Delta_\phi,0}^{(1)}(x,x')_{\mu\nu'}&\equiv\left\langle j_\mu^{\rm mat}(x)j_{\nu'}^{\rm mat}(x')\right\rangle\\
    &=2G_{\Delta_\phi}G_{\Delta_\phi}'\nabla_\mu\nabla_{\nu'}u+2\left[G_{\Delta_\phi}G_{\Delta_\phi}''-(G_{\Delta_\phi}')^2\right]\nabla_\mu u\,\nabla_{\nu'}u\,.
\end{aligned}
\end{equation}
We seek a decomposition of the bubble in Proca propagators,
\begin{equation}
    B_{\Delta_\phi,0}^{(1)}(x,x')_{\mu\nu'}=\sum_{n=0}^{\infty}b_n(\Delta_\phi)\,\Pi_{2\Delta_\phi+2n+1}(x,x')_{\mu\nu'}\,,\quad x\neq x'\,.
\end{equation}
Expanding both sides at large $u$, one can match the coefficients $b_n(\Delta_\phi)$ recursively. One finds
\begin{equation}
    b_n(\Delta_\phi)=\frac{(n+1)(2\Delta_\phi+2n)^2(2\Delta_\phi+n-1)}{\pi\left[(2\Delta_\phi+2n)^2-1\right]}\,.
\end{equation}
The spectral function of the bubble is then formally given by
\begin{equation}
    \widetilde B_{\Delta_\phi,0}^{(1)}(\nu)=\sum_{n=0}^{\infty}\frac{b_n(\Delta_\phi)}{\nu^2+(2\Delta_\phi+2n)^2}\,.
\end{equation}
The corresponding spectral sum is UV divergent and is defined only up to a contact term. We fix the remaining local ambiguity by imposing the Ward identity: upon gauging the matter $U(1)$ symmetry the corresponding photon must be massless, which implies \cite{Ankur:2023lum}
\begin{equation}
    \widetilde B_{\Delta_\phi,0}^{(1)}\left(0\right)=0\,.
\end{equation}
With this prescription, the bubble spectral function is found to be
\begin{equation}\label{eq:B1-scalar}
    \widetilde B_{\Delta_\phi,0}^{(1)}(\nu)=\frac{(2\Delta_\phi-3)\nu^2}{8\pi(\nu^2+1)}+\frac{\nu\left[4(\Delta_\phi-1)^2+\nu^2\right]}{16\pi i(\nu^2+1)}\left[\psi\left(\Delta_\phi-\frac{i\nu}{2}\right)-\psi\left(\Delta_\phi+\frac{i\nu}{2}\right)\right]\,.
\end{equation}
This agrees with \cite{Ankur:2023lum}, after translating conventions.

\subsection{Fermion bubbles}
\label{subsec:fermion-bubbles}
We now compute the scalar and spin-one bubbles entering the quadratic large-$N$ effective action. In this subsection all bubbles refer to a single Dirac fermion. The decompositions below hold in either quantization.

\paragraph{Scalar bubble.} The connected two-point function of the scalar bilinear is
\begin{equation}
    B_{\Delta_\psi,1/2}^{(0)}(x,y)\equiv\left\langle\overline\Psi\Psi(x)\,\overline\Psi\Psi(y)\right\rangle_{\rm c}=-\operatorname{tr}\left[S_\eta(x,y)S_\eta(y,x)\right]\,.
\end{equation}
Using the identities \eqref{eq:spinortransport}, this reduces to
\begin{equation}
    B_{\Delta_\psi,1/2}^{(0)}(u)=-2\left[A_1(u)^2-A_2(u)^2\right]\,.
\end{equation}
The result admits a decomposition in terms of scalar propagators,
\begin{equation}
    B_{\Delta_\psi,1/2}^{(0)}(u)=\sum_{n=0}^{\infty}c_n(\Delta_\psi)\,G_{2\Delta_\psi+2n+1}(u)\,,
\end{equation}
with the coefficients $c_n$ determined to be
\begin{equation}
    c_n(\Delta_\psi)=\frac{2(n+1)(2\Delta_\psi+n-1)}{\pi}\,.
\end{equation}
Its spectral function is then formally given by
\begin{equation}
    \widetilde B_{\Delta_\psi,1/2}^{(0)}(\nu)=\sum_{n=0}^{\infty}\frac{c_n(\Delta_\psi)}{\nu^2+(2\Delta_\psi+2n)^2}\,.
\end{equation}
However, this sum is linearly divergent. There is no Ward identity fixing the corresponding local ambiguity, which amounts to a renormalization of the quadratic coupling of the auxiliary scalar. Using the same parity-preserving Pauli--Villars prescription as for the effective potential, with regulator masses shifted by $\pm M_{\rm PV}$, the regulated bubble takes the form
\begin{equation}
    \widetilde B_{\Delta_\psi,1/2}^{(0),{\rm PV}}(\nu)=\frac{M_{\rm PV}}{\pi}+\widetilde B_{\Delta_\psi,1/2}^{(0)}(\nu)+O(M_{\rm PV}^{-1})\,,
\end{equation}
where the finite part is
\begin{equation}
    \widetilde B_{\Delta_\psi,1/2}^{(0)}(\nu)=i\frac{4(\Delta_\psi-1)^2+\nu^2}{8\pi\nu}\left[\psi\left(\Delta_\psi+\frac{i\nu}{2}\right)-\psi\left(\Delta_\psi-\frac{i\nu}{2}\right)\right]-\frac{2\Delta_\psi-1}{4\pi}\,.
\end{equation}
The divergent term is absorbed into the renormalized quadratic coupling $\lambda_{\rm F}$.

\paragraph{Spin-one bubble.} We next consider the $U(1)$ matter current
\begin{equation}
    j_\mu^{\rm mat}=i\overline\Psi\Gamma_\mu\Psi\,.
\end{equation}
Its two-point function is
\begin{equation}
\begin{aligned}
    B_{\Delta_\psi,1/2}^{(1)}(x,x')_{\mu\nu'}&\equiv\left\langle j_\mu^{\rm mat}(x)j_{\nu'}^{\rm mat}(x')\right\rangle\\
    &=+\operatorname{tr}\left[\Gamma_\mu\left(\eta A_1+A_2\slashed n\right)\Gamma_\rho\left(\eta A_1-A_2\slashed n\right)\right]g^\rho{}_{\nu'}\,,
\end{aligned}
\end{equation}
where we have used parallel transport to bring the spinor structures at $x'$ to $x$. Using the three-dimensional Clifford algebra gives
\begin{equation}
\begin{aligned}
    B_{\Delta_\psi,1/2}^{(1)}(x,x')_{\mu\nu'}&=2\left(A_1^2+A_2^2\right)g_{\mu\nu'}+4A_2^2n_\mu n_{\nu'}-4i\eta A_1A_2\,\epsilon_{\mu\alpha\rho}n^\alpha g^\rho{}_{\nu'}\\
    &\equiv B_{\Delta_\psi,1/2}^{(1,E)}(x,x')_{\mu\nu'} + B_{\Delta_\psi,1/2}^{(1,O)}(x,x')_{\mu\nu'}\,,
\end{aligned}
\end{equation}
where we separated the parity-even and parity-odd structures as
\begin{equation}
\begin{aligned}
    B_{\Delta_\psi,1/2}^{(1,E)}(x,x')_{\mu\nu'}&=- 2\left(A_1^2+A_2^2\right)\nabla_\mu\nabla_{\nu'}u+2\left[\frac{A_1^2}{u+2}+\frac{A_2^2}{u}\right]\nabla_\mu u\,\nabla_{\nu'}u\,,\\
    B_{\Delta_\psi,1/2}^{(1,O)}(x,x')_{\mu\nu'}&=-\frac{4i\eta A_1A_2}{\sqrt{u(u+2)}}\mathcal E_{\mu\nu'}\,.
\end{aligned}
\end{equation}
We find a decomposition of the parity-even part in terms of Proca propagators,
\begin{equation}
    B_{\Delta_\psi,1/2}^{(1,E)}=\sum_{n=0}^{\infty}d_n^E(\Delta_\psi)\,\Pi_{2\Delta_\psi+2n}\,,
\end{equation}
with coefficients determined recursively to be
\begin{equation}
    d_n^E(\Delta_\psi)=\frac{(2\Delta_\psi+2n-1)^2\left[2\Delta_\psi^2+\Delta_\psi(2n-3)+n(n-1)+1\right]}{4\pi(\Delta_\psi+n-1)(\Delta_\psi+n)}\,.
\end{equation}
The parity-even spectral function is formally given by
\begin{equation}
    \widetilde B_{\Delta_\psi,1/2}^{(1,E)}(\nu)=\sum_{n=0}^{\infty}\frac{d_n^E(\Delta_\psi)}{\nu^2+(2\Delta_\psi+2n-1)^2}\,.
\end{equation}
Like the spin-one bubble of scalars, it is defined up to a contact term. Fixing this ambiguity by the $U(1)$ Ward identity gives
\begin{equation}
\begin{aligned}
    \widetilde B_{\Delta_\psi,1/2}^{(1,E)}(\nu)={}&-\frac{(\Delta_\psi-1)\nu^2}{4\pi(1+\nu^2)}\\
    &+\frac{\nu\left[-4\Delta_\psi^2+8\Delta_\psi+\nu^2-3\right]}{16\pi i(1+\nu^2)}\left[\psi\left(\Delta_\psi-\frac{1+i\nu}{2}\right)-\psi\left(\Delta_\psi-\frac{1-i\nu}{2}\right)\right]\,.
\end{aligned}
\end{equation}

For the parity-odd structure, it is convenient to introduce
\begin{equation}
    \Pi_{\Delta}^{O}\equiv -\frac{\mathcal{C}}{(\Delta-1)^2}\Pi_\Delta =-\int_{\mathbb R}d\nu\,\frac{\nu \, \Omega_\nu^O}{(\Delta-1)^2\left[\nu^2+(\Delta-1)^2\right]}\,.
\end{equation}
In particular, at separated points,
\begin{equation}
    \mathcal{C}\Pi_\Delta^O(x,x')_{\mu\nu'}=\Pi_\Delta(x,x')_{\mu\nu'}\,,\quad x\neq x'\,.
\end{equation}
We then seek a decomposition of the form
\begin{equation}
    B_{\Delta_\psi,1/2}^{(1,O)}=i\eta\sum_{n=0}^{\infty}d_n^O(\Delta_\psi)\,\Pi_{2\Delta_\psi+2n}^{O}\,,\quad x\neq x'\,.
\end{equation}
Taking the curl and expanding at large $u$ allows us to determine
\begin{equation}
    d_n^O(\Delta_\psi)=-\frac{(\Delta_\psi-1)(2\Delta_\psi+2n-1)^4}{4\pi(\Delta_\psi+n-1)(\Delta_\psi+n)}\,.
\end{equation}
It follows that the parity-odd spectral function is, up to local contact terms,
\begin{equation}
\widetilde B_{\Delta_\psi,1/2}^{(1,O)}(\nu)=-i\eta\nu\sum_{n=0}^{\infty}\frac{d_n^O(\Delta_\psi)}{(2\Delta_\psi+2n-1)^2\left[\nu^2+(2\Delta_\psi+2n-1)^2\right]}\,.
\end{equation}
The sum is convergent and regular at $\nu=0$. The remaining ambiguity, which corresponds to a local Chern--Simons counterterm, is fixed by the choice of regularization scheme: in a parity-preserving scheme, like parity-preserving Pauli--Villars, this term is absent. One then obtains
\begin{equation}
\widetilde B_{\Delta_\psi,1/2}^{(1,O)}(\nu)=\frac{i\eta\nu}{4\pi(1+\nu^2)}-\eta\frac{(\Delta_\psi-1)\nu^2}{4\pi(1+\nu^2)}\left[\psi\left(\Delta_\psi-\frac{1+i\nu}{2}\right)-\psi\left(\Delta_\psi-\frac{1-i\nu}{2}\right)\right]\,.
\end{equation}
In particular,
\begin{equation}\label{eq:CSlevelshift}
\widetilde B_{\Delta_\psi,1/2}^{(1,O)}(\nu)=\frac{i\eta\nu}{4\pi}+O(\nu^3)\,,
\end{equation}
corresponding to an induced Chern--Simons level shift of $\eta/2$ \cite{Redlich:1983dv}.

For later use, we also introduce the helicity-plus combination
\begin{equation}
\widetilde B_{\Delta_\psi,1/2}^{(1,+)}(\nu)\equiv\widetilde B_{\Delta_\psi,1/2}^{(1,E)}(\nu)+\widetilde B_{\Delta_\psi,1/2}^{(1,O)}(\nu)\,.
\end{equation}

\section{Bulk-channel expansions}
\label{app:bulk-channel}
In this appendix, we explain how the bulk-channel expansions of the scalar singlet and topological-current correlators determine the BCFT data reported in section~\ref{sec:conformal-points}. We follow the spectral approach of \cite{Carmi:2018qzm}, adapting it to our more general setup.

Bulk conformality requires a convergent bulk OPE. For a two-point function at invariant distance $u$, the limit $u\to0$ brings the insertions together away from the boundary. At a bulk conformal point, which is Weyl-equivalent to a BCFT on half-space, the correlator must then admit a bulk-channel conformal block expansion. Unlike the boundary-channel expansion, which follows from the AdS isometries alone, this constrains the bulk spectrum and OPE coefficients. After subtracting the identity block, the remaining short-distance terms must assemble into bulk primaries and their descendants, so the perturbative large-$|\nu|$ expansion must match spectral transforms of bulk blocks, modulo contact terms. Contributions regular at $u=0$ can be invisible in this expansion, which therefore captures only part of the bulk OPE. While this matching alone does not establish bulk conformality, the agreement between the bulk data extracted with different boundary conditions and with independent flat-space results is an important consistency check on the conformal interpretation of these points.

\subsection{Scalar two-point functions}
\label{subapp:scalar-bulk-expansion}
Let $\mathcal O$ be a scalar bulk primary of dimension $\Delta$. Using the conventions of section~\ref{sec:conformal-points}, taking the expectation value of the bulk OPE gives
\begin{equation}
F_{\mathcal O\mathcal O}(u)\equiv\langle\mathcal O(X)\mathcal O(Y)\rangle=\frac{C_{\mathcal O}}{(2u)^\Delta}+\sum_{\mathcal O'\neq\mathbf1}a_{\mathcal O'}C_{\mathcal O\mathcal O}{}^{\mathcal O'}f_{\mathcal O\mathcal O}^{\mathcal O'}(u)\,,
\end{equation}
where the sum runs over scalar primaries with nonzero one-point functions. Each exchanged primary determines the product $a_{\mathcal O'}C_{\mathcal O\mathcal O}{}^{\mathcal O'}$. Extracting the OPE coefficient separately therefore requires knowledge of the one-point coefficient. For connected correlators, the disconnected contribution $a_{\mathcal O}^{\,2}$ is subtracted from this expression. In these conventions, the block of a scalar primary of dimension $\Delta'$ is
\begin{equation}
f_{\mathcal O\mathcal O}^{\mathcal O'}(u)=(2u)^{-\Delta+\Delta'/2}{}_2F_1\!\left(\frac{\Delta'}2,\frac{\Delta'}2;\Delta'-\frac d2+\frac12;-\frac u2\right)\,.
\end{equation}
Its leading behavior is
\begin{equation}
f_{\mathcal O\mathcal O}^{\mathcal O'}(u)=(2u)^{-\Delta+\Delta'/2}\bigl[1+O(u)\bigr]\,.
\end{equation}
The identity block is recovered by setting $\Delta'=0$.

To implement the same procedure in spectral space, we use the scalar harmonics normalized as in appendix~\ref{app:harmonic-functions},
\begin{equation}
F(u)=\int_{-\infty}^{+\infty}d\nu\,\widetilde F(\nu)\Omega_\nu(u)\,.
\end{equation}
Their convolution identity gives the inverse transform
\begin{equation}
\widetilde F(\nu)=\frac{\operatorname{Vol}(S^d)}{\Omega_\nu(0)}\int_0^\infty du\,[u(u+2)]^{(d-1)/2}\,\Omega_\nu(u)F(u)\,,\qquad \operatorname{Vol}(S^d)=\frac{2\pi^{(d+1)/2}}{\Gamma((d+1)/2)}\,.
\end{equation}
Local scalar contact terms correspond to polynomials in $\nu^2+d^2/4$, as discussed in appendix~\ref{subapp:parity-odd-harmonics}.

The basic ingredient is the transform of a single power,
\begin{equation}
F_p(u)=\left(\frac u2\right)^{-p}\,,
\end{equation}
which reads
\begin{equation}
\widetilde F_p(\nu)=(4\pi)^{(d+1)/2}\frac{\Gamma\!\left(\frac{d+1}{2}-p\right)}{\Gamma(p)}\frac{\Gamma\!\left(p-\frac d2+i\nu\right)\Gamma\!\left(p-\frac d2-i\nu\right)}{\Gamma\!\left(\frac12+i\nu\right)\Gamma\!\left(\frac12-i\nu\right)}\,.
\end{equation}
For generic $p$, its algebraic large-$|\nu|$ expansion starts as
\begin{equation}
\widetilde F_p(\nu)\sim(4\pi)^{(d+1)/2}\frac{\Gamma\!\left(\frac{d+1}{2}-p\right)}{\Gamma(p)}|\nu|^{2p-d-1}\left[1+O(\nu^{-2})\right]\,.
\end{equation}
Exceptional values of $p$ are treated by analytic continuation, subtracting contact divergences where required. Since $f_{\mathcal O\mathcal O}^{\mathbf1}=4^{-\Delta}F_\Delta$, the identity transform is
\begin{equation}
\widetilde{f}_{\mathcal O\mathcal O}^{\mathbf1}(\nu)=4^{-\Delta}\widetilde F_\Delta(\nu)\,.
\end{equation}
In particular, in $\mathrm{AdS}_3$,
\begin{equation}
\left.\widetilde f_{\mathcal O\mathcal O}^{\mathbf1}(\nu)\right|_{\Delta=1}=\frac{2\pi^2}{\nu}\coth(\pi\nu)\,,\qquad \left.\widetilde f_{\mathcal O\mathcal O}^{\mathbf1}(\nu)\right|_{\Delta=2}=-\pi^2\nu\coth(\pi\nu)\,.
\end{equation}
These formulas fix the scalar normalizations quoted in section~\ref{sec:conformal-points} directly from the leading large-$|\nu|$ coefficients. For a general exchanged primary, expanding the hypergeometric function and transforming each power gives the formal prescription
\begin{equation}
\widetilde f_{\mathcal O\mathcal O}^{\mathcal O'}(\nu)\underset{\rm alg}{\sim}2^{\Delta'}4^{-\Delta}\sum_{n=0}^{\infty}\frac{(-1)^n(\Delta'/2)_n^2}{n!\left(\Delta'-\frac d2+\frac12\right)_n}\widetilde F_{\Delta-\Delta'/2-n}(\nu)\,.
\end{equation}
Here and in the following $\underset{\rm alg}\sim$ means an algebraic asymptotic expansion at large real $|\nu|$. Each successive term starts two powers of $|\nu|$ lower, so only finitely many terms are needed at a given order.

Generically,
\begin{equation}
\widetilde f_{\mathcal O\mathcal O}^{\mathcal O'}(\nu)\underset{\rm alg}{\sim}2^{\Delta'}4^{-\Delta}(4\pi)^{\frac{d+1}{2}}\frac{\Gamma\!\left(\frac{d+1}{2}-\Delta+\frac{\Delta'}2\right)}{\Gamma\!\left(\Delta-\frac{\Delta'}2\right)}|\nu|^{2\Delta-\Delta'-d-1}\,.
\end{equation}
At special exchanged dimensions, the algebraic transform can vanish; dimension derivatives of such blocks will be discussed in appendix~\ref{subapp:derived-blocks}.

Finally, the algebraic expansion does not capture all contributions to the correlator. A useful example is the regular image term
\begin{equation}
R_\Delta(u)=\frac{1}{[2(u+2)]^\Delta}\,.
\end{equation}
Its Taylor expansion around $u=0$ contains only nonnegative integer powers. Their term-by-term transforms give no algebraic large-$|\nu|$ contribution, owing to the factor $1/\Gamma(p)$ at $p=0,-1,\ldots$. Nevertheless, the exact transform is nonzero:
\begin{equation}
\widetilde R_\Delta(\nu)=4^{-\Delta}(4\pi)^{(d+1)/2}\frac{\Gamma\!\left(\Delta-\frac d2+i\nu\right)\Gamma\!\left(\Delta-\frac d2-i\nu\right)}{\Gamma(\Delta)\Gamma\!\left(\Delta-\frac d2+\frac12\right)}\,,
\end{equation}
and behaves as
\begin{equation}
\widetilde R_\Delta(\nu)\propto|\nu|^{2\Delta-d-1}e^{-\pi|\nu|}\,.
\end{equation}
Thus regular terms can carry bulk-channel information that is absent from the algebraic spectral expansion.

\subsection{Transverse spin-one two-point functions}
\label{subapp:transverse-spin-one-bulk-expansion}
We now consider a conserved bulk current $J$ of dimension $\Delta_J=d$. In embedding space, we write
\begin{equation}
J(X,W_X)=W_X\cdot J(X)\,,\qquad W_X\cdot X=0\,.
\end{equation}
We first discuss the parity-even part of its two-point function in generic AdS$_{d+1}$; the additional parity-odd structure in $\mathrm{AdS}_3$ is treated in appendix~\ref{subapp:parity-odd-bulk-expansion}. Taking the expectation value of the bulk OPE selects scalar exchanged primaries and gives
\begin{equation}
F_{JJ}^{\perp}=C_J\,\mathcal F_{JJ}^{\perp,\mathbf1}+\sum_{\mathcal O'\neq\mathbf1}a_{\mathcal O'}C_{JJ}^{\perp}{}^{\mathcal O'}\mathcal F_{JJ}^{\perp,\mathcal O'}\,.
\end{equation}
Here $C_J$ is the bulk current normalization, and the one-point coefficients follow the convention introduced in appendix~\ref{subapp:scalar-bulk-expansion}. The parity-even correlator admits two equivalent tensor decompositions,
\begin{equation}
\begin{aligned}
F_{JJ}^{\perp}(X,Y;W_X,W_Y)&=(W_X\cdot W_Y)F_0(u)+(W_X\cdot Y)(W_Y\cdot X)F_1(u)\\
&=\frac{1}{(2u)^d}\left[P(u)\mathcal I(X,Y;W_X,W_Y)+Q(u)\mathcal X(X,Y;W_X,W_Y)\right]\,,
\end{aligned}
\end{equation}
where
\begin{equation}
\begin{aligned}
\mathcal I(X,Y;W_X,W_Y)&=W_X\cdot W_Y+\frac{(W_X\cdot Y)(W_Y\cdot X)}{u}\,,\\
\mathcal X(X,Y;W_X,W_Y)&=\frac{(W_X\cdot Y)(W_Y\cdot X)}{u(u+2)}\,.
\end{aligned}
\end{equation}
Thus
\begin{equation}
F_0(u)=\frac{P(u)}{(2u)^d}\,,\qquad F_1(u)=\frac{1}{(2u)^d}\left[\frac{P(u)}u+\frac{Q(u)}{u(u+2)}\right]\,.
\end{equation}
The tensor $\mathcal I$ reduces to the usual inversion tensor in the short-distance limit. Current conservation relates the two coefficient functions. Defining
\begin{equation}
\Pi(u)=P(u)+Q(u)\,,
\end{equation}
we obtain
\begin{equation}
Q(u)=\frac{u(u+2)}d\,\partial_u\Pi(u)\,,\qquad P(u)=\Pi(u)-Q(u)\,.
\end{equation}
Therefore, a single function $\Pi$ determines the conserved parity-even correlator at separated points. The identity block is
\begin{equation}
\mathcal F_{JJ}^{\perp,\mathbf1}=\frac{\mathcal I}{(2u)^d}\,,\qquad \Pi_{\mathbf1}(u)=1\,.
\end{equation}
For a scalar exchanged primary of dimension $\Delta'$, we use the block generated by \cite{Herzog:2017xha}
\begin{equation}
\Pi_{\mathcal O'}(u)=\frac d{\Delta'}(2u)^{\Delta'/2}{}_2F_1\!\left(\frac{\Delta'}2,1+\frac{\Delta'}2;\Delta'-\frac d2+\frac12;-\frac u2\right)\,.
\end{equation}
Together with the conservation relations above, this fixes the normalization of $\mathcal F_{JJ}^{\perp,\mathcal O'}$, and hence of $C_{JJ}^{\perp}{}^{\mathcal O'}$. The identity is defined separately.

Using the transverse harmonics of appendix~\ref{app:harmonic-functions}, we write
\begin{equation}
F_{JJ}^{\perp}(X,Y;W_X,W_Y)=\int_{-\infty}^{+\infty}d\nu\,\widetilde F_{JJ}^{\perp}(\nu)\Omega_\nu^{(1)}(X,Y;W_X,W_Y)\,.
\end{equation}
The convolution identity of the transverse harmonics yields the inverse transform
\begin{equation}
\begin{aligned}
\widetilde F_{JJ}^{\perp}(\nu)={}&\operatorname{Vol}(S^d)\int_0^\infty du\,\frac{[u(u+2)]^{(d-1)/2}}{(2u)^d}\Bigg[\left(\mathsf F_\nu^{(0)}(u)-\frac{u+2}{d+1}\mathsf F_\nu^{(1)}(u)\right)P(u)\\
&\hspace{5cm}-\frac{1}{d+1}\mathsf F_\nu^{(1)}(u)Q(u)\Bigg]\,.
\end{aligned}
\end{equation}

As in the scalar case, the algebraic large-$|\nu|$ expansion can be obtained by expanding the generating function $\Pi(u)$ around $u=0$ and transforming each term. For the power
\begin{equation}
\Pi_q(u)=\left(\frac u2\right)^q\,,
\end{equation}
denote the resulting transverse spectral coefficient by $\widetilde F_q^\perp(\nu)$. In $\mathrm{AdS}_3$, it takes the compact form
\begin{equation}
\widetilde F_q^\perp(\nu)=-\frac{\pi^{3/2}\nu^2}{2}\frac{\Gamma\!\left(q+\frac12\right)}{\Gamma(2-q)}\frac{\Gamma(-q+i\nu)\Gamma(-q-i\nu)}{\Gamma\!\left(\frac12+i\nu\right)\Gamma\!\left(\frac12-i\nu\right)}\,.
\end{equation}
At generic $q$,
\begin{equation}
\widetilde F_q^\perp(\nu)\underset{\rm alg}{\sim}-\frac{\pi^{3/2}}2\frac{\Gamma\!\left(q+\frac12\right)}{\Gamma(2-q)}|\nu|^{1-2q}\left[1+O(\nu^{-2})\right]\,.
\end{equation}
Setting $q=0$ gives the identity transform,
\begin{equation}
\widetilde{\mathcal F}_{JJ}^{\perp,\mathbf1}(\nu)=-\frac{\pi^2}{2}\nu\coth(\pi\nu)\,.
\end{equation}
Consequently, the leading term of the current correlator is
\begin{equation}
\widetilde F_{JJ}^{\perp}(\nu)\sim-\frac{\pi^2}{2}C_J|\nu|\,,
\end{equation}
which determines the two-point function normalization $C_J$. For a scalar exchange, the hypergeometric series gives the prescription
\begin{equation}
\widetilde{\mathcal F}_{JJ}^{\perp,\mathcal O'}(\nu)\underset{\rm alg}{\sim}\frac{2^{\Delta'}d}{\Delta'}\sum_{n=0}^{\infty}\frac{(-1)^n\left(\frac{\Delta'}2\right)_n\left(1+\frac{\Delta'}2\right)_n}{n!\left(\Delta'-\frac d2+\frac12\right)_n}\widetilde F_{\Delta'/2+n}^{\perp}(\nu)\,,
\end{equation}
where the explicit power transform above applies at $d=2$. In particular, its generic leading behavior in $\mathrm{AdS}_3$ is
\begin{equation}
\widetilde{\mathcal F}_{JJ}^{\perp,\mathcal O'}(\nu)\underset{\rm alg}{\sim}-\frac{2^{\Delta'}\pi^{3/2}}{\Delta'}\frac{\Gamma\!\left(\frac{\Delta'+1}{2}\right)}{\Gamma\!\left(2-\frac{\Delta'}2\right)}|\nu|^{1-\Delta'}\,.
\end{equation}

\subsection{Parity-odd spin-one blocks in $\mathrm{AdS}_3$}
\label{subapp:parity-odd-bulk-expansion}
We now specialize to $\mathrm{AdS}_3$, so that $d=2$. The transverse structure discussed in appendix~\ref{subapp:transverse-spin-one-bulk-expansion} describes the parity-even part of the current correlator, and we denote it by $E$ in this subsection. There is also an independent parity-odd structure,
\begin{equation}
F_{JJ}^{O}(X,Y;W_X,W_Y)=\epsilon(X,W_X,Y,W_Y)\,F_2(u)=\epsilon(X,W_X,Y,W_Y)\,\frac{\Pi^{O}(u)}{(2u)^2}\,.
\end{equation}
We use the embedding-space orientation and epsilon contraction defined in section~\ref{subsec:boundary-limits}. This structure is automatically transverse at separated points, without imposing a differential constraint on $\Pi^O$. For a spin-zero primary $\mathcal O'$ of dimension $\Delta'$, we write the parity-odd bulk block as
\begin{equation}
\mathcal F_{JJ}^{O,\mathcal O'}=\frac{\epsilon(X,W_X,Y,W_Y)}{(2u)^2}\Pi_{\mathcal O'}^O(u)\,.
\end{equation}
The bulk Casimir equation, with eigenvalue $\Delta'(\Delta'-3)$, reduces to
\begin{equation}
\left[2u^2(u+2)\frac{d^2}{du^2}+2u(3u+1)\frac{d}{du}+2u-2-\Delta'(\Delta'-3)\right]\Pi_{\mathcal O'}^O(u)=0\,.
\end{equation}
Selecting the solution with $\Pi_{\mathcal O'}^O(u)\sim(2u)^{(\Delta'-1)/2}$ as $u\to0$, we obtain
\begin{equation}
\Pi_{\mathcal O'}^O(u)=(2u)^{(\Delta'-1)/2}{}_2F_1\!\left(\frac{\Delta'+1}{2},\frac{\Delta'+1}{2};\Delta'-\frac12;-\frac u2\right)\,.
\end{equation}
This implies the normalization of the odd OPE coefficient used in section~\ref{sec:conformal-points}. In particular, the contribution of a given scalar primary to the full correlator is
\begin{equation}
a_{\mathcal O'}\left[C_{JJ}^{E}{}^{\mathcal O'}\mathcal F_{JJ}^{E,\mathcal O'}+C_{JJ}^{O}{}^{\mathcal O'}\mathcal F_{JJ}^{O,\mathcal O'}\right]\,.
\end{equation}
The labels $E,O$ refer to the two tensor structures of the $JJ\mathcal O'$ three-point function. They do not correspond to an intrinsic parity of $\mathcal O'$: when parity is broken, the same primary can contribute through both structures. The parity-odd spectral representation reads
\begin{equation}
F_{JJ}^{O}(X,Y;W_X,W_Y)=\int_{-\infty}^{+\infty}d\nu\,\widetilde F_{JJ}^{O}(\nu)\Omega_\nu^O(X,Y;W_X,W_Y)\,.
\end{equation}
The convolution identity of appendix~\ref{app:harmonic-functions} gives the inverse transform
\begin{equation}
\widetilde F_{JJ}^{O}(\nu)=-\frac{4\pi\nu}{3}\int_0^\infty du\,[u(u+2)]^{3/2}{}_2F_1\!\left(2+i\nu,2-i\nu;\frac52;-\frac u2\right)F_2(u)\,.
\end{equation}
For the elementary structure generated by
\begin{equation}
\Pi_q^O(u)=\left(\frac u2\right)^q\,,\qquad F_{2,q}(u)=\frac{1}{(2u)^2}\left(\frac u2\right)^q\,,
\end{equation}
the transform is
\begin{equation}
\widetilde F_q^O(\nu)=\frac{2}{\nu}\widetilde F_q^\perp(\nu)=-\pi^{3/2}\nu\frac{\Gamma\!\left(q+\frac12\right)}{\Gamma(2-q)}\frac{\Gamma(-q+i\nu)\Gamma(-q-i\nu)}{\Gamma\!\left(\frac12+i\nu\right)\Gamma\!\left(\frac12-i\nu\right)}\,.
\end{equation}
Its generic algebraic asymptotics are
\begin{equation}
\widetilde F_q^O(\nu)\underset{\rm alg}{\sim}-\pi^{3/2}\frac{\Gamma\!\left(q+\frac12\right)}{\Gamma(2-q)}\operatorname{sgn}(\nu)|\nu|^{-2q}\left[1+O(\nu^{-2})\right]\,.
\end{equation}
Expanding the block around $u=0$ gives
\begin{equation}
\widetilde{\mathcal F}_{JJ}^{O,\mathcal O'}(\nu)\underset{\rm alg}{\sim}2^{\Delta'-1}\sum_{n=0}^{\infty}\frac{(-1)^n\left(\frac{\Delta'+1}{2}\right)_n^2}{n!\left(\Delta'-\frac12\right)_n}\widetilde F_{\frac{\Delta'-1}{2}+n}^{O}(\nu)\,.
\end{equation}
In particular, for generic $\Delta'$,
\begin{equation}
\widetilde{\mathcal F}_{JJ}^{O,\mathcal O'}(\nu)\underset{\rm alg}{\sim}-2^{\Delta'-1}\pi^{3/2}\frac{\Gamma(\Delta'/2)}{\Gamma\!\left(\frac{5-\Delta'}2\right)}\operatorname{sgn}(\nu)|\nu|^{1-\Delta'}\,.
\end{equation}

\subsection{Perturbative expansion and conformal blocks}
\label{subapp:derived-blocks}
The exact bulk OPE is organized in ordinary conformal blocks. Expanding the conformal data at large $N$, however, also generates derivatives of these blocks. For an exchanged scalar primary with
\begin{equation}
\Delta'=\Delta'_0+\frac{\gamma'}N+O(N^{-2})\,,
\end{equation}
we have
\begin{equation}
f_{\cal O\cal O}^{\cal O'}=\left.f_{\cal O \cal O}^{\cal O'}\right|_{\Delta'=\Delta'_0}+\frac{\gamma'}N\left.\partial_{\Delta'}f_{\cal O \cal O}^{\cal O'}\right|_{\Delta'=\Delta'_0}+O(N^{-2})\,.
\end{equation}
We refer to the dimension derivative as a derived block. It accounts for the perturbative shift of an exchanged dimension; corrections to OPE and one-point coefficients instead multiply the ordinary block.

The order at which a derived block contributes depends on the scaling of the full coefficient. For example, if
\begin{equation}
a_{\mathcal O'}C_{\mathcal O\mathcal O}{}^{\mathcal O'}=N A^{(0)}+A^{(1)}+O(N^{-1})\,,
\end{equation}
then
\begin{equation}
\begin{aligned}
a_{\mathcal O'}C_{\mathcal O\mathcal O}{}^{\mathcal O'}f_{\cal O \cal O}^{\cal O'}={}&\left.N A^{(0)}f_{\cal O \cal O}^{\cal O'}\right|_{\Delta'=\Delta'_0}+A^{(1)}\left.f_{\cal O \cal O}^{\cal O'}\right|_{\Delta'=\Delta'_0}\\
&+A^{(0)}\gamma'\left.\partial_{\Delta'}f_{\cal O \cal O}^{\cal O'}\right|_{\Delta'=\Delta'_0}+O(N^{-1})\,.
\end{aligned}
\end{equation}
An $O(1/N)$ anomalous dimension can thus contribute at order one when the one-point function contains products of $O(\sqrt N)$ condensates. If several primaries share the leading dimension $\Delta'_0$, the derived-block coefficient is instead the weighted sum $\sum_i A_i^{(0)}\gamma_i$, evaluated in a basis that diagonalizes the dilatation operator. Extracting individual anomalous dimensions then requires additional information about the contributing operators and their mixing.

In spectral space, the algebraic transform of a derived block follows by differentiating the ordinary-block expansion. For identical external scalars of dimension $\Delta$, the leading term from appendix~\ref{subapp:scalar-bulk-expansion} gives, at generic dimensions,
\begin{equation}
\begin{aligned}
\partial_{\Delta'}\widetilde f_{\mathcal O\mathcal O}^{\mathcal O'}(\nu)\underset{\rm alg}{\sim}{}&\frac{2^{\Delta'}(4\pi)^{(d+1)/2}}{4^\Delta}\frac{\Gamma\!\left(\frac{d+1}{2}-\Delta+\frac{\Delta'}2\right)}{\Gamma\!\left(\Delta-\frac{\Delta'}2\right)}|\nu|^{2\Delta-\Delta'-d-1}\\
&\times\left[-\log|\nu|+\log2+\frac12\psi\!\left(\frac{d+1}{2}-\Delta+\frac{\Delta'}2\right)+\frac12\psi\!\left(\Delta-\frac{\Delta'}2\right)\right]\,.
\end{aligned}
\end{equation}
Here $\psi$ is the digamma function, and the derivative is taken at fixed external dimension. Subleading terms follow by differentiating the full algebraic series, including the descendant contributions. At the special dimensions $\Delta'_0=2\Delta+2n$, $n\geq0$, the ordinary scalar block has vanishing algebraic transform. Its dimension derivative nevertheless survives. Taking the limit of the complete expression above yields
\begin{equation}
\left.\partial_{\Delta'}\widetilde f_{\mathcal O\mathcal O}^{\mathcal O'}(\nu)\right|_{\Delta'=2\Delta+2n}\underset{\rm alg}{\sim}(-1)^{n+1}2^{2n-1}(4\pi)^{(d+1)/2}n!\,\Gamma\!\left(\frac{d+1}{2}+n\right)|\nu|^{-d-1-2n}\,.
\end{equation}
The logarithmic leading term disappears because its coefficient vanishes at these dimensions. Thus a conformal family can first become visible in the algebraic spectral expansion through its anomalous dimension, while its ordinary block remains present in position space.

\subsection{Extracting bulk data at conformal points}
\label{subapp:extracting-bulk-data}
We now apply the preceding formulas to the correlators of section~\ref{sec:conformal-points}. We display the terms determining the bulk normalizations and leading scalar exchanges, together with a representative derived-block contribution. All spectral expansions below refer to large real $|\nu|$, with contact terms treated separately. The scalar block transforms involving the identity, the dimension-one scalar $\boldsymbol\sigma$, and the dimension-two scalar $\boldsymbol{\chi}$ that we will need satisfy
\begin{equation}\label{eq:scalar-block-transforms}
\widetilde f_{\sigma\sigma}^{\mathbf1}(\nu)\underset{\rm alg}{\sim}\frac{2\pi^2}{|\nu|}\,,\qquad \widetilde f_{\chi\chi}^{\mathbf1}(\nu)\underset{\rm alg}{\sim}-\pi^2|\nu|\,,\qquad \widetilde f_{\sigma\sigma}^{\sigma}(\nu)\underset{\rm alg}{\sim}\frac{4\pi}{\nu^2}\,.
\end{equation}
For the current correlators, the leading terms in all the theories considered here are
\begin{equation}
\begin{aligned}
\widetilde G_E^{\rm top}(\nu)&\sim-\frac{\pi^2}{2}C_{J^{\rm top}}(\kappa)|\nu|+\cdots\,,\\
\widetilde G_O^{\rm top}(\nu)&\sim-4\pi i\kappa C_{J^{\rm top}}(\kappa)\nu+\cdots\,.
\end{aligned}
\end{equation}
The even term fixes the current normalization \eqref{eq:CJtop}, while the displayed odd term is a contact contribution. The nonlocal subleading terms determine scalar exchanges through the blocks of appendices~\ref{subapp:transverse-spin-one-bulk-expansion} and~\ref{subapp:parity-odd-bulk-expansion}. For the exchanged scalars of dimension one ($\boldsymbol\sigma$) and two ($\boldsymbol\chi$), their leading transforms are
\begin{equation}\label{eq:JJ-block-transforms}
\widetilde{\mathcal F}_{JJ}^{E,\sigma}\underset{\rm alg}{\sim}-4\pi\,,\qquad \widetilde{\mathcal F}_{JJ}^{O,\sigma}\underset{\rm alg}{\sim}-\pi^2\operatorname{sgn}\nu\,,\qquad \widetilde{\mathcal F}_{JJ}^{E,\chi}\underset{\rm alg}{\sim}-\frac{\pi^2}{|\nu|}\,,\qquad \widetilde{\mathcal F}_{JJ}^{O,\chi}\underset{\rm alg}{\sim}-4\pi\frac{\operatorname{sgn}\nu}{|\nu|}\,.
\end{equation}

\paragraph{Critical scalar QED$_3$.} The ordinary and normal scalar singlet propagators coincide,
\begin{equation}
\widetilde G_{\chi\chi}(\nu)=-8\nu\coth\frac{\pi\nu}{2}\underset{\rm alg}{\sim}-8|\nu|\,.
\end{equation}
Matching the identity gives \eqref{eq:crit-bulk-data}. There are no further algebraic terms in this correlator; its remaining bulk-channel information is invisible in the algebraic expansion.

The current correlator detects the $\chi$ exchange. For ordinary boundary conditions, the first nonlocal corrections are
\begin{equation}
\begin{aligned}
\widetilde G_E^{\rm top}(\nu)&\underset{\rm alg}{\sim}-\frac{\pi^2}{2}C_{J^{\rm top}}(\kappa)|\nu|-\frac{\pi^2C_{J^{\rm top}}(\kappa)}{2|\nu|}\frac{1-(8\kappa/\pi)^2}{1+(8\kappa/\pi)^2}\,,\\
\widetilde G_O^{\rm top}(\nu)&\underset{\rm alg}{\sim}-4\pi i\kappa C_{J^{\rm top}}(\kappa)\nu-\frac{8\pi i\kappa C_{J^{\rm top}}(\kappa)}{1+(8\kappa/\pi)^2}\frac{\operatorname{sgn}\nu}{|\nu|}\,.
\end{aligned}
\end{equation}
The leading even term corresponds to the identity, while the leading odd term is a contact contribution. The subleading terms correspond to dimension-two blocks. Using \eqref{eq:JJ-block-transforms},
\begin{equation}
\begin{aligned}
a_\chi^{\rm ord}C^{E,{\rm crit}}_{J^{\rm top}J^{\rm top}}{}^{\chi}&=\frac{C_{J^{\rm top}}(\kappa)}2\frac{1-(8\kappa/\pi)^2}{1+(8\kappa/\pi)^2}\,,\\
a_\chi^{\rm ord}C^{O,{\rm crit}}_{J^{\rm top}J^{\rm top}}{}^{\chi}&=\frac{2i\kappa C_{J^{\rm top}}(\kappa)}{1+(8\kappa/\pi)^2}\,.
\end{aligned}
\end{equation}
Both products are multiplied by $-3$ for normal boundary conditions. Since $a_\chi^{\rm norm}=-3a_\chi^{\rm ord}$, the extracted bulk OPE coefficients agree. After subtracting the complete $\boldsymbol{\chi}$ blocks, the current remainder generically begins at order $|\nu|^{-3}$. Its successive terms admit an organization into derived even blocks and ordinary odd blocks at leading dimensions $\Delta'=4,6,\dots$. These dimensions allow contributions from both scalar composites and current double traces.

\paragraph{Tricritical scalar QED$_3$.} For all three boundary conditions, the leading scalar expansion can be written uniformly as
\begin{equation}
\widetilde G_{\sigma\sigma}(\nu)\underset{\rm alg}{\sim}\frac1{8|\nu|}+\frac{2\Sigma^{\rm AdS}}{\nu^2}\left(1-\frac h{64}\right)+\frac{h(h-128)(\Sigma^{\rm AdS})^2}{128|\nu|^3}+\cdots\,,
\end{equation}
with $\Sigma^{\rm AdS}$ the value of $\Sigma$ on the corresponding saddle. The identity term gives \eqref{eq:tri-bulk-data}. Using \eqref{eq:scalar-block-transforms}, we further find
\begin{equation}
a_\sigma C^{{\rm tri}}_{\sigma\sigma}{}^\sigma=\frac{\Sigma^{\rm AdS}}{2\pi}\left(1-\frac h{64}\right)\,.
\end{equation}
The $|\nu|^{-3}$ term supplies a simple example of the mechanism discussed in appendix~\ref{subapp:derived-blocks}. Neither the identity nor the dimension-one exchange contributes at this power, while
\begin{equation}
\left.\partial_{\Delta'}\widetilde f_{\sigma\sigma}^{\cal O'}(\nu)\right|_{\Delta'=2}\underset{\rm alg}{\sim}-\frac{2\pi^2}{|\nu|^3}\,.
\end{equation}
The coefficient of this derived block is proportional to the squared condensate, which is consistent with the enhanced one-point function of a scalar composite. Subsequent terms can be organized into ordinary blocks at dimensions $\Delta'=3,5,\dots$ and derived blocks at dimensions $\Delta'=4,6,\dots$.

For the current correlator, using \eqref{eq:JJ-block-transforms} and matching the corresponding terms for ordinary and special boundary conditions gives
\begin{equation}
\begin{aligned}
a_\sigma C^{E,{\rm tri}}_{J^{\rm top}J^{\rm top}}{}^{\sigma}&=\beta C_{J^{\rm top}}(\kappa)\frac{1-(8\kappa/\pi)^2}{1+(8\kappa/\pi)^2}\,,\\
a_\sigma C^{O,{\rm tri}}_{J^{\rm top}J^{\rm top}}{}^{\sigma}&=\frac{64i\kappa\beta}{\pi^2}\frac{C_{J^{\rm top}}(\kappa)}{1+(8\kappa/\pi)^2}\,.
\end{aligned}
\end{equation}
The normal boundary condition correlator gives the same bulk OPE coefficients, once we account for the different one-point coefficients. The remaining current expansion includes ordinary exchanges at leading dimensions two and three. At higher dimensions, the even structure admits derived blocks at $\Delta'=4,6,\dots$ and ordinary blocks at $\Delta'=5,7,\dots$; the odd structure admits the mirrored pattern. Their coefficients can receive contributions from several degenerate composite families.

\paragraph{Critical fermionic theories.} In the regular theory,
\begin{equation}
\widetilde G_{\chi\chi}(\nu)=-\frac{\nu}{8}\coth\frac{\pi\nu}{2}\underset{\rm alg}{\sim}-\frac{|\nu|}{8}\,,
\end{equation}
which gives \eqref{eq:reg-bulk-data}. As in the critical scalar singlet correlator, the algebraic expansion contains only the identity contribution.

The first nonlocal terms of the current correlator are
\begin{equation}
\begin{aligned}
\widetilde G_E^{\rm top}(\nu)&\underset{\rm alg}{\sim}-\frac{\pi^2}{2}C_{J^{\rm top}}(\kappa)|\nu|+32\eta\kappa\,C_{J^{\rm top}}(\kappa)\frac{1}{1+\left(\frac{8\kappa}{\pi}\right)^2}\frac1{|\nu|}\,,\\
\widetilde G_O^{\rm top}(\nu)&\underset{\rm alg}{\sim}-4\pi i\kappa C_{J^{\rm top}}(\kappa)\nu-2\pi i\eta\,C_{J^{\rm top}}(\kappa)\frac{1-\left(\frac{8\kappa}{\pi}\right)^2}{1+\left(\frac{8\kappa}{\pi}\right)^2}\frac{\operatorname{sgn}\nu}{|\nu|}\,.
\end{aligned}
\end{equation}
Matching the two structures with \eqref{eq:JJ-block-transforms} gives
\begin{equation}
\begin{aligned}
a_\chi^{\rm reg}C^{E,{\rm reg}}_{J^{\rm top}J^{\rm top}}{}^\chi&=-\frac{32\eta\kappa}{\pi^2}\,C_{J^{\rm top}}(\kappa)\frac{1}{1+\left(\frac{8\kappa}{\pi}\right)^2}\,,\\
a_\chi^{\rm reg}C^{O,{\rm reg}}_{J^{\rm top}J^{\rm top}}{}^\chi&=\frac{i\eta}{2}\,C_{J^{\rm top}}(\kappa)\frac{1-\left(\frac{8\kappa}{\pi}\right)^2}{1+\left(\frac{8\kappa}{\pi}\right)^2}\,.
\end{aligned}
\end{equation}
After subtracting the complete $\boldsymbol{\chi}$ block, the higher terms admit derived even blocks and ordinary odd blocks at dimensions $\Delta'=4,6,\dots$.

For the Gross--Neveu family, the algebraic expansions of the bubbles are polynomial in $\hat\Delta_\psi-1$, so the expansions below hold for both the standard and the alternate saddles, with $\calM_\eta\to\calM_{\eta,-}$ in the latter. The scalar expansion is
\begin{equation}
\widetilde G_{\sigma\sigma}(\nu)\underset{\rm alg}{\sim}\frac8{|\nu|}-\frac{128y\calM_\eta}{\nu^2}-\frac{32\calM_\eta^2(1-64y^2)}{|\nu|^3}+\cdots\,.
\end{equation}
The first term gives \eqref{eq:GN-bulk-data}, and the second, using \eqref{eq:scalar-block-transforms}, gives
\begin{equation}
a_\sigma^{\rm GN} C^{{\rm GN}}_{\sigma\sigma}{}^\sigma=-\frac{32y\calM_\eta}{\pi}\,.
\end{equation}
The higher terms starting from $|\nu|^{-3}$ follow the same pattern of decomposition in blocks as in the tricritical scalar family.

For the current correlator, the first subleading terms are
\begin{equation}
\begin{aligned}
\widetilde G_E^{\rm top}(\nu)&\underset{\rm alg}{\sim}-\frac{\pi^2}{2}C_{J^{\rm top}}(\kappa)|\nu|+32\pi\kappa\calM_\eta\,C_{J^{\rm top}}(\kappa)\frac{1}{1+\left(\frac{8\kappa}{\pi}\right)^2}\,,\\
\widetilde G_O^{\rm top}(\nu)&\underset{\rm alg}{\sim}-4\pi i\kappa C_{J^{\rm top}}(\kappa)\nu-2\pi^2 i\calM_\eta\,C_{J^{\rm top}}(\kappa)\frac{1-\left(\frac{8\kappa}{\pi}\right)^2}{1+\left(\frac{8\kappa}{\pi}\right)^2}\operatorname{sgn}\nu\,.
\end{aligned}
\end{equation}
Using \eqref{eq:JJ-block-transforms}, we obtain
\begin{equation}
\begin{aligned}
a_\sigma^{\rm GN}C^{E,{\rm GN}}_{J^{\rm top}J^{\rm top}}{}^\sigma&=-8\kappa\calM_\eta\,C_{J^{\rm top}}(\kappa)\frac{1}{1+\left(\frac{8\kappa}{\pi}\right)^2}\,,\\
a_\sigma^{\rm GN}C^{O,{\rm GN}}_{J^{\rm top}J^{\rm top}}{}^\sigma&=2i\calM_\eta\,C_{J^{\rm top}}(\kappa)\frac{1-\left(\frac{8\kappa}{\pi}\right)^2}{1+\left(\frac{8\kappa}{\pi}\right)^2}\,.
\end{aligned}
\end{equation}
Higher terms can contain contributions from several degenerate composite families and determine only the corresponding weighted OPE data.

\bibliographystyle{JHEP}
\bibliography{biblio}
\end{document}